\documentclass[useAMS,usenatbib,usegraphicx,usedcolumn]{mn2e}
\usepackage{txfonts}
\usepackage{lscape}

\newcommand{\ion}[2]{\ensuremath{\mbox{#1~{\sc #2}}}}
\newcommand{\pedix}[2]{\ensuremath{#1_{\,\mbox{\scriptsize #2}}}}
\newcommand{\apix}[2]{\ensuremath{#1^{\,\mbox{\scriptsize #2}}}}
\newcommand{\pedap}[3]{\ensuremath{#1_{\,\mbox{\scriptsize #2}}^{\,\mbox{\scriptsize #3}}}}
\newcommand{\pedSM}[2]{\ensuremath{#1_{\,\mbox{\tiny #2}}}}

\newcommand{\pedapSM}[3]{\ensuremath{#1_{\,\mbox{\tiny #2}}^{\,\mbox{\tiny #3}}}}
\newcommand{\errUD}[2]{\ensuremath{^{+#1}_{-#2}}}
\newcommand{\ie}{{i.e.}}
\newcommand{\eg}{{e.g.}}
\newcommand{\myast}{\ensuremath{\ast}}
\newcommand{\mydag}{\ensuremath{\mbox{\dag}}}
\newcommand{\myddag}{\ensuremath{\mbox{\ddag}}}
\newcommand{\hb}{\ensuremath{\mbox{H}\beta}}

\newcommand{\nev}{\ensuremath{\mbox{[\ion{Ne}{v}]}}}
\newcommand{\feka}{\ensuremath{\mbox{Fe~K}\alpha}}
\newcommand{\fekb}{\ensuremath{\mbox{Fe~K}\beta}}
\newcommand{\nika}{\ensuremath{\mbox{Ni~K}\alpha}}
\newcommand{\xmm}{{XMM-\emph{Newton}}}
\newcommand{\galex}{{\sc GALEX}}
\newcommand{\sdss}{{\emph{SDSS}}}
\newcommand{\spitzer}{{\emph{Spitzer}}}
\newcommand{\lum}{\ensuremath{\mbox{ergs~s}^{-1}}}
\newcommand{\flux}{\ensuremath{\mbox{ergs~cm}^{-2}\mbox{~s}^{-1}}}
\newcommand{\fluxA}{\ensuremath{\mbox{ergs~cm}^{-2}\mbox{~s}^{-1}\mbox{~\AA}^{-1}}}
\newcommand{\fluxHz}{\ensuremath{\mbox{ergs~cm}^{-2}\mbox{~s}^{-1}\mbox{~Hz}^{-1}}}
\newcommand{\nh}{\ensuremath{\mbox{cm}^{-2}}}
\newcommand{\nhsym}{\ensuremath{N_{\mbox{\scriptsize H}}}}
\newcommand{\kev}{\ensuremath{\,\mbox{\scriptsize keV}}}
\newcommand{\mum}{\ensuremath{\,\mu\mbox{\scriptsize m}}}
\newcommand{\arcdeg}{\ensuremath{^{\circ}}}
\renewcommand{\arcmin}{\ensuremath{^{\prime}}}
\renewcommand{\arcsec}{\ensuremath{^{\prime\prime}}}
\newcommand{\chidof}{\ensuremath{\chi^2/\mbox{d.o.f.}}}
\newcommand{\dchidof}{\ensuremath{\Delta\chi^2/\Delta\mbox{d.o.f.}}}
\newcommand{\normGAUSS}{\ensuremath{\mbox{photons~cm}^{-2}\mbox{~s}^{-1}}}
\newcommand{\llqsoa}{XBSJ052128.9-253032}
\newcommand{\llqsob}{XBSJ080411.3+650906}
\newcommand{\nsrc}{\ensuremath{14}}

\newcounter{numtab}
\newcounter{numfig}
\newcounter{numfigsed}
\newcounter{numtabdata}
\title[The XBS XQSO2s: X-ray and accretion properties]{The \xmm\ Bright Survey sample of absorbed quasars: X-ray and accretion properties}
\author[L.~Ballo et al.]{
  L.~Ballo$^{1}$\thanks{E-mail: lucia.ballo@brera.inaf.it (LB)},
  P.~Severgnini$^{1}$,
  R.~Della~Ceca$^{1}$,
  A.~Caccianiga$^{1}$,
  C.~Vignali$^{2,3}$,
  F.J.~Carrera$^{4}$,
  \newauthor 
  A.~Corral$^{5}$,
  and
  S.~Mateos$^{4}$\\
  $^{1}$Osservatorio Astronomico di Brera (INAF), via Brera 28, I-20121, Milano (Italy) \\
  $^{2}$Dipartimento di Fisica e Astronomia, Universit\`a degli Studi di Bologna,
 viale Berti Pichat 6/2, I-40127, Bologna (Italy) \\
  $^{3}$Osservatorio Astronomico di Bologna (INAF), Via Ranzani 1, I-40127, Bologna (Italy) \\
  $^{4}$Instituto de F\'\i{}sica de Cantabria (CSIC-UC), Avenida de los Castros, E-39005 Santander (Spain) \\
  $^{5}$Institute for Astronomy, Astrophysics, Space Applications \& Remote
  Sensing, National Observatory of Athens, Lofos Nymfon, Thiseio, P.O. Box 20048 \\
  GR-11810 Athens (Greece)
}

\begin{document}

\date{Accepted 2014 August 7.  Received 2014 June 30; in original form 2014 April 24}

\pagerange{\pageref{firstpage}--\pageref{lastpage}} \pubyear{2014}

\maketitle

\label{firstpage}

\begin{abstract}
Although absorbed quasars are extremely important 
for our understanding of the energetics of the Universe, 
the main physical parameters of their central engines
are still poorly known.
In this work we present and study
a complete sample of \nsrc\ quasars (QSOs) that are absorbed in the X-rays
(column density $\nhsym > 4\times10^{21}\,$\nh\ and X-ray luminosity $\pedix{L}{2-10\kev} > 10^{44}\,$\lum; XQSO2)
belonging to the \xmm\ Bright Serendipitous Survey (XBS).
From the analysis of their ultraviolet-to-mid-infrared spectral energy distribution
we can separate the nuclear emission from the host galaxy contribution, obtaining a
measurement of the fundamental nuclear parameters, like
the mass of the central supermassive black hole and the value of Eddington ratio, \pedix{\lambda}{Edd}.

Comparing the properties of XQSO2s with 
those previously obtained for the X-ray unabsorbed QSOs in the XBS,
we do not find any evidence that the two samples are drawn from different populations.
In particular, the two samples span the same range in Eddington ratios, up to $\pedix{\lambda}{Edd}\sim0.5$; this implies that 
our XQSO2s populate the ``forbidden region'' in the so-called ``effective Eddington limit paradigm''.
A combination of low grain abundance,
presence of stars inwards of the absorber,
and/or anisotropy of the disk emission, can explain this result. 
\end{abstract}


\section{Introduction}\label{sect:intro}

Active Galactic Nuclei (AGN) are 
believed to be powered by accretion of matter onto
a Supermassive Black Hole \citep[SMBH, $\pedix{M}{BH}>10^{6}\,\pedix{M}{\sun}$;][]{salpeter64,lyndenbell69}. 
The 
discovery of quiescent ``non-active'' SMBHs
in the nuclei of nearby bulge galaxies, along with the presence of scaling relations between
the central black hole (BH) mass and galaxy properties \citep[\eg, bulge luminosity/mass and velocity
dispersion; among others,][]{kormendy95,ferrarese00,marconi03,ferrarese06,cattaneo09}
clearly indicate that AGN are fundamental actors in the formation and evolution of galaxies
and, in general, of cosmic structures in the Universe 
\citep[see][for two recent reviews on different aspects of the SMBH-host galaxy coevolution]{kormendy13,heckman14}. 
Essential points to study the galaxy evolution 
are the identification of AGN in all states of accretion and obscuration and an accurate assessment of their physical properties,
in order to infer their influence (``feedback'') on the evolution of the host galaxy.

The ``first order'' model proposed to explain their characteristics, known as the Unified Model
\citep{antonucci93}, assumes that all the AGN have a similar internal structure.
The basic idea is that most of the observed differences are caused by the angle between our
line of sight and the plane of the system.
In particular, the toroidal obscuring material supposed to partially cover the nuclear engine
along particular lines of sight 
regulates the possibility of observing in the optical spectrum both broad and narrow (optically unobscured, or type~1, AGN)
or only narrow (optically obscured, or type~2, AGN) emission lines.
Confirmations of this schematic picture come from spectropolarimetric studies of optically obscured
AGN, and from X-ray observations \citep[for a recent review, see][]{bianchi12}.

However, several authors suggested a possible evolution of the obscuration with cosmic time, connected with the SMBH growth
(\citealt{hopkins06}; \citealt{lapi14}, recently updating the original model presented by \citealt{granato04})
in addition to the effects of different viewing angles. 
The main phase of efficient accretion would be characterized by AGN heavily obscured by gas and dust; this obscuring matter would 
be then swept away by the nuclear radiation, and the accreting SMBH would shine as an optically luminous quasar (QSO) 
for a brief period, until it runs out of gas. 
The final product of the sequence is a ``dead QSO'' inside a passively evolving galaxy. 
In this 
scenario, unobscured AGN are a subsequent phase of obscured QSO. 

In the framework of the Unified Model, we expect that absorbed and unabsorbed accreting nuclei share 
the same intrinsic properties (\eg, luminosity, BH mass, and accretion rate).
Conversely, an evolutionary pattern for the two classes of AGN would imply a difference in their physical properties.
Recently, a possible link between accretion and obscuration has been presented by \citet{fabian08} and \citet{fabian09}. 
By discussing the effects of accretion driven feedback on gas clouds in the innermost region of
an AGN, the authors proposed that dusty absorbing structures around an AGN can be stable only 
for column densities $\nhsym \gtrsim 5 \times 10^{23}\pedix{\lambda}{Edd}\,$[\nh], 
where \pedix{\lambda}{Edd} is the AGN Eddington ratio, defined as the ratio between the bolometric luminosity and 
the Eddington luminosity\footnote{The Eddington luminosity,
defined as $\pedix{L}{Edd} = 4\pi G\pedix{M}{BH}\pedix{m}{p}c/\pedix{\sigma}{T}\simeq1.3 \times 10^{38} \pedix{M}{BH}/\pedix{M}{\sun}\,$[\lum], 
represents the exact balance between inward gravitational force and
outward radiation force acting on the gas, assumed to be of ionized hydrogen in a spherical configuration.}.

\begin{figure*}
 \centering
 \resizebox{0.33\hsize}{!}{\includegraphics{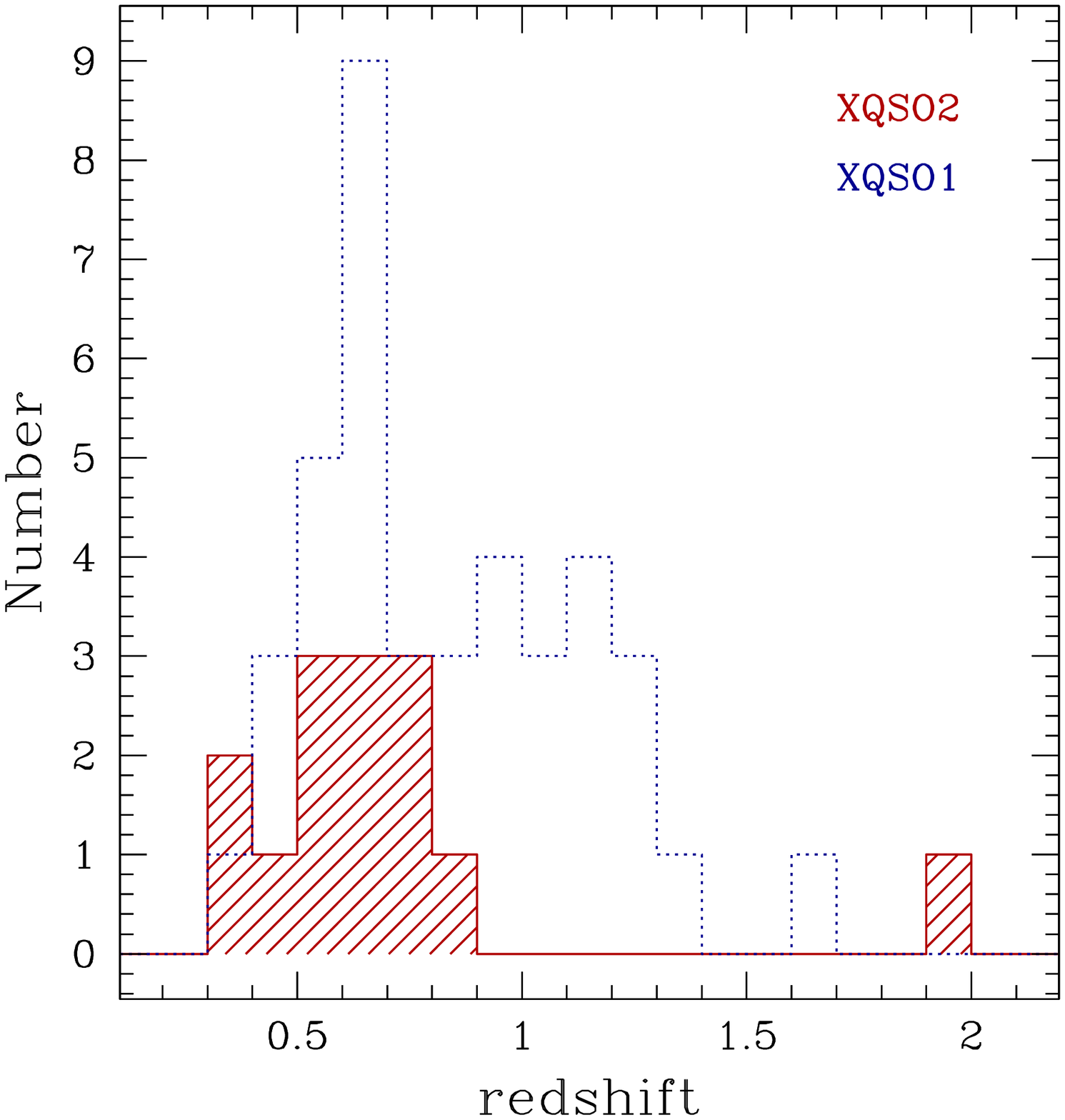}}
 \resizebox{0.33\hsize}{!}{\includegraphics{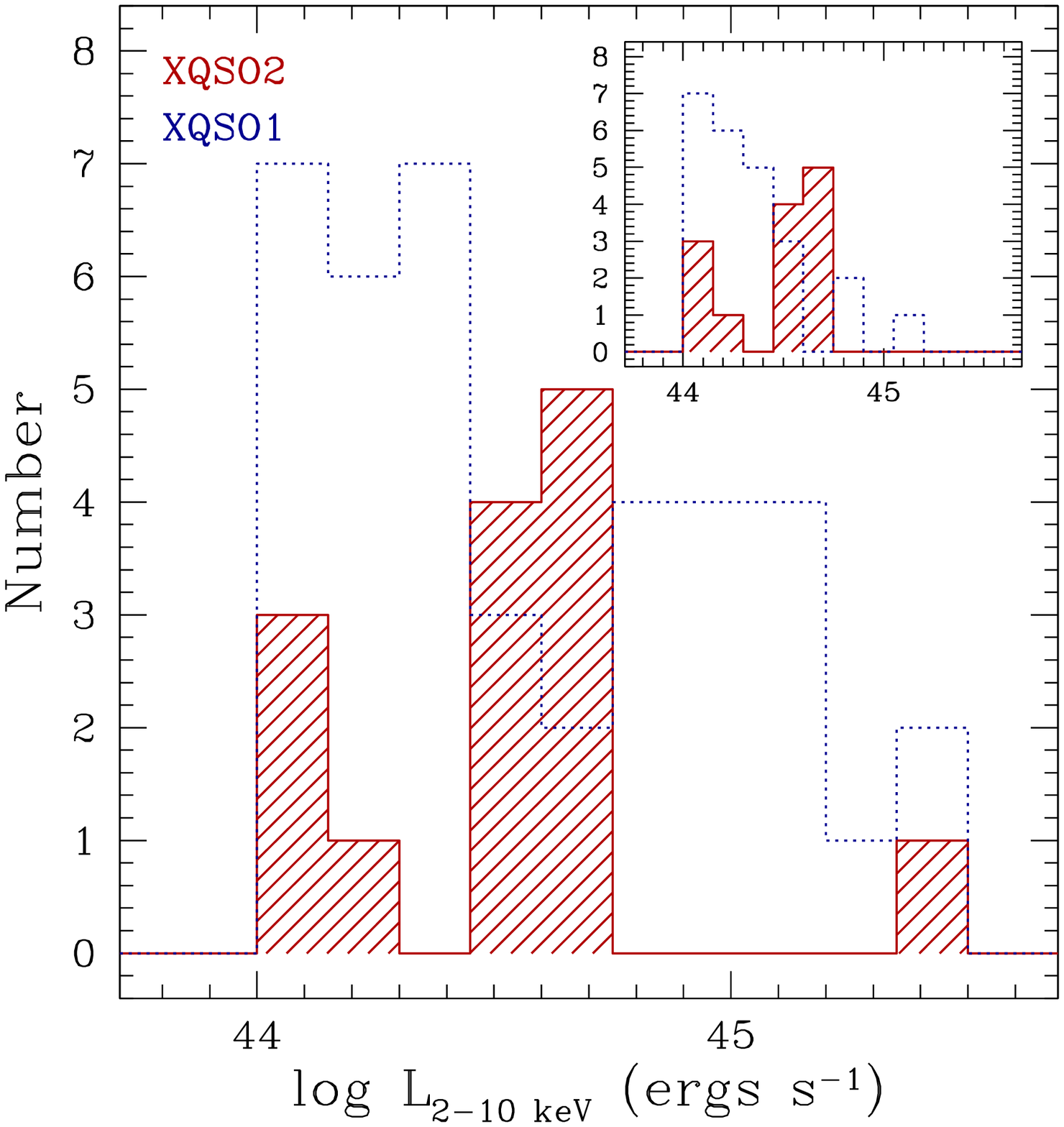}}
 \resizebox{0.33\hsize}{!}{\includegraphics{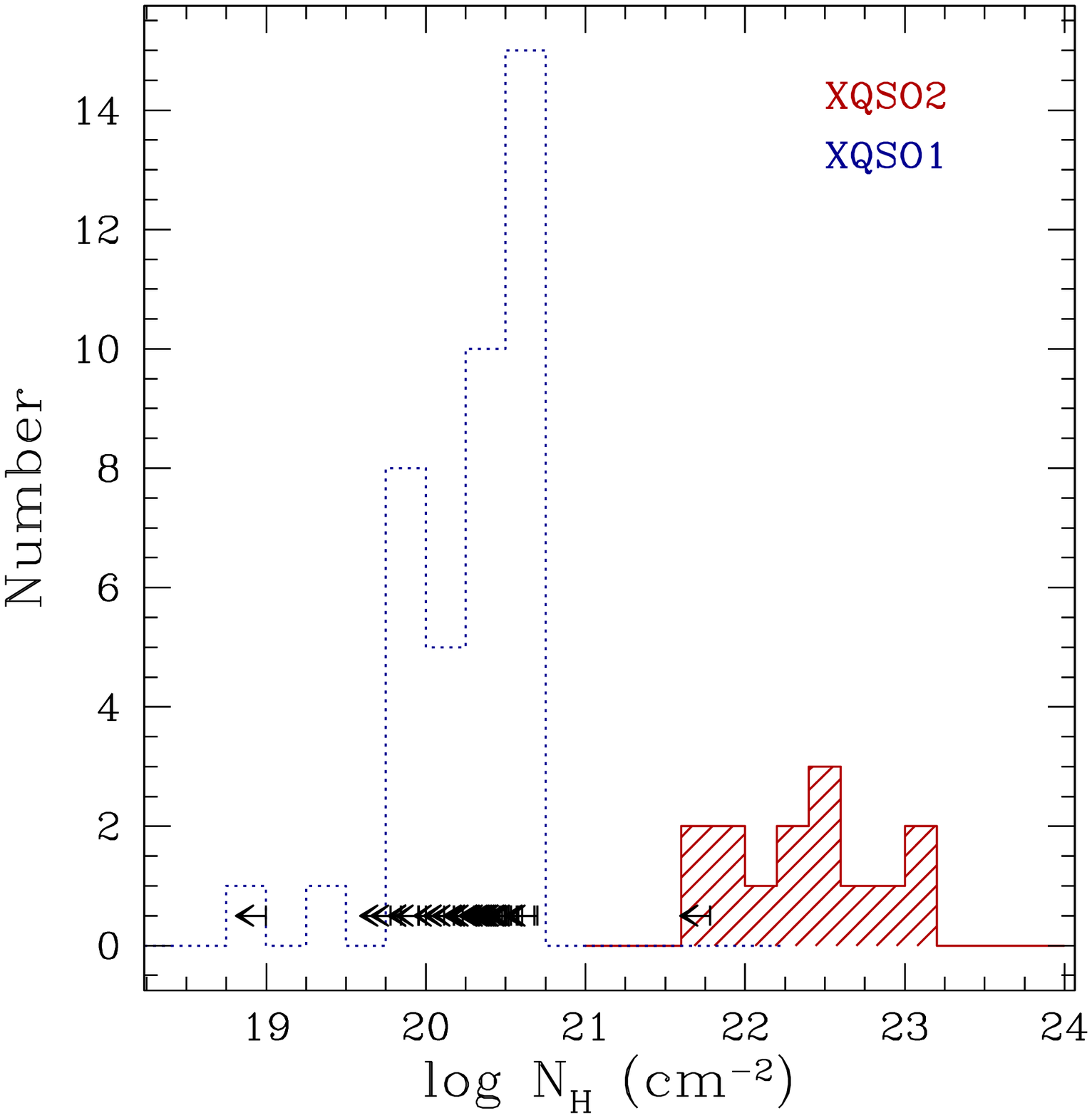}}

 \caption{Distribution in redshift (left-hand panel), intrinsic X-ray luminosity (central panel) and column density (right-hand panel) of the XQSO2s in the XBS (red shaded 
 histogram); the blue dotted lines 
 show the distributions for the XQSO1s 
 (see Sect.~\ref{sect:nucl}).
 In the inset (central panel), we show the distributions of \pedix{L}{2-10\kev} obtained when only sources with $z<0.85$ are considered 
 (see Sect.~\ref{sect:xqso12}).
 The right-pointing arrows in the right-hand panel represent upper limits to \nhsym.}
 \label{fig:znhdist}%
\end{figure*}
\addtocounter{numfig}{1}
\addtocounter{numfigsed}{1}

The assessment of the nuclear properties of absorbed AGN is an issue that becomes more and
more difficult as the amount of circumnuclear obscuring medium along the line of sight
increases and does not permit a clear and direct view to the central energy
source.
In addition, at low nuclear luminosities a host contribution at optical-infrared (IR) wavelengths
is expected to be increasingly noticeable, making even harder to isolate the nuclear component.

X-ray surveys are one of the best approaches to search for obscured AGN, since in this band the emission 
is less affected by absorption, if compared with the ultraviolet (UV)-optical domain.
In this paper we discuss the broad-band physical properties of all the 
X-ray absorbed QSOs 
($\nhsym > 4\times 10^{21}\,$\nh, intrinsic $\pedix{L}{2-10\kev} > 10^{44}\,$\lum; hereafter, XQSO2) in the 
\xmm\ Bright Serendipitous Survey\footnote{The \xmm\ Bright Serendipitous
Survey is a programme conducted by the \xmm\ Survey Science Center (SSC, see
http://xmmssc-www.star.le.ac.uk), a consortium of $10$ institutes in the ESA
community; a full description of the role of the SSC is given in
\citet{watson01}.} (XBS), a complete sample of $400$ bright X-ray
sources.
The XQSO2 sample considered here is unique at the
moment: it consists of \nsrc\ X-ray and optically bright sources at intermediate redshift ($0.35<z<0.9$, plus one object at $z\sim2$), belonging to a complete and well-defined
survey, for which multi-wavelength data
have been already collected.
Proprietary or archival photometry from \spitzer, the Wide-field Infrared Survey Explorer \citep[WISE;][]{wise}, the Sloan Digital Sky Surveys 
\citep[\sdss;][]{abazajian09} and the Digitized Sky Survey (DSS), and good quality optical 
spectra, in addition to X-ray spectroscopic observations from \xmm, allow us to properly characterize
the sources in our sample.
The major goal of this paper is to estimate the nuclear properties of the XQSO2s, in particular
black hole masses and Eddington ratios, and to compare them with those of 
the sample of X-ray unabsorbed QSOs in the XBS.

The paper is organized as follows: in Sect.~\ref{sect:xbs} we briefly describe the XBS and the XQSO2 sub-sample.
Sect.~\ref{sect:data} reports on the multiwavelength data used to construct the UV-to-IR Spectral Energy Distribution (SED) for the XQSO2s studied here.
We present the method adopted to deconvolve the host and AGN contributions in Sect.~\ref{sect:sed}, while the details on the recovering of the nuclear 
properties are described in Sect.~\ref{sect:nucl}.
We discuss the results obtained and their consequences, and compare the properties recovered for absorbed and unabsorbed sources, in Sect.~\ref{sect:disc}.
Finally, the main results are summarized in
Sect.~\ref{sect:summ}.
The analysis of new \xmm\ observations is presented in Appendix~\ref{sect:xmm}, while details on the UV-to-mid-IR properties of the 
individual sources are reported in Appendix~\ref{sect:note}.
Throughout this paper, for consistency with the previous works on the XBS sample,
a concordance cosmology with $\pedix{H}{0}=65\;$km s$^{-1}$ Mpc$^{-1}$,
$\pedix{\Omega}{\ensuremath{\Lambda}}=0.7$, and 
$\pedix{\Omega}{m}=0.3$ is adopted; when necessary, data and relations taken from the literature have been converted to our 
cosmology.
The energy spectral index, $\alpha$, is defined such that
$\pedix{F}{\ensuremath{\nu}} \propto \apix{\nu}{-\ensuremath{\alpha}}$. 
The X-ray photon index, defined as 
$\pedix{N}{\ensuremath{\epsilon}}\propto\apix{\epsilon}{-\ensuremath{\Gamma}}$, 
is $\Gamma=\alpha+1$.
Finally, X-ray surveys demonstrated the existence of a small number of AGN 
showing a mismatch between the optical and X-ray based methods of classification in terms of obscuration.
In this paper, we define a source ``absorbed'' or ``unabsorbed'' on the basis of its X-ray properties (irrespective of the optical properties), 
adopting as threshold\footnote{For a Galactic dust-to-gas ratio \citep{bohlin78}, 
the adopted threshold corresponds to a dust extinction of $\pedix{A}{V}\sim2\,$mag, which corresponds to the adopted dividing line 
between optically absorbed and unabsorbed sources \citep{caccianiga08}.} an intrinsic absorbing column density $\nhsym = 4\times 10^{21}\,$\nh.


\section{The XBS sample of X-ray absorbed QSOs}\label{sect:xbs}

The XBS consists of two flux-limited samples of
X-ray selected serendipitous sources at high Galactic latitude ($|b|>20\arcdeg$; total sky area $\sim28\,$sq. deg): 
the XMM Bright Serendipitous Survey
sample (BSS; $389$ sources selected in the $0.5-4.5\,$keV energy band, identification level of $98$\%) 
and the XMM Hard Bright Serendipitous Survey sample (HBSS; $67$ sources selected in the $4.5-7.5\,$keV energy band, identification 
level of $100$\%; $56$ sources are in common with
the BSS sample).
The flux limit is $\sim 7\times10^{-14}\,$\flux\ in both the selection energy bands.

%
\begin{table*}
\begin{minipage}[t]{1\textwidth}
 \caption{Basic information for the XQSO2s found in the XBS and X-ray spectral results from the observations used in the XBS catalogue. }
 \label{tab:xbs}
 \begin{center}
\resizebox{1\textwidth}{3cm} 
{
 \footnotesize
  \begin{tabular}{r@{\extracolsep{0.2cm}} r@{\extracolsep{0.cm}:}c@{\extracolsep{0.cm}:}l@{\extracolsep{0.2cm}} r@{\extracolsep{0.cm}:}c@{\extracolsep{0.cm}:}l@{\extracolsep{0.2cm}} c@{\extracolsep{0.2cm}} c@{\extracolsep{0.2cm}} r@{\extracolsep{0.2cm}} l@{\extracolsep{0.2cm}} c@{\extracolsep{0.2cm}} r@{\extracolsep{0.cm}.}l@{\extracolsep{0.2cm}} r@{\extracolsep{0.cm}.}l@{\extracolsep{0.2cm}} c@{\extracolsep{0.2cm}} c@{\extracolsep{0.2cm}}}
   \hline\hline       
    \multicolumn{1}{c}{Name} & \multicolumn{3}{c}{R.A.} & \multicolumn{3}{c}{Dec.} & $z$ & Opt. Class & Counts & Sample & \pedSM{N}{H,\,Gal} &  \multicolumn{2}{l}{$\Gamma$} & \multicolumn{2}{c}{\nhsym} & \pedSM{\log F}{2-10\kev} & \pedSM{\log L}{2-10\kev} \\
    \multicolumn{1}{c}{} & \multicolumn{3}{c}{} & \multicolumn{3}{c}{} &  &  &  &  & ($10^{20}\,$\nh) & \multicolumn{2}{l}{} & \multicolumn{2}{c}{($10^{22}\,$\nh)} & (\flux) & (\lum) \\
    \multicolumn{1}{c}{(1)} & \multicolumn{3}{c}{(2)} & \multicolumn{3}{c}{(3)} & (4) & (5) & (6) & (7) & (8) & \multicolumn{2}{l}{(9)} & \multicolumn{2}{c}{(10)} & (11) & (12)  
    \vspace{0.1cm} \\
   \hline
    \vspace{-0.2cm} \\
    XBSJ000100.2-250501 & $00$&$01$&$00.0$& $-25$&$05$&$03.9$& $0.850$ & AGN1    & $830 $ & BSS      & $1.88$ & $1$&$67\errUD{0.14}{0.18}$ & $ 0$&$71\errUD{ 0.22}{0.24}$ & $-12.86$ & $44.66$ \\
    $\myast$XBSJ013240.1-133307 & $01$&$32$&$40.3$& $-13$&$33$&$07.2$& $0.562$ & AGN2 & $323 $ & HBSS,BSS & $1.64$ & $1$&$9^{f}		    $ & $ 2$&$55\errUD{0.7 }{0.57}$ & $-12.75$ & $44.43$ \\ 
     & \multicolumn{3}{c}{} & \multicolumn{3}{c}{} & & & & & & $1$&$47\errUD{0.09}{0.13}$ & $ 3$&$22\errUD{0.40}{0.49}$ & $-12.42$ & $44.64$ \\ 
    XBSJ021642.3-043553 & $02$&$16$&$42.3$& $-04$&$35$&$52.9$& $1.985$ & AGN2    & $880 $ & BSS      & $2.42$ & $1$&$91\errUD{0.18}{0.17}$ & $ 4$&$20\errUD{ 1.20}{1.00}$ & $-12.99$ & $45.49$ \\
    XBSJ022707.7-050819 & $02$&$27$&$07.8$& $-05$&$08$&$16.0$& $0.358$ & AGN2  & $505 $ & BSS      & $2.63$ & $1$&$68\errUD{0.48}{0.37}$ & $ 1$&$31\errUD{ 0.80}{0.60}$ & $-12.59$ & $44.09$ \\
    XBSJ050536.6-290050 & $05$&$05$&$36.7$& $-29$&$00$&$50.8$& $0.577$ & AGN2  & $1000$ & HBSS,BSS & $1.49$ & $1$&$85\errUD{0.20}{0.18}$ & $ 0$&$61\errUD{ 0.21}{0.17}$ & $-12.88$ & $44.30$ \\
    $^{\myddag}$XBSJ051413.5+794343 & $05$&$14$&$13.5$& $+79$&$43$&$47.0$ & $0.766$ & AGN2 & $^{m}442$ & BSS & $8.01$ & $1$&$58\errUD{0.24}{0.23}$ & $<0$&$60$ & $-12.78$ & $44.61$ \\  
    XBSJ052128.9-253032 & $05$&$21$&$28.9$& $-25$&$30$&$32.4$& $0.586$ & Elusive & $70  $ & HBSS     & $1.92$ & $1$&$9^{f}		$ & $13$&$80\errUD{ 7.50}{4.57}$ & $-12.85$ & $44.49$ \\
    $^{\myddag}$XBSJ080411.3+650906 & $08$&$04$&$11.1$& $+65$&$09$&$07.0$& $0.604$ & Elusive  & $410$ & HBSS,BSS  & $4.32$ & $1$&$62\errUD{0.45}{0.38}$ &  $2$&$47\errUD{ 1.49}{1.33}$ & $-12.66$ & $44.53$ \\ 
    $\myast$XBSJ113148.7+311358 & $11$&$31$&$48.7$& $+31$&$14$&$00.9$& $0.500$ & AGN2 & $526 $ & HBSS,BSS & $2.02$ & $1$&$9^{f}		    $ & $ 2$&$9 \errUD{0.66}{0.60}$ & $-12.55$ & $44.52$ \\ 
     & \multicolumn{3}{c}{} & \multicolumn{3}{c}{} & & & & & & $1$&$59\errUD{0.39}{0.40}$ & $ 4$&$05\errUD{2.39}{2.84}$ & $-12.93$ & $44.00$ \\ 
    XBSJ122656.5+013126 & $12$&$26$&$56.5$& $+01$&$31$&$24.2$& $0.733$ & AGN2  & $1454$ & HBSS,BSS & $1.84$ & $1$&$61\errUD{0.24}{0.21}$ & $ 2$&$40\errUD{ 0.65}{0.56}$ & $-12.65$ & $44.73$ \\
    XBSJ134656.7+580315 & $13$&$46$&$56.8$& $+58$&$03$&$15.7$& $0.373$ & Elusive & $1236$ & HBSS     & $1.28$ & $1$&$9^{f}	       $ & $ 9$&$27\errUD{ 4.23}{2.91}$ & $-12.83$ & $44.04$ \\
    $^{\mydag}$XBSJ144021.0+642144 & $14$&$40$&$21.0$& $+64$&$21$&$44.1$& $0.720$ & AGN1  & $720 $ & BSS      & $1.68$ & $1$&$88\errUD{0.22}{0.19}$ & $ 0$&$67\errUD{ 0.34}{0.25}$ & $-12.93$ & $44.48$ \\
    $\myast$XBSJ160645.9+081525 & $16$&$06$&$45.7$& $+08$&$15$&$25.0$& $0.618$ & AGN2 & $211 $ & HBSS,BSS & $4.01$ & $1$&$68\errUD{0.77}{0.65}$ & $13$&$8 \errUD{6.7 }{5.15}$ & $-12.41$ & $44.92$ \\
     & \multicolumn{3}{c}{} & \multicolumn{3}{c}{} & & & & & & $2$&$02\errUD{0.56}{0.52}$ & $29$&$62\errUD{9.63}{8.01}$ & $-12.75$ & $44.77$ \\ 
    XBSJ204043.4-004548 & $20$&$40$&$43.5$& $-00$&$45$&$50.0$& $0.615$ & AGN2  & $302 $ & HBSS,BSS & $6.72$ & $1$&$9^{f}		 $ & $ 3$&$28\errUD{ 1.35}{0.97}$ & $-12.76$ & $44.52$
  \end{tabular}
 }
 \end{center}       
 {\footnotesize   {\sc Note:} The X-ray spectral results are from \citet{corral11}, except for the new optical identifications, marked with $^{\myddag}$ (this work;
{\sc OBSID} of the \xmm\ observations:  0094400101 and 0094400301 for XBSJ051413.5+794343 and XBSJ080411.3+650906, respectively; a paper 
reporting an update of the spectroscopic identifications is in preparation). 
Fluxes and luminosities refer to the MOS calibration.
 For all but one sources, the best-fitting model is an absorbed power law.
 $^{\mydag}\,$For this source, a best fit is not found 
 \citep[i.e., null hypothesis probability $<10$\% with all the model tested; see][]{corral11}; we adopt the absorbed power law 
 as best fit, as none of the additional components significantly improves the fit.
 ${\myast}\,$mark the XQSO2s for which we obtained new \xmm\ observations: for these sources, more complex models (applied to the new data) are tested in 
 Appendix~\ref{sect:xmm} and the results are reported in Table~\ref{tab:xbf}. 
Here, we quote the values derived for the most significant properties (column density, photon index, observed flux, and intrinsic luminosity; second row). 
 $^{m}\,$For this source, only MOS cameras have been considered, 
 because in the pn the source falls in a CCD gap.
 \vspace{0.1cm}\\
 \footnotesize Column (1): source name in the XBS sample.
 \footnotesize Columns (2) and (3): J2000 coordinates of the X-ray object.
 \footnotesize Column (4): source redshift.
 \footnotesize Column (5): optical classification, from \citet{caccianiga08} except for XBSJ144021.0+642144 \citep[see][]{corral11} and for the new optical identifications; 
 a paper reporting an update of the spectroscopic identifications is in preparation. 
 \footnotesize Column (6): total EPIC counts ($0.3-10\,$keV).
 \footnotesize Column (7): sample to which the source belongs.
 \footnotesize Column (8): Galactic column density, from \citet{nh}.
 \footnotesize Column (9): source photon index, and $90$\% confidence errors; $f=$~parameter fixed during the fit.
 \footnotesize Column (10): intrinsic column density.
 \footnotesize Column (11): observed flux (deabsorbed by our Galaxy) in the $2-10\,$keV energy band.
 \footnotesize Column (12): intrinsic luminosity in the $2-10\,$keV energy band.
 }
\end{minipage}
\end{table*}
\addtocounter{numtab}{1}
\addtocounter{numtabdata}{1}

The details on the \xmm\ fields selection strategy and the source selection
criteria of the XMM BSS and HBSS samples are discussed in \citet{rdc04}.
A description of the optical data and analysis, of the optical classification
scheme, and of the optical properties of the extragalactic sources identified 
so far is presented in \citet{caccianiga07,caccianiga08}; 
a paper presenting an update of the spectroscopic identifications (mostly based on data taken at the TNG, GTC and VLT telescopes)
of the XBS survey is in preparation.
The identification statistics below take into account these new identifications, along with the new X-ray spectral analysis in \citet{corral11}.

Regarding the AGN population, the entire XBS sample comprises $320$ AGN divided into
$281$ type~1, $32$ type~2, $6$ BL~Lacs, and $1$ unclassified AGN
\citep[following the criteria described in][]{caccianiga08}.
For the subsample of $40$
``optically elusive AGN'' \citep[i.e. sources for which the X-ray data are the only way to
identify the AGN nature;][]{caccianiga07}, the type~1/type~2 classification is assigned on the
basis of the intrinsic absorption in their X-ray spectra ($14$ absorbed AGN, $25$ unabsorbed AGN, and $1$ 
AGN for which the upper limit on \nhsym\ is not stringent enough to allow a classification).
All but one of the type~1 AGN belong to the BSS sample, while a selection at higher energies confirms its effectiveness in 
finding obscured sources, with $\sim 62$\% of type~2 AGN detected in the HBSS, as opposed to only $\sim 15$\% of type~1 AGN.
The cosmological and statistical properties of the HBSS AGN sample are extensively discussed in \citet{rdc08}.

For most of the type~1 sources in the XBS AGN sample, the optical/UV SED
has been build, and the bolometric luminosity has been estimated by fitting a multicolour disk model
\citep{marchese12}. 
\citet{caccianiga13} report their black hole masses, estimated using the ``single epoch'' 
spectral method (based on measuring the broad line widths and the continuum emission).

The X-ray properties of the XBS AGN population are presented in
\citet{corral11}.
From their analysis, the authors identify $15$ QSOs absorbed in the X-ray
(XQSO2s,
\ie\ sources with intrinsic absorption
$\nhsym > 4\times
10^{21}\,$\nh\ and X-ray luminosity $\pedix{L}{2-10\kev} > 10^{44}\,$\lum).
In the work presented here, 
we excluded $3$ sources, optically classified as unobscured objects, having \nhsym\ with error bars crossing the threshold 
(XBSJ102417.5+041656, XBSJ213719.6-433347, and XBSJ213820.2-142536).
We included two more XQSO2s, identified after the publication of the X-ray analysis of the sample
(XBSJ051413.5+794343 and XBSJ080411.3+650906);
the data analysis has been performed following the 
same prescriptions as in \citet{corral11}.
Note that for XBSJ051413.5+794343
the quality of the X-ray spectra 
allows us to obtain only an upper limit to the intrinsic column density \nhsym.
Nevertheless, this object has been included in the sample, since the optical classification as AGN2 
suggests the presence of obscuration.

The final sample is then composed by \nsrc\ objects;
$2$ are found only in the HBSS, $5$ belong only to the BSS, and $7$ are
present in both samples.
We report the distributions in redshift, X-ray luminosity, and column density of the XQSO2s in Fig.~\ref{fig:znhdist}.
The XQSO2 sample covers a redshift range 
$0.35<z<0.9$, with only one object at higher redshift, $z=1.985$ \citep[XBSJ021642.3-043553, an X-ray selected Extremely Red 
Object discussed in details by][]{severgnini06}.
Basic information is listed in Table~\ref{tab:xbs}, together with a summary of the results of the analysis performed by
\citet{corral11}, except for the two new optical identifications, presented here for the first time.
The X-ray and optical spectra for all the sources are shown in Fig.~\ref{fig:sed} (top and central panels):
for the majority of the XQSO2s, the optical spectra are 
typical of optically type~2 AGN, with only $3$ optically elusive AGN
(XBSJ052128.9-253032, XBSJ080411.3+650906, and XBSJ134656.7+580315) and
$2$ objects (XBSJ000100.2-250501 and XBSJ144021.0+642144, $14$\% of the XQSO2s) classified as type~1 AGN, in agreement with the fraction 
of ``mixed'' classification of about $\sim 20$\% reported in literature \citep[\eg,][]{trouille09,merloni14}.
For an extensive discussion on the X-ray versus optical absorption in the XBS, we refer the reader to section~4.1 in \citet{corral11};
here we note that the only two optically classified type~1 sources have the lowest \nhsym, 
near the adopted threshold, and the observed mismatch is probably due to the assumption of the Galactic dust-to-gas ratio as mean $\pedix{A}{V}/\nhsym$.

\addtocounter{numfig}{1}
\renewcommand{\thefigure}{\arabic{numfig}}

As a follow-up programme of specific XBS sources, we obtained new \xmm\ observations for $3$ out of the \nsrc\ XQSO2s,
namely XBSJ013240.1-133307, XBSJ113148.7+311358, and XBSJ160645.9+081525 
({\sc OBSID} $0550960101$, $0550960301$, and $0550960601$, respectively; PI Della~Ceca).
A detailed analysis of the new data is reported in Appendix~\ref{sect:xmm}. 
In the following, for these three sources we will assume the \nhsym\ and \pedix{L}{X} reported in Table~\ref{tab:xbf} for the new \xmm\ data; 
for each source, the most significant parameters are quoted also in Table~\ref{tab:xbs} (second row).


\section{UV-to-IR SED}\label{sect:data}

\addtocounter{numtab}{2}
\renewcommand{\thetable}{\arabic{numtab}}

In order to investigate the physical properties of the QSOs, we constructed their broad-band SED, from UV to IR wavelengths.

\subsection{\spitzer\ observations and data reduction}\label{sect:spit}

\spitzer\ data for $6$ XQSO2s were obtained in the Cycle-$3$, as part of a study of a statistically
complete sample of optically absorbed AGN drawn from the XMM HBSS (Programme ID:$30606$; PI P.~Severgnini).
The observations were performed with the IRAC ($3.6$, $4.5$, $5.8$, and $8\,\mu$m) and MIPS ($24$ and $70\,\mu$m) 
instruments in photometry mode.
We used a frame time of $12\,$sec with a dither pattern of $16$ points for the IRAC observations, and a frame time of
$10\,$sec with a $5-20$ cycles and small field pattern for the MIPS ones.
To estimate the flux densities in the IRAC and at the MIPS-$24\,\mu$m bands, we used the final combined
post-basic calibrated data (BCD) mosaics produced by the \spitzer\ Science Center (SSC) pipeline. 
At $70\,\mu$m, we looked at both the filtered and unfiltered post-BCD mosaics and if we saw negative side-lobes around
our target in the filtered data, we re-filtered our BCD data by Ge Reprocessing Tools (GeRT). 
The GeRT is a contributed software from the MIPS Instrument Support Team made available to the scientific community to
carry out custom offline data reduction of MIPS-Ge ($70$ and $160\,\mu$m) data. 
In those cases in which the re-filtering offline 
process was applied, we have co-added and background subtracted 
the updated BCDs with the SSC Spitzer Mosaiker - MOPEX software \citep{makovoz05}. 
The resulting mosaics were made with $\sim 4\arcsec$ pixel size. 

The flux densities of our targets have been measured using the IRAF package and the Sextractor software. 
For point-like sources, we estimated the aperture photometry at the position of the sources and we applied aperture
corrections as derived by using the {\it Point Response Function}. 
The chosen aperture radii are $\sim 6\arcsec$ for the IRAC bands, and $\sim 13\arcsec$ at $24\,\mu$m. 
Only in the presence of nearby optical and/or IR sources smaller apertures have been used. 
For all of those sources which appear extended in the IRAC and/or MIPS images we used the curve of growth by masking 
possible sources nearby the target. 

All the sources are detected up to $24\,\mu$m.
We find flux densities ranging from $170\,\mu$Jy to $4.4\,$mJy in the IRAC bands, and from $1$ to $7\,$mJy at
$24\,\mu$m.
At $70\,\mu$m, we measure a significant emission (from $5$ to $135\,$mJy) for $4$ of our targets.
For one additional source, we were able to estimate an upper limit to the flux using the weakest source detected in the field as limit.

Among the $8$ sources for which we do not have our own \spitzer\ data, XBSJ021642.3-043553 and XBSJ022707.7-050819 fall in one 
of the fields covered by the \spitzer\ Wide Area Infrared Extragalactic Survey \citep[SWIRE;][]{lonsdale03};
from the SWIRE catalogue\footnote{http://irsa.ipac.caltech.edu/applications/Gator}, we retrieved the 
flux densities obtained with aperture \#4, corresponding to a radius of $4.1\arcsec$ and $10.5\arcsec$ for the IRAC and MIPS 
data, respectively.
Both sources are detected up to $24\,\mu$m.

Images in the IRAC and MIPS bands for XBSJ122656+013126 were retrieved from the archive and reduced as was done for the proprietary data; 
the source is detected up to  $24\,\mu$m.
We found imaging at $24\,\mu$m for one more object, XBSJ144021.0+642144;
for the remaining $4$ sources in our sample,  we did not find either IRAC or MIPS data
in the \spitzer\ archive. 

Summarizing, including both proprietary and archival observations, we have $9$ and $10$ detections in the IRAC bands 
(at $3.6$, $4.5$, $5.8$ and $8\,\mu$m) and at $24\,\mu$m, 
respectively, while at $70\,\mu$m we have $4$ detections and $1$ upper limit; the results of the \spitzer\ data analysis are reported in Table~\ref{tab:ir}.

\begin{figure*}
\begin{center}
 \resizebox{0.33\hsize}{!}{\includegraphics[]{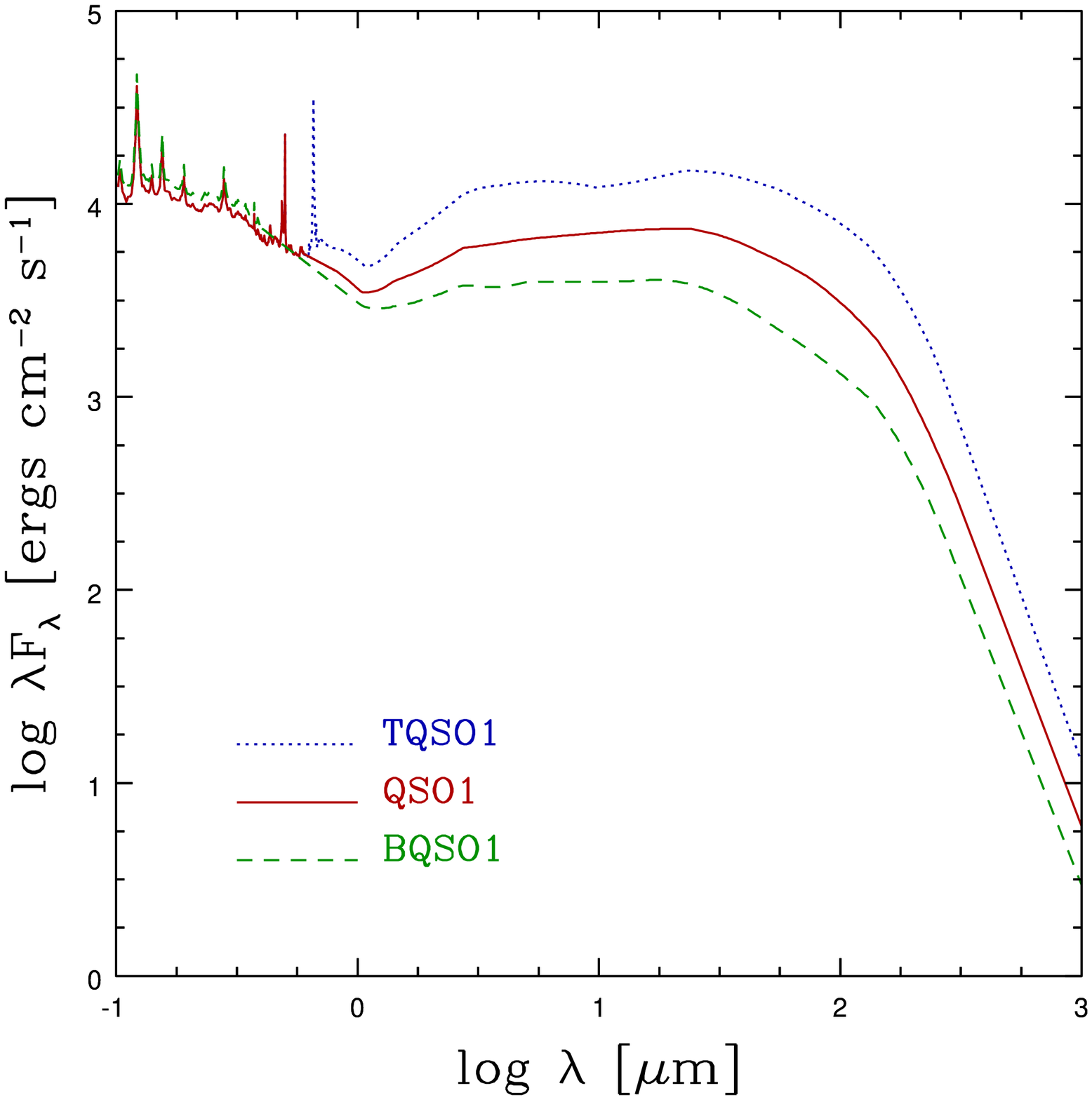}}
 \resizebox{0.33\hsize}{!}{\includegraphics[]{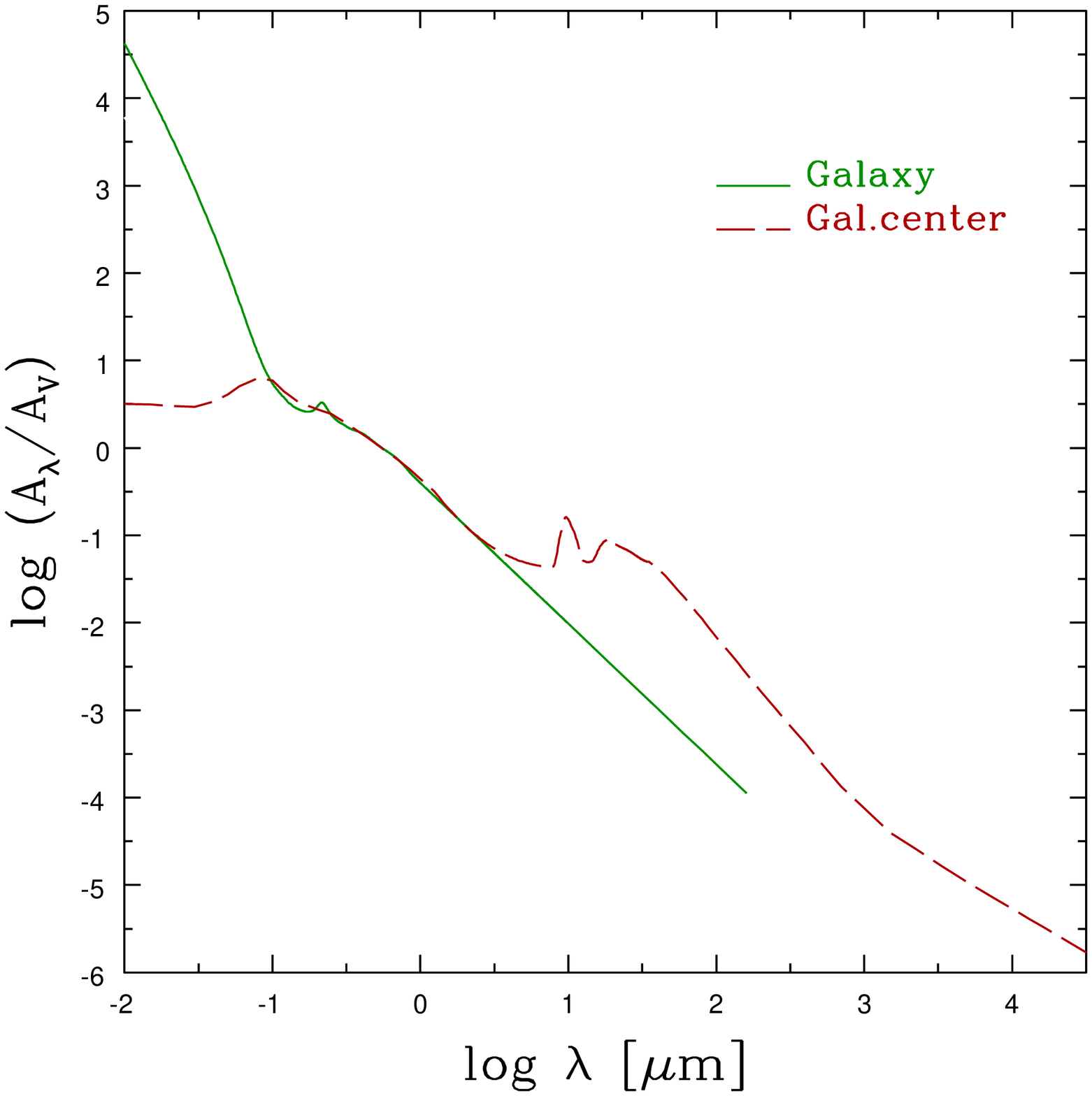}}
 \resizebox{0.33\hsize}{!}{\includegraphics[]{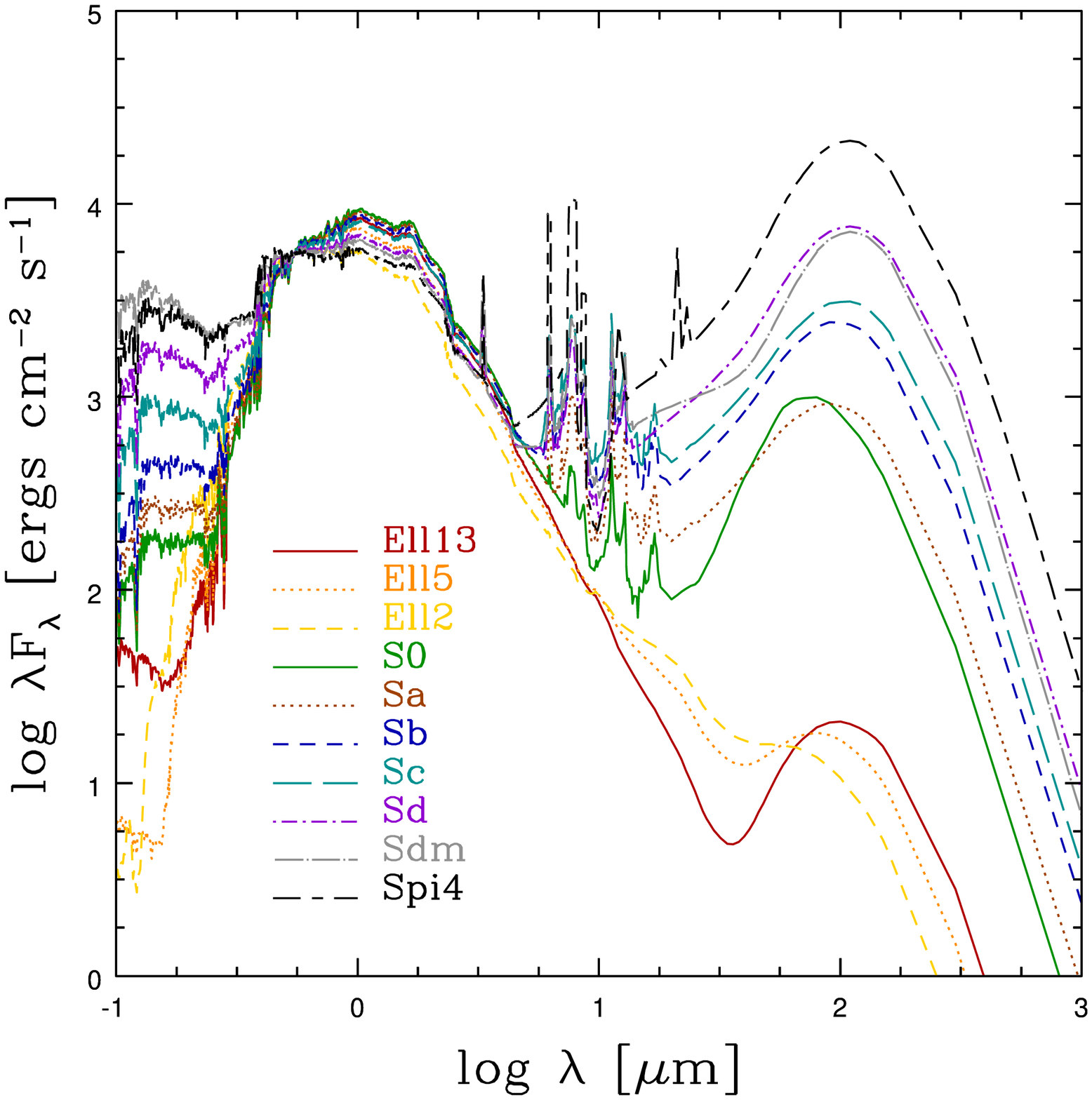}}
\caption{Templates adopted in this work for the SED modelling (see Sect.~\ref{sect:sed}).
{\it Left-hand panel}: nuclear templates ($0.1-1000\,\mu$m) of optically-selected luminous AGN with different values of the IR/optical ratio, from \citet{polletta07}. 
{\it Central panel}: extinction curves of our Galaxy (green continuos line) and of the Galactic Centre \citep[red dashed line;][]{chiar06}.
{\it Right-hand panel}: Templates of $3$ elliptical galaxies with different ages, 
$3$ early spirals,
and $4$ late spirals and irregulars 
(as labelled), generated with the GRASIL code \citep{silva98}, from \citet{polletta07}.}
\label{fig:templ}
\end{center}
\end{figure*}
\addtocounter{numfig}{1}

\subsection{WISE}\label{sect:wise}
 
WISE is a NASA satellite
that imaged the whole sky in four mid-IR photometric bands, centred at $3.4$, $4.6$,
$12$ and $22\,\mu$m.
We refer to these bands as W1, W2, W3 and W4, respectively. 
The angular resolution is $6.1\arcsec$, $6.4\arcsec$, $6.5\arcsec$, and $12.0\arcsec$, respectively. 
We use the WISE All-Sky Data Release (2012 March), which includes all
observations obtained during the fully cryogenic mission (2010 January 7 to 2010 August 6).
The WISE All-Sky source catalogue has been searched by using the X-ray source position and 
assuming an impact parameter equal to $4\arcsec$; 
we considered as detection only WISE magnitudes with signal-to-noise ratio $S/N>3$ in the W1, W2, and W3 bands, 
and $S/N>5$ in the W4 band; for lower $S/N$, we assumed the corresponding $2\sigma$ upper limits.
A posteriori we found that the selected
WISE objects were closer than $2\arcsec$ from the input source position.
Using a 
simulated
sample 
of $\sim1600$ objects, we estimated that the
probability to have a random WISE source into a circle of $2\arcsec$
radius is $\sim 1.1$\%, implying that all the associations between the WISE objects and our XQSO2s are 
likely real.

In Table~\ref{tab:ir}, we report the flux densities, computed from the
profile-fit photometry (obtained from the WISE All-Sky source catalogue)
by assuming the magnitude zero points of the Vega system corresponding
to a power-law spectrum ($\pedix{f}{\ensuremath{\nu}} \propto \apix{\nu}{-\ensuremath{\alpha}}$) with $\alpha=1$. 
The differences in the computed flux densities expected using flux correction factors that correspond to $\alpha=-1$, $0$ or $2$ 
(lower than $0.8$\%, $0.6$\%, $6$\%, and $0.7$\% in W1, W2, W3 and W4, respectively) have been linearly added to the
catalogued flux errors. 
Following the prescription in \citet{wise},
an additional $10$\% uncertainty was linearly added to the $12$ and $22\,\mu$m fluxes.

\subsection{Other data}\label{sect:odata}

In Table~\ref{tab:optir} we report the \sdss\ $u$, $g$, $r$, $i$, and $z$ magnitudes found for $7$ XQSO2s.
Magnitudes in the $R$-band were collected from the
literature for $7$ XQSO2s \citep{fiore03,caccianiga08}; a DSS magnitude was available for one source, while 
we found a magnitude in the $g^{\prime}$-band for one more object \citep{stalin10}.
From the \emph{Galaxy Evolution Explorer} (\galex) catalogue we retrieved fluxes in the near-UV (at $2310\,$\AA) for $3$ XQSO2s, one of them having also an entry in the far-UV (at $1539\,$\AA);
for XBSJ013240.1-133307 a magnitude in the $U$-band was obtained from the Optical Monitor (OM) telescope onboard \xmm\ (see Appendix~\ref{sect:xmm}).
At longer wavelengths, in addition to \spitzer\ and WISE photometry, a $K$-band magnitude for 
XBSJ021642.3-043553 is reported by \citet{severgnini06}.


\section{SED deconvolution}\label{sect:sed}

To decompose the galaxy and AGN contributions, we adopted a very simple phenomenological 
approach.
We have constraints from both photometry and optical spectra; to account for all of them, we made use of
\vspace{-0.13cm}
\begin{description}
\item[{\it (a)}] an AGN spectral template, composed by a combination of 
broad line and continuum emission, covering the optical energy range \citep{francis91,elvis94};
\item[{\it (b)}] a template of AGN narrow lines, 
constructed adopting  the prescriptions in \citet{krolik98} for the relative intensities of the lines;
\item[{\it (c)}] a set of templates of QSOs and galaxies, extending from the UV up to the mid-IR wavelength bands,
chosen from the template library of \citet{polletta07}.
The broad-band QSO templates have been derived by combining the \sdss\ QSO composite spectrum and rest-frame IR data of a sample
of optically-selected type~1 QSOs observed in the SWIRE programme.
We considered three templates with the same optical spectrum but with three different IR SEDs: a mean IR spectrum, obtained
from the average fluxes of all the measurements (``QSO1''), a template with high IR/optical flux ratio,
obtained from the highest $25$\% measurements per bin (``TQSO1''), and a low-IR emission SED obtained from the lowest
$25$\% measurements per bin (``BQSO1'').
The three templates are shown in Fig.~\ref{fig:templ}, left-hand panel.
Regarding the galaxy emission in the UV-IR, we considered $10$ templates, each of them 
covering the wavelength range between $1000\,$\AA\ and $1000\,\mu$m.
They include $3$ ellipticals and $7$ spirals (from early to late types, S0-Sdm; see Fig.~\ref{fig:templ}, right-hand panel).
\end{description}
\vspace{-0.13cm}

To extinguish the nuclear emission, both the extinction curves of our Galaxy, and of the Galactic Centre \citep{chiar06}, have been considered;
a comparison between the two curves is shown in Fig.~\ref{fig:templ}, central panel.

First, we concentrated on the optical spectra, to obtain a guess for the dust extinction.
XBSJ021642.3-043553 has been discussed in details in \citet{severgnini06}; for the remaining $13$ sources,
 the optical AGN template (previously labelled as {\it a)} has been absorbed
and then summed up to the narrow line component (previously labelled as {\it b)};
the total AGN template was redshifted to the redshift of the source and summed with a redshifted galaxy template. 
We varied the amount of dust extinction, 
\pedix{A}{V}, and the relative AGN/galaxy normalization until a good reproduction of the available optical spectrum is reached.
An estimate of \pedix{A}{V} ($6$ objects), or a lower limit to the same parameter ($7$ objects), 
have been thus obtained.

We then moved to the analysis of the UV-to-mid-IR SED.
In this modelling, the \pedix{A}{V} (or the lower limit to \pedix{A}{V}) obtained from the spectral analysis has been used as a first guess for the photometric dust extinction;
the latter has been left free to vary during the SED modelling.
Similarly, as guess for the host type, we considered the Ca break at $4000\,$\AA\ obtained from the optical spectra.
After shifting the photometric data to the rest frame,
each SED has been reproduced by the sum of a galaxy plus a QSO template (both previously labelled as {\it c)}.
In the modelling, the latter was absorbed to reproduce the intrinsic obscuration at the source redshift before summing up with the galaxy component.
For each source and for each QSO template, different combinations of nuclear \pedix{A}{V} and galaxy templates have been tested; 
we selected the combination that, from a visual match, was able to better reproduce most of the photometric points.
In the SED analysis described here, we did not take into account information derived from the X-ray data. 
In particular, we did not assume any relation between the \nhsym\ and the dust extinction.

Note that, since a significant fraction of the flux in the UV and optical bands has been absorbed in our AGN template, the UV and
optical photometric data are most-likely dominated by the host galaxy template.
In this sense, among the photometric points, the \sdss\ data are the most relevant ones to constrain the galaxy type,
at least for the optically type~2 XQSO2s.

In discussing the results of our analysis, a couple of considerations about the adopted method of modelling must be taken into account.
First, here we are assuming that the emission of type~2 QSOs can be reproduced using the mean disk$+$torus emission of the QSO1s, 
after screening both components with a nuclear extinction, reproducing the absorption and auto-absorption associated with the torus.
A more physical approach would require a more sophisticated parametrization of the AGN torus, to properly describe the auto-absorption and re-emission
balance and its dependence on the angle of view or the dust properties. 
In particular, observational evidences suggest that the dust within the torus is arranged in clumps instead of being smoothly distributed 
\citep[and references therein]{krolik88,tacconi94,tacconi96,markowitz14}.
However, our data coverage and spectral resolution are not sufficient to constrain the huge number of parameters present in
the torus models incorporating clumpy gas (\eg, \citealt{nenkova02,hoenig06,nenkova08a,nenkova08b,schartmann08,heymann12}; 
see \citealt{hoenig13} for a review).
Instead, an approach like the one adopted here is more than adequate for sparse photometric data point.

Moreover, in the mid/far-IR band the empirical QSO1 templates from \citet{polletta07} in principle could suffer of contamination 
due to the 
host galaxy emission.
However, a comparison with the results presented by \citet{mateos12,mateos13} on a sample of X-ray selected type~1 AGN SED
strongly suggests that at luminosities greater than $10^{44}\,$\lum\ such a contribution, if present, would not affect our results.
The selection technique applied by the authors, based on the mid-IR colours, suggests a total 
mid-IR emission dominated by the AGN component.
The median SED of the Mateos et al. type~1 AGN is well represented by the same templates adopted here, therefore
we expect that a similar result can be applied to our XQSO2s.

In Figure~\ref{fig:sed}, bottom panel, we report the resulting SEDs normalized to the de-absorbed rest frame $2-10\,$keV luminosity
derived from the X-ray spectral analysis (as in Table~\ref{tab:xbs}).
For most of the sources, each of the tested templates of QSO (\ie, ``TQSO1'', ``QSO1'' and ``BQSO1'') provides an acceptable description 
of the observed SED, with slightly different 
relative normalizations of the templates and amount of dust extinction: of course, this implies different values for the physical 
parameters estimated from the modelling (see Sect.~\ref{sect:nucl}).
Details on the SED modelling of each object are reported in  Appendix~\ref{sect:note}; in the following, for each parameter we 
will quote the value estimated assuming the mean template, ``QSO1'', while the associated range corresponds to the minimum 
and maximum values obtained when all the possible modellings (with all the QSO templates and extinction curves) are considered.

For \llqsoa\ and \llqsob, the available data do not allow us to 
infer both the intensity of the nuclear emission and the level of optical absorption.
In these cases, we carried out the modelling several times, with \pedix{A}{V} fixed to different values ranging between 
the lower limit recovered by the analysis of the optical spectrum (\pedix{A}{V,opt}) and the highest value obtained from the column density 
provided by the X-ray analysis (\pedix{A}{V,X}), assuming a 
Galactic dust-to-gas ratio, $\pedix{A}{V}/\nhsym=5.27\times 10^{-22}\,$mag/cm$^{-2}$ \citep[][]{bohlin78} to investigate the effects of different levels of obscuration.
For each physical parameter, we assume the value provided by the modelling with \pedix{A}{V,opt} as reference.

When the spectral analysis provides estimates of \pedix{A}{V}, these differ from the values derived from the SED
by a factor lower than $1.5$.
Regarding the $7$ sources for which we obtained only a lower limit to the dust extinction, in $4$ cases the SED-derived values agree 
with these limits.
For three sources, namely XBSJ022707.7-050819, XBSJ134656.7+580315, and XBSJ160645.9+081525
(which optical spectra suggest $\pedix{A}{V}>2-3\,$mag), 
the SEDs require an \pedix{A}{V} in good agreement with the \nhsym\ derived from the X-ray analysis (see Table~\ref{tab:der}). 
For what concerns the two optically type~1 AGN, XBSJ000100.2-250501 and XBSJ144021.0+642144, 
both the spectral analysis and the SED modelling give us results in agreement with the optical classification.
At the same time, the quite different \pedix{A}{V} recovered reflects the difference observed in the optical spectra 
(see Fig.~\ref{fig:sed}), with XBSJ000100.2-250501 ($\pedix{A}{V}=1.5\,$mag, as derived from the SED) being clearly redder than 
XBSJ144021.0+642144 (for which we find $\pedix{A}{V}=0.2\,$mag).

From our analysis, $9$ out of the \nsrc\ XQSO2s reside in ellipticals or early spirals (S0 or Sa).
Our findings are quite in agreement with previous results on luminous type~2 AGN, such as the \sdss\ sample by 
\citet{zakamska06}, who found mainly elliptical hosts.
Recently, \citet{mignoli13} linked the lower fraction of ellipticals found in their sample of \nev-selected $z\sim 1$ obscured AGN to the range 
in nuclear luminosity covered, extended to an order of magnitude lower than \sdss\ type~2 QSOs.
For $5$ XQSO2, instead, the SED modelling suggests a late-type host galaxy (Sd and Sdm). 
In three cases, this result is quite robust thanks to the availability of \sdss\ and/or \galex\ data. 
Although any comparison should be done with caution, due to the different methods of classification (X-ray based or optically-based),
this is an interesting result since a paucity of AGN hosted in late-type spiral and irregular morphologies has been claimed, 
at least in optically selected samples \citep[\eg,][]{mignoli13}.


\section{Recovering the nuclear properties}\label{sect:nucl}

Disentangling the AGN and host galaxy emission as described
in Sect.~\ref{sect:sed} provides detailed information on the
host-galaxy bulge and nuclear components. 
In this section, we detail how we use this information to derive the BH masses, 
the bolometric luminosities, and the Eddington ratios for the XQSO2s.

\begin{figure*}
\begin{center}
 \resizebox{0.33\hsize}{!}{\includegraphics[]{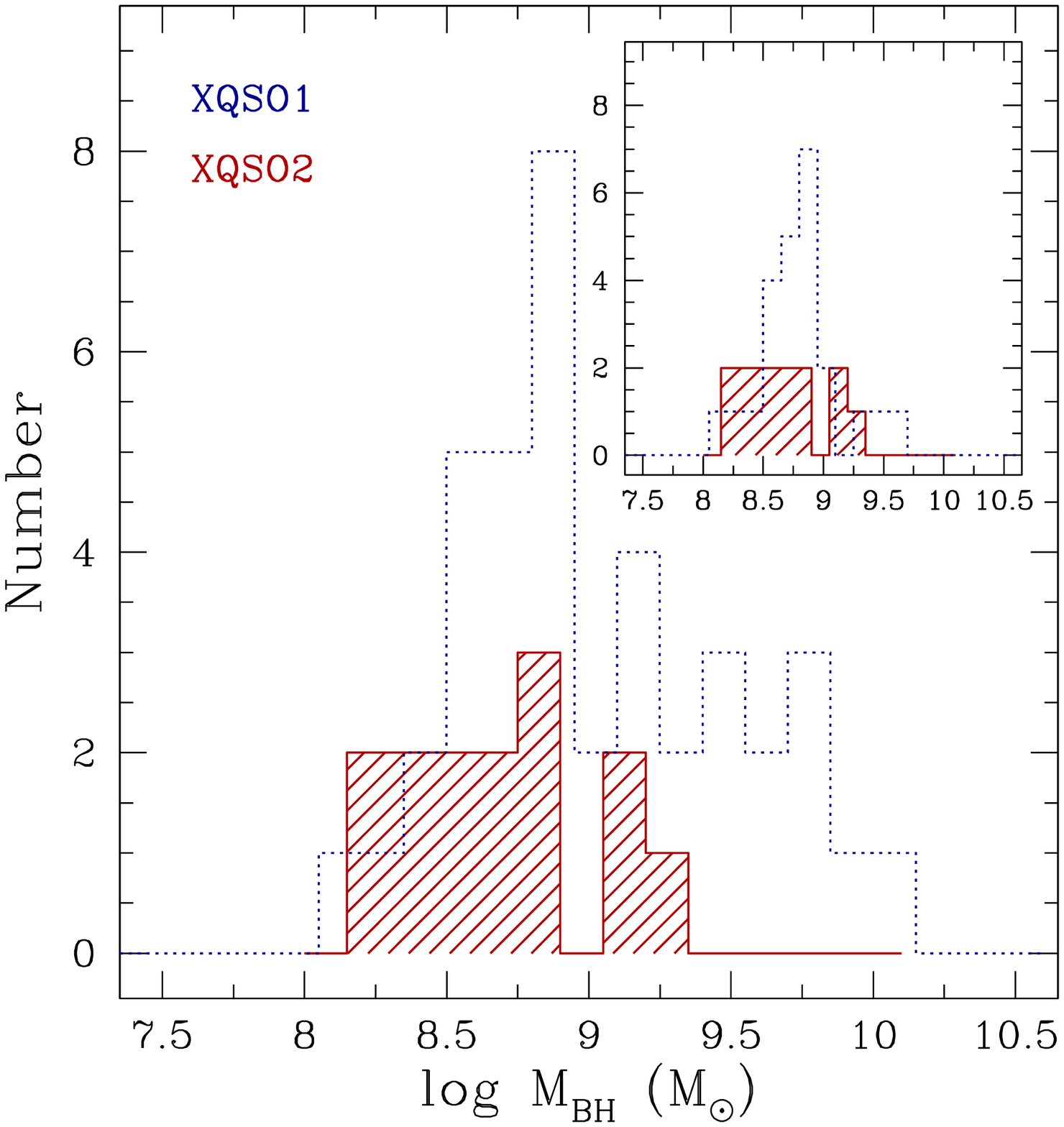}}
 \resizebox{0.33\hsize}{!}{\includegraphics[]{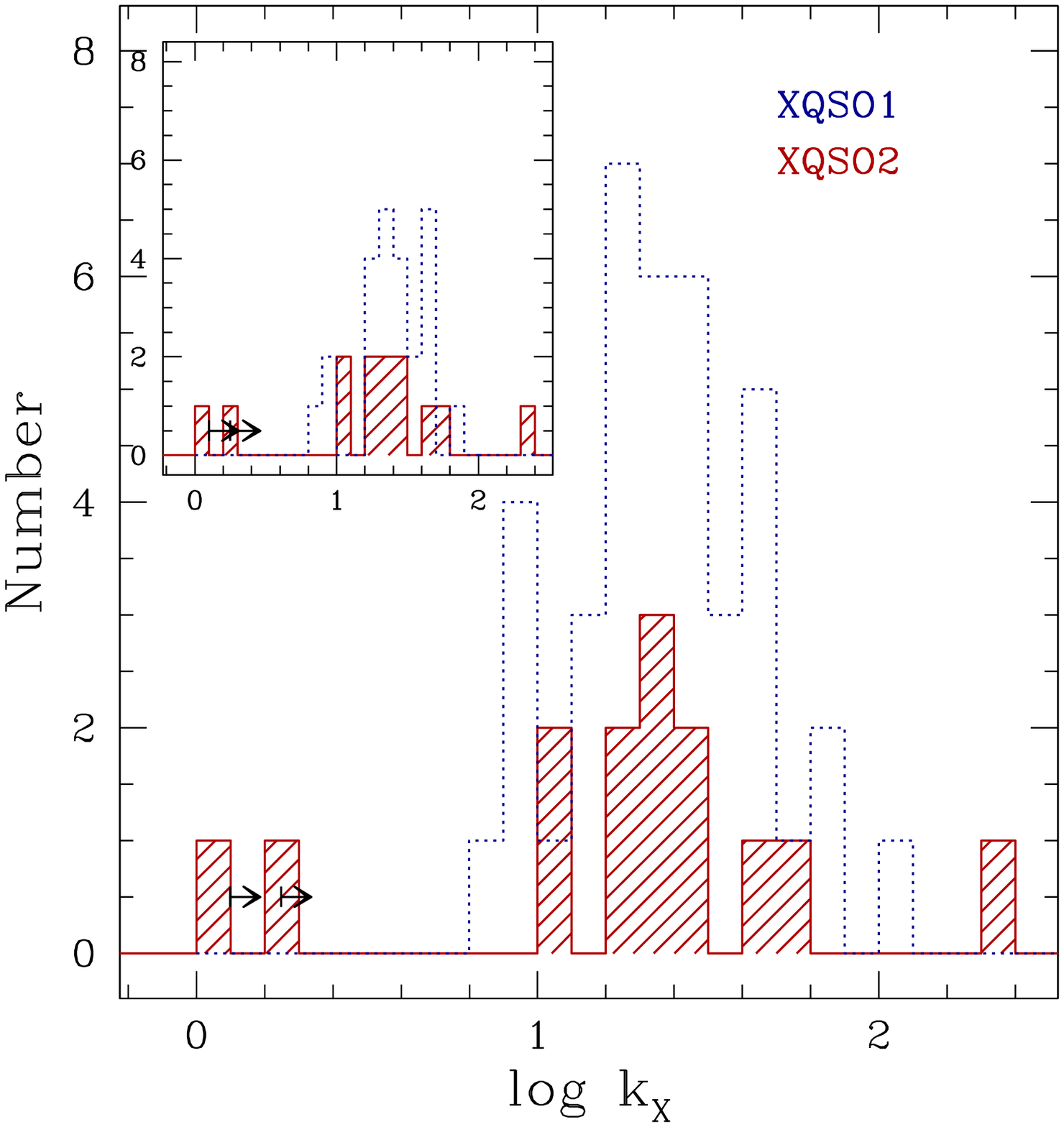}}
 \resizebox{0.33\hsize}{!}{\includegraphics[]{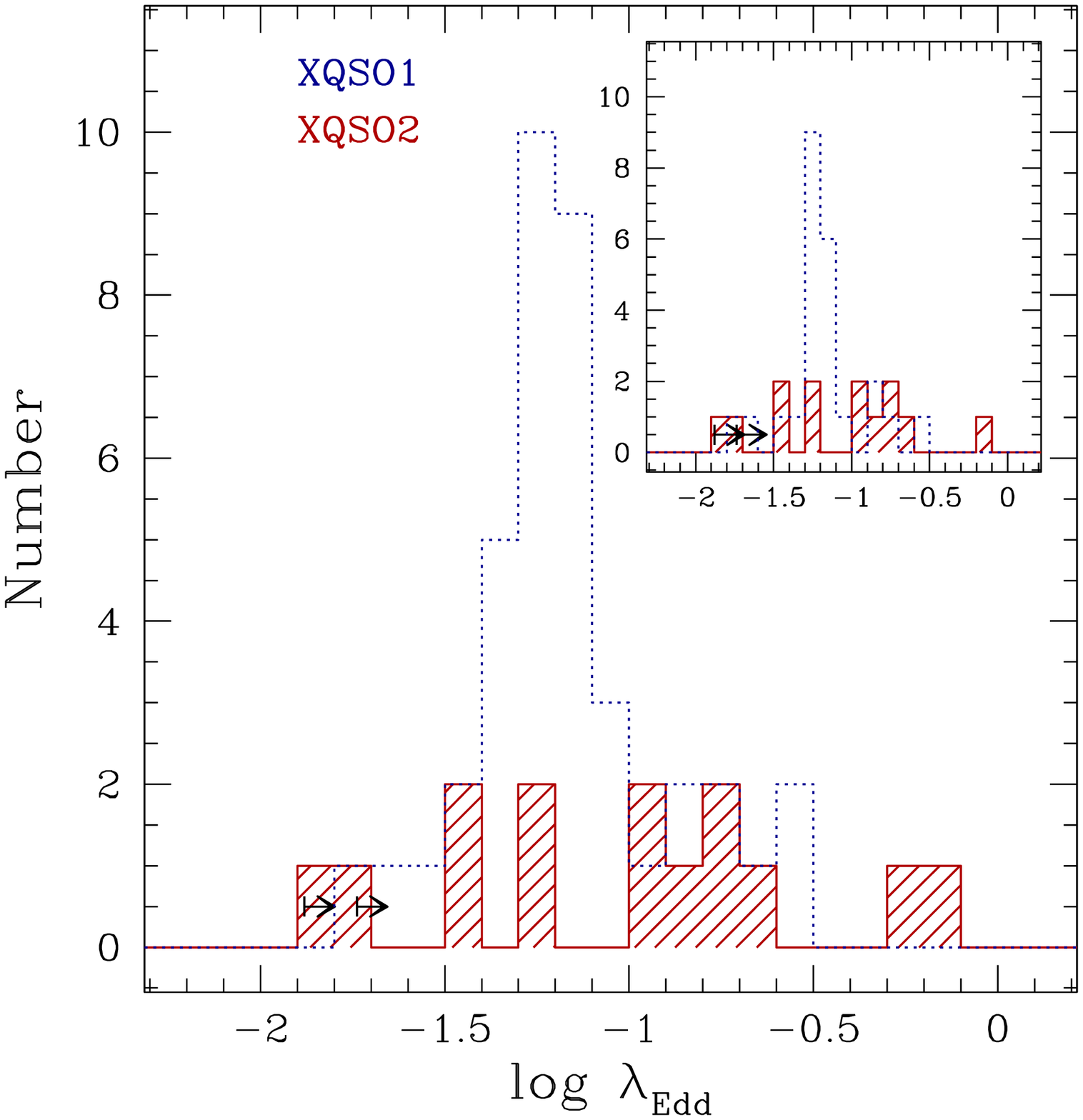}}
\caption{Nuclear properties of the XQSO2s studied here, derived as described in Sect.~\ref{sect:nucl} 
(red shaded histograms): distributions in BH mass (left-hand panel), X-ray bolometric correction (central panel), and 
Eddington ratio (right-hand panel).
The arrows mark the two objects for which we are able to put only limits to the nuclear emission 
(namely, \llqsoa\ and \llqsob).
The blue dotted lines show the corresponding distributions for the XQSO1s.
 In the insets, we show the corresponding distributions obtained when only sources with $z<0.85$ are considered 
 (see Sect.~\ref{sect:xqso12}).
 }
\label{fig:disres}
\end{center}
\end{figure*}
\addtocounter{numfig}{1}

\subsection{Black hole masses}\label{sect:mbh}
Starting from the absolute magnitude of the host galaxy only,
we obtain an estimate of the mass of the central compact object,
using the observed local relationship between BH mass and
bulge component luminosity. 
Several versions of this relation
have been proposed with the bulge luminosity evaluated in different 
electromagnetic bands ranging from $B$ to $K$ 
\citep[see \eg,][]{kormendy01,mclure02,marconi03}.
Here, we adopt the $\pedix{M}{BH}-\pedix{L}{bulge,K}$ relation discussed in \citet{graham07}, corrected for our 
different cosmology:
\begin{equation}\label{eq:mbh}
 \log\pedix{M}{BH}=-0.37(\pedix{M}{K}+24)+8.25
\end{equation}
where \pedix{M}{BH} is given in units of solar masses and \pedix{M}{K} is the absolute $K$-band magnitude of the bulge; the
total scatter associated is $\sim0.33\,$dex.
In order to determine the $K$-band luminosity of the bulge component, we combined the total luminosity
of the host galaxy, as estimated from our SED fitting, with the bulge-to-disk flux 
ratio in the $K$-band as a function of galaxy type reported by \citet[][]{graham08}.

Note that adopting these relations implies assuming
that the absolute magnitude of the 
bulge strictly mirrors the
mass in old stars, \pedix{M}{bulge}, which is the quantity primarily related
to the BH mass.
Moreover, we are implicitly assuming that the \pedix{M}{bulge}-\pedix{M}{BH} relation is already imprinted at
$z \sim1$, driving the main episode of accretion.

We found rather high masses, narrowly distributed
between $\sim 1.7\times 10^8$ and $\sim 1.6\times 10^9\,$\pedix{M}{\sun} (see Table~\ref{tab:der} and Fig.~\ref{fig:disres}, left-hand panel).

For the two optically type~1 objects in our sample (\ie, XBSJ000100.2-250501 and XBSJ144021.0+642144) we 
compared the mass of the central BH recovered from the scaling relations with the value estimated 
via the ``single epoch'' method. 
Taking into account the redshift of the sources ($z=0.85$ and $z=0.72$ for XBSJ000100.2-250501 and 
XBSJ144021.0+642144, respectively), we used the $\mbox{\ion{Mg}{ii}}\,\lambda 2798\,$\AA\ emission line, adopting the 
relation presented in \citet{shen11}:
\begin{equation}
 \log\pedix{M}{BH}=6.74+2\log\left[\frac{FWHM(\mbox{\ion{Mg}{ii}})}{1000\,\mbox{km~s}^{-1}}\right]+
   0.62\log\left[\frac{\lambda\pedix{L}{$\lambda$}}{10^{44}\,\lum}\right]
\end{equation}
for $\lambda=3000\,$\AA.
The width of the line has been measured as described in \citet{caccianiga13}, while the monochromatic continuum 
luminosity has been recovered from our SED modelling, after correcting the nuclear template for the absorption.
We obtain values of mass of $\log\pedix{M}{BH}=8.73\pedix{M}{\sun}$ and $\log\pedix{M}{BH}=9.26\pedix{M}{\sun}$
for XBSJ000100.2-250501 and XBSJ144021.0+642144, respectively.
For one more object, XBSJ122656.5+013126, showing a low-level of dust extinction in the optical spectrum,
\citet{shen11} derived an estimate of the BH mass of
$\log\pedix{M}{BH}=8.88\pedix{M}{\sun}$ (from the FWHM of the \hb\ line and the continuum luminosity at $5100\,$\AA,
adopting their equation~5).
In all cases, the virial estimators give values for the BH mass in relatively good agreement with the values obtained 
from equation~\ref{eq:mbh} (see Table~\ref{tab:der}), considering that the uncertainties  associated with the ``single epoch'' masses are of the order of
$0.35-0.46\,$dex \citep{park12}.

\subsection{Bolometric luminosities and Eddington ratios}\label{sect:lboleddr}
The next step 
is to investigate the nuclear bolometric luminosity of the XQSO2s.
In unabsorbed sources, this quantity 
can be computed 
as the sum\footnote{In this case, the IR emission is not taken into account: being re-processed emission mainly from
the UV, its inclusion would mean counting part of the emission twice, overestimating
the derived bolometric luminosities.} of the accretion disk optical/UV luminosity
integrated 
at wavelengths shorter than $1\,\mu$m and the intrinsic X-ray luminosity between $0.1$ and $500\,$keV rest frame,
$\pedix{L}{bol}=\pedix{L}{opt-UV}$+\pedix{L}{X}.

However, in obscured sources \pedix{L}{opt-UV} is almost completely absorbed due to the high dust extinction.
Thus any estimate of the accretion disk luminosity 
would strongly depend on the template adopted to reproduce the intrinsic optical-UV emission of the QSO.
Moreover, from the analysis of the optical/UV SED of the type~1 AGN in the XBS presented in 
\citet{marchese12}, we found that 
the effective disk temperature ranges from $kT\sim1$ to $kT\sim 8\,$eV, 
and on average 
the disk emission has a $\langle kT \rangle \sim 3.7\,$eV.
Therefore, the limit at $0.1\,\mu$m of the template of AGN would provide a change in an estimated disk luminosity lower than the 
intrinsic one by 
a factor up to $\sim 57$\%.

We therefore adopted a different approach:
we self-calibrated the intrinsic nuclear emission using the SED of the X-ray unabsorbed QSOs drawn from the XBS.

We considered the $40$ AGN in the XBS (hereafter, XQSO1s) with {\it (i)} intrinsic absorption (detected, or upper limit) low enough 
to minimize the possible effects of 
absorption on the estimate of the \pedix{L}{bol} \citep[\nhsym\ or upper limit to \nhsym\ lower than $5\times10^{20}\,$\nh;][]{fanali13}; 
and {\it (ii)} X-ray luminosity 
above the threshold assumed for the XQSO2s ($\pedix{L}{2-10\kev}  > 10^{44}\,$\lum).
The redshift, X-ray luminosity, and column density distributions of the XQSO1s are reported as dashed blue lines in Fig.~\ref{fig:znhdist}.
A Kolmogorov-Smirvov (KS) test indicates that this subsample is not statistically
different from the sample of $95$ X-ray unabsorbed ($\nhsym<4\times10^{21}\,$\nh) AGN
with $\pedix{L}{2-10\kev} > 10^{44}\,$\lum\ belonging to the XBS for
what concerns the distributions in redshift, X-ray and bolometric luminosities, BH mass and Eddington ratio.

We want to use this sample of $40$ XQSO1s, for which a bolometric luminosity has been already computed, to obtain an 
IR bolometric correction to be applied to the sample of XQSO2. 
In particular, we use the WISE W3 magnitude since in this band
{\it (a)} observations with good $S/N$ are available, and {\it (b)} we are confident that in luminous unobscured QSOs the data are 
dominated by the AGN 
(note that all the XQSO1s are also optically classified as type~1).
From our SED modelling, 
when the QSO emission is corrected for the absorption (thus reproducing the SED of an unabsorbed QSO),
we expect that the galaxy contributes less than $10$\% to the total luminosity observed in the W3 band.

As done for the XQSO2s, we looked for a WISE counterpart of the 
$40$ XQSO1s, finding an IR detection in WISE for all but one; here we considered the $36$
WISE counterparts with $S/N>3$ in W3.
For these XQSO1s, we computed the luminosity in the W3 band, $\pedix{L}{WISE,3}=\pedix{\nu}{W3}\times \pedix{L}{W3}$, 
where \pedix{L}{W3} and \pedix{\nu}{W3} are the monochromatic luminosity (derived from the catalogued WISE magnitude as 
detailed in Sect.~\ref{sect:data}) and the central frequency correspondent to the W3 band, 
observed frame (corresponding to $\sim 4.5-9.1\,\mu$m rest frame, depending on the redshift).
Since early studies based on IRAS data \citep[\eg][]{spinoglio89}, it is known that the AGN flux emitted at these wavelengths 
is a fraction almost constant of the bolometric flux.

\begin{figure}
\begin{center}
 \resizebox{1\hsize}{!}{\includegraphics[]{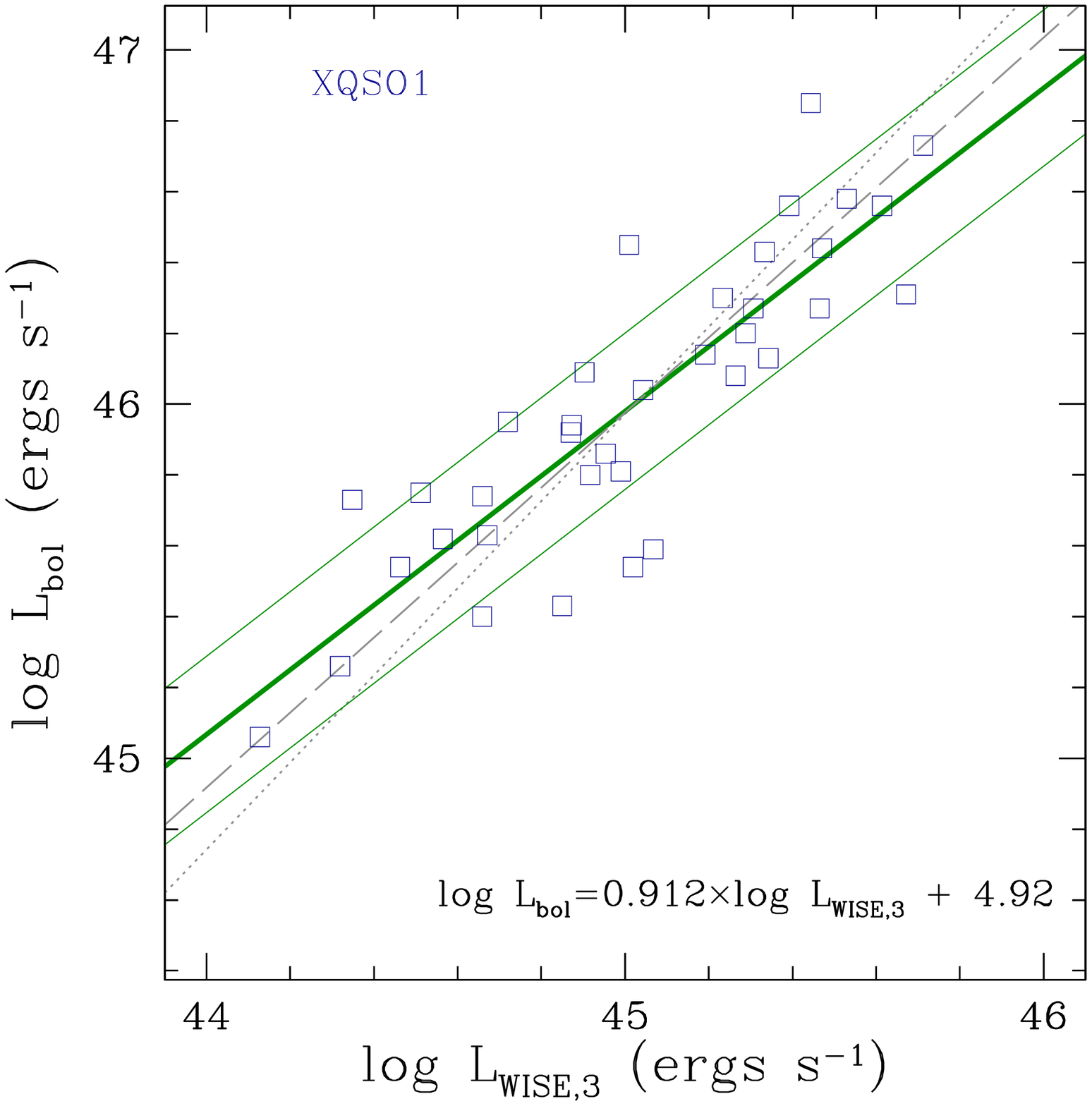}}
\caption{
Comparison between the bolometric luminosity 
\citep[estimated from the optical/UV SED by][]{marchese12} and the luminosity derived from the WISE W3 data
for the XQSO1s with a detection in WISE.
The green thick solid line is the result of the least-squares fitting when \pedix{L}{bol} is assumed as dependent variable 
(corresponding to the relation quoted in the plot; the thinnest solid green lines mark the $1\sigma$ dispersion).
The grey dotted line corresponds to the least-squares fitting when \pedix{L}{WISE,3} is assumed as dependent variable, 
while the long-dashed line is the bisector of the two least-squares fitting lines. 
}
\label{fig:cflbol}
\end{center}
\end{figure}

\begin{figure}
\begin{center}
 \resizebox{1\hsize}{!}{\includegraphics[]{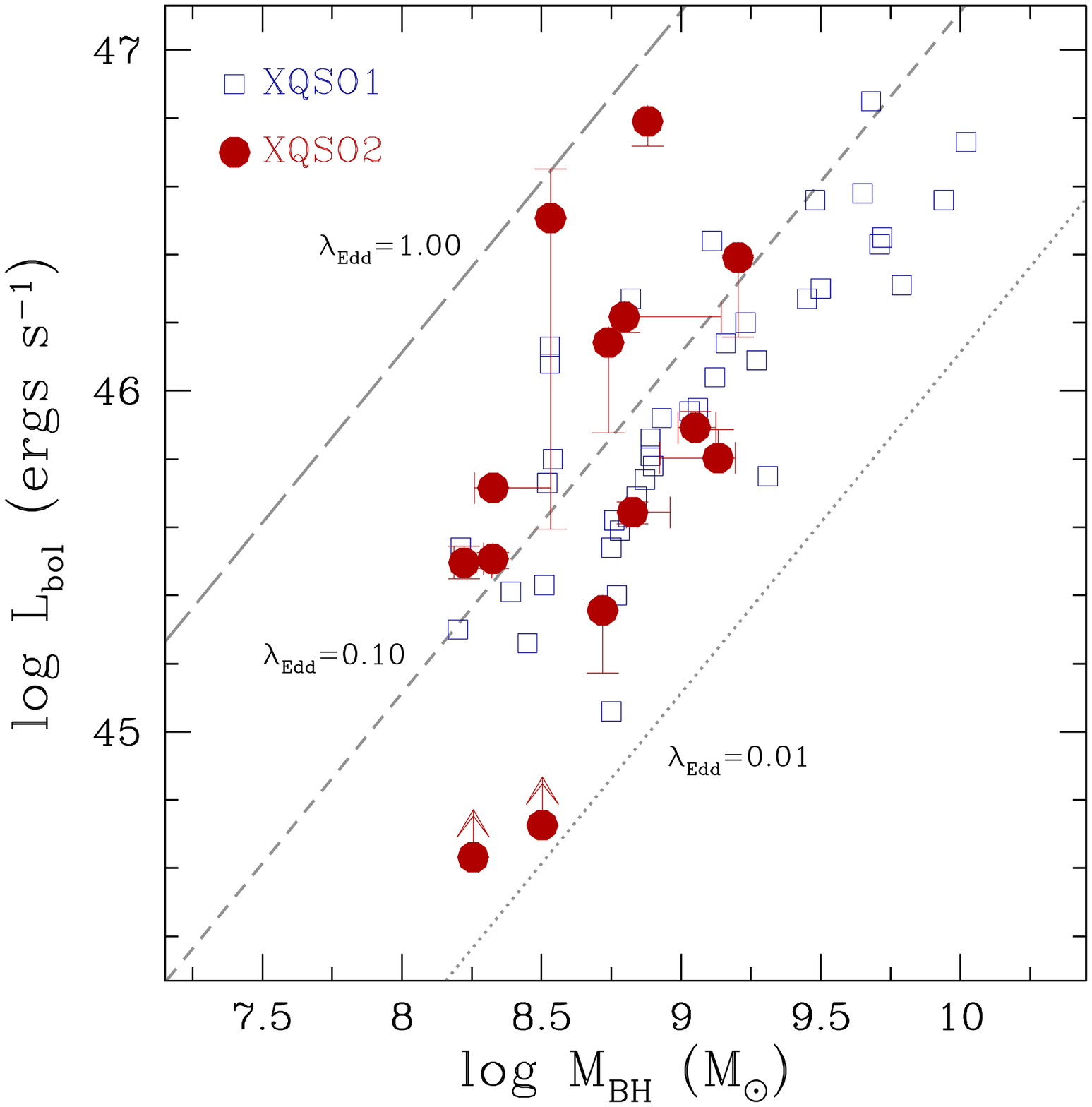}}
\caption{Bolometric luminosities vs. black hole masses for the sample of
XQSO2s (red filled circles); the bars identify the ranges of values obtained 
when all the possible modelling (with all the QSO templates and extinction curves) are considered.
The arrows mark the two objects for which we are able to put only limits to the nuclear emission 
(namely, \llqsoa\ and \llqsob).
The blue open squares mark the positions of the XQSO1s.
Grey long-dashed, dashed and dotted lines define the locus
for sources emitting at the Eddington limit, at $1/10$ of it, and at $1/100$ of it.}
\label{fig:mbhlbol}
\end{center}
\end{figure}

We then compared \pedix{L}{WISE,3} with the \pedix{L}{bol}; the latter has been derived by
\citet{marchese12} from an optical/UV SED modelling (therefore the IR emission is not included in the calculation).
The ratio between \pedix{L}{bol} and \pedix{L}{WISE,3} is approximately constant, with a mean ``WISE W3'' bolometric correction 
$\langle\pedix{k}{WISE,3}\rangle=10.77$  (see Fig.~\ref{fig:cflbol}).
In these high-luminosity QSOs the bolometric and W3 luminosities are well-correlated over $\sim 2\,$dex.
The best-fitting relations using an ordinary least squares (OLS) analysis, treating as independent variable 
$\log\pedix{L}{WISE,3}$ [OLS(Y$|$X)] or $\log\pedix{L}{bol}$ [OLS(X$|$Y)],
as well as the OLS bisector\footnote{This is the line that bisects the smaller of the two angles between the OLS(Y$|$X) and OLS(X$|$Y) lines.},
are as follows:
\begin{eqnarray}
 \label{eq:lirlbol1}\quad\log\pedix{L}{bol} = 0.912\cdot \log\pedix{L}{WISE,3}+4.92 &\qquad&\mbox{[OLS(Y$|$X)]}  \\
 \label{eq:lirlbol2}\log\pedix{L}{bol} = 1.229\cdot \log\pedix{L}{WISE,3}-9.34 & &\mbox{[OLS(X$|$Y)]}  \\
 \label{eq:lirlbol3}\log\pedix{L}{bol} = 1.058\cdot \log\pedix{L}{WISE,3}-1.65 & &\mbox{[bisector]}  
\end{eqnarray}
The bolometric luminosities of the XQSO2s have been recovered from equation~\ref{eq:lirlbol1}; the 
\pedix{L}{WISE,3} 
has been derived from the nuclear component, as obtained in our deconvolution of the SED, after correcting for the estimated absorption.
The estimated \pedix{L}{bol} are reported in Table~\ref{tab:der} and are plotted versus the \pedix{M}{BH} in Fig.~\ref{fig:mbhlbol}
(where the diagonal lines represent the trend between \pedix{L}{bol} and \pedix{M}{BH} at different fractions of the Eddington luminosity).
The distribution of X-ray bolometric corrections, $\pedix{k}{X}\equiv\pedix{L}{bol}/\pedix{L}{2-10\kev}$, is reported in Fig.~\ref{fig:disres} (central panel).
We note again that our framework is the basic Unified Model; in particular, we are assuming that the covering factors of absorbed and unabsorbed objects 
are the same.

\addtocounter{numfig}{1}

From the black hole mass and the bolometric luminosity, we can estimate the accretion rate normalized to the Eddington luminosity,
\pedix{\lambda}{Edd} (as defined in Sect.~\ref{sect:intro}).
The values are reported in Table~\ref{tab:der}; as shown in Fig.~\ref{fig:disres} (right-hand panel), \pedix{\lambda}{Edd} ranges from $0.01$ and $0.73$;
for $6$ out of $14$ QSOs (among them, the two sources for which we are only able to put limits to the nuclear emission) we 
found $\pedix{\lambda}{Edd}<0.1$.

%
\begin{table*}
\begin{minipage}[h]{1\textwidth}
 \caption{Derived properties for the XQSO2s in the XBS, estimated assuming the mean nuclear template, ``QSO1''. In parentheses, we report the range of values obtained 
when all the possible modelling (with all the QSO templates and extinction curves) are considered.}
 \label{tab:der}
 \begin{center}
 \renewcommand{\footnoterule}{}  
 {\footnotesize
  \begin{tabular}{l@{\extracolsep{0.2cm}} r@{\extracolsep{0.4cm}} r@{\extracolsep{0cm}.}l@{\extracolsep{-0.4cm}} r@{\extracolsep{0.1cm}} c@{\extracolsep{0.1cm}-} c@{\extracolsep{0.3cm}}  r@{\extracolsep{0.1cm}}  c@{\extracolsep{0.1cm}-} c@{\extracolsep{0.3cm}} l@{\extracolsep{0.3cm}} r@{\extracolsep{0.1cm}} c@{\extracolsep{0.1cm}-} c@{\extracolsep{0.3cm}} r@{\extracolsep{0.1cm}} c@{\extracolsep{0.1cm}\,\,-} c}
   \hline\hline       
     \multicolumn{1}{c}{Name} & \pedix{A}{V} & \multicolumn{2}{l}{\pedap{A}{V}{exp. from \nhsym}} & \multicolumn{3}{c}{\pedSM{k}{X}} & \multicolumn{3}{c}{$\log\pedSM{L}{bol}$} & Host & \multicolumn{3}{c}{$\log\pedSM{M}{BH}$} &  \multicolumn{3}{c}{$\log\pedSM{\lambda}{Edd}$} \\
     \cline{5-7}\cline{8-10}\cline{12-14}\cline{15-17}
     \multicolumn{1}{c}{(1)}  & (2) & \multicolumn{2}{c}{(3)} & \multicolumn{3}{c}{(4)} & \multicolumn{3}{c}{(5)} & (6)  & \multicolumn{3}{c}{(7)} & \multicolumn{3}{c}{(8)} 
    \vspace{0.1cm} \\
   \hline   
    \vspace{-0.2cm} \\
  XBSJ000100.2-250501 & $ 1.5$ & $  3$&$7$ & $ 11.4$ & ($ 11.4 $ & $  11.4$) & $45.7$ & ($45.7 $ & $ 45.7$) & Ell2 & $8.3$ & ($8.3 $ & $ 8.5$) & $-0.7$ & ($-0.9 $ & $ -0.7$) \\
  XBSJ013240.1-133307 & $ 1.9$ & $ 16$&$8$ & $ 17.7$ & ($ 17.7 $ & $  19.7$) & $45.9$ & ($45.9 $ & $ 45.9$) & Sa   & $9.1$ & ($9.0 $ & $ 9.1$) & $-1.3$ & ($-1.3 $ & $ -1.2$) \\
  XBSJ021642.3-043553 & $ 5.0$ & $ 22$&$1$ & $ 20.0$ & ($ 16.9 $ & $  20.0$) & $46.8$ & ($46.7 $ & $ 46.8$) & Sd   & $8.9$ & ($8.9 $ & $ 8.9$) & $-0.2$ & ($-0.3 $ & $ -0.2$) \\
  XBSJ022707.7-050819 & $15.0$ & $  6$&$9$ & $ 25.5$ & ($ 22.8 $ & $  28.5$) & $45.5$ & ($45.4 $ & $ 45.5$) & S0   & $8.2$ & ($8.2 $ & $ 8.3$) & $-0.8$ & ($-0.9 $ & $ -0.8$) \\
  XBSJ050536.6-290050 & $ 3.6$ & $  3$&$2$ & $ 16.1$ & ($ 15.1 $ & $  16.8$) & $45.5$ & ($45.5 $ & $ 45.5$) & Sdm  & $8.3$ & ($8.3 $ & $ 8.3$) & $-0.9$ & ($-1.0 $ & $ -0.9$) \\
  XBSJ051413.5+794345$^{\myddag}$ & $ 2.5$ & $  3$&$2$ & $ 10.7$ & ($  9.9 $ & $  11.5$) & $45.6$ & ($45.6 $ & $ 45.7$) & Ell2 & $8.8$ & ($8.8 $ & $ 9.0$) & $-1.3$ & ($-1.5 $ & $ -1.2$) \\
  XBSJ052128.9-253032$^{\mydag}$ & $ 1.6$ & $ 72$&$7$ & $  1.7$ & ($  1.7 $ & $  22.4$) & $44.7$ & ($44.7 $ & $ 45.8$) & S0   & $8.5$ & ($8.5 $ & $ 8.5$) & $-1.9$ & ($-1.9 $ & $ -0.8$) \\
  XBSJ080411.3+650906$^{\myddag}$$^{\mydag}$ & $ 3.1$ & $ 13$&$0$ & $  1.3$ & ($  1.1 $ & $   2.7$) & $44.6$ & ($44.6 $ & $ 45.0$) & Sa   & $8.3$ & ($8.3 $ & $ 8.3$) & $-1.7$ & ($-1.8 $ & $ -1.4$) \\
  XBSJ113148.7+311358 & $14.0$ & $ 16$&$5$ & $ 22.0$ & ($ 14.4 $ & $  23.0$) & $45.4$ & ($45.2 $ & $ 45.4$) & Sd   & $8.7$ & ($8.7 $ & $ 8.7$) & $-1.5$ & ($-1.7 $ & $ -1.5$) \\
  XBSJ122656.5+013126 & $ 1.6$ & $ 12$&$6$ & $ 30.7$ & ($ 27.7 $ & $  30.7$) & $46.2$ & ($46.2 $ & $ 46.2$) & Sdm  & $8.8$ & ($8.8 $ & $ 9.1$) & $-0.7$ & ($-1.1 $ & $ -0.7$) \\
  XBSJ134656.7+580315 & $25.0$ & $ 48$&$9$ & $224.3$ & ($131.0 $ & $ 228.1$) & $46.4$ & ($46.2 $ & $ 46.4$) & Ell5 & $9.2$ & ($9.2 $ & $ 9.2$) & $-0.9$ & ($-1.2 $ & $ -0.9$) \\
  XBSJ144021.0+642144 & $ 0.2$ & $  3$&$5$ & $ 21.0$ & ($ 21.0 $ & $  25.5$) & $45.8$ & ($45.8 $ & $ 45.9$) & Sa   & $9.1$ & ($8.9 $ & $ 9.2$) & $-1.4$ & ($-1.4 $ & $ -1.1$) \\
  XBSJ160645.9+081525 & $62.0$ & $145$&$1$ & $ 60.0$ & ($  7.3 $ & $  83.7$) & $46.5$ & ($45.6 $ & $ 46.7$) & S0   & $8.5$ & ($8.5 $ & $ 8.5$) & $-0.1$ & ($-1.1 $ & $  0.0$) \\
  XBSJ204043.4-004548 & $18.0$ & $ 17$&$3$ & $ 41.9$ & ($ 22.7 $ & $  41.9$) & $46.1$ & ($45.9 $ & $ 46.1$) & Sd   & $8.7$ & ($8.7 $ & $ 8.7$) & $-0.7$ & ($-1.0 $ & $ -0.7$) \\
  \end{tabular}
 }
 \end{center}       
 {\footnotesize   {\sc Note:} $^{\myddag}\,$ New optical identifications. $^{\mydag}\,$The two objects for which we are able to put only limits to the nuclear emission.
 \vspace{0.1cm}\\
 \footnotesize Col. (1): Source name in the XBS sample.
 \footnotesize Col. (2): Dust extinction (in mag).
 \footnotesize Col. (3): Dust extinction (in mag) expected from the observed \nhsym, for a Galactic dust-to-gas ratio.
 \footnotesize Col. (4): X-ray bolometric correction.
 \footnotesize Col. (5): Nuclear bolometric luminosity, recovered as described in Sect.~\ref{sect:nucl}, in units of \lum.
 \footnotesize Col. (6): Galactic template used in the SED modelling.
 \footnotesize Col. (7): Mass of the central BH, in units of \pedix{M}{\sun}.
 \footnotesize Col. (8): Eddington ratio, defined as $\pedSM{\lambda}{Edd}\equiv\pedSM{L}{bol}/\pedSM{L}{Edd}$}.
\end{minipage}
\end{table*}
\addtocounter{numtab}{1}

\addtocounter{numfig}{1}

\section{Discussion}\label{sect:disc}

\subsection{XQSO1s vs. XQSO2s in the XBS}\label{sect:xqso12}

In the previous sections we described how we recovered \pedix{L}{bol}, \pedix{M}{BH} and \pedix{\lambda}{Edd}
for the {\it complete} sample of XQSO2s in the XBS.
Thanks to the analysis previously performed for the XQSO1s \citep{marchese12,caccianiga13,fanali13}, 
we are now able to compare the physical properties of X-ray absorbed and X-ray unabsorbed QSOs 
on a statistical basis.

Using the project for statistical computing\footnote{http://www.r-project.org/index.html} {\it R} \citep{rstat}, 
we performed a set of two-samples KS tests; we compared for various parameters of interest 
the cumulative distributions obtained for the $40$ XQSO1s and for the $12$ XQSO2s for which we were able to infer the 
intensity of the nuclear emission (\ie\ excluding \llqsoa\ and \llqsob). 
There is no evidence that the two samples are drawn from different populations
in terms of any of the tested quantities (see the first row in Table~\ref{tab:ks}, where the results are summarized), 
although the KS probability for the Eddington ratio is quite low, $p\sim 4$\%.

While the lower end of the two redshift distributions are similar ($z=0.32$ and $z=0.35$ for XQSO1s and XQSO2s, respectively),
all but one the XQSO2s have $z<0.85$, while $40$\% of the XQSO1s have a redshift between $0.85$ and $1.64$ (see Fig.~\ref{fig:znhdist}, left-hand panel).
In Table~\ref{tab:ks}, second row, we quote the results of the KS tests performed on the two subsamples with $z<0.85$ 
(their distributions in terms of intrinsic X-ray luminosity, black hole mass, X-ray bolometric correction and Eddington ratio are reported in the insets of
Fig.~\ref{fig:znhdist}, central panel, and Fig.~\ref{fig:disres}):
also with this cut applied, the hypothesis of the same original population is confirmed, although with a lower significance in \pedix{L}{2-10\kev}.
The KS probability for \pedix{M}{BH} increases significantly when we restrict to $z<0.85$;
$10$ out of the $13$ XQSO1s with $\pedix{M}{BH}\geq 1.6\times10^{9}\,\pedix{M}{\sun}$, that is, the maximum value found for the XQSO2s
(see Fig.~\ref{fig:disres}, left-hand panel, and Fig.~\ref{fig:mbhlbol}) have $z>1$.

\begin{table}
\begin{minipage}[h]{0.47\textwidth}
 \caption{Results of the KS tests between the XQSO1s and the XQSO2s with well-constrained nuclear component.}
 \label{tab:ks}
 \begin{center}
 \renewcommand{\footnoterule}{}  
 {
   \begin{tabular}{r|| c| c| c| c| c| c}
     & \multicolumn{6}{c}{Probability}  \\
     \cline{2-7}
      & $z$ & $\pedix{L}{X}$ & $\pedix{M}{BH}$  & $\pedix{L}{bol}$ & $\pedix{k}{X}$  & $\pedix{\lambda}{Edd}$ \\
   \hline\hline       
      All $z$ & 0.284 & 0.412 & 0.284 & 0.999 & 0.988 & 0.036 \\
      $z<0.85$ & 0.860 & 0.126 & 0.716 & 0.594 & 0.942 & 0.072 
  \end{tabular}
 }
 \end{center}       
 {\footnotesize   {\sc Note:} For each physical parameter, the probability 
 for the null hypothesis (\ie\ the two cumulative distributions are drawn from the same parent population)
 is reported (first and second rows, for the whole samples and the subsamples with $z<0.85$, respectively).
}
\end{minipage}
\end{table}
\addtocounter{numtab}{1}

\subsection{IR emission and \pedix{L}{bol}}\label{sect:cflbol}

\begin{figure}
\begin{center}
 \resizebox{1\hsize}{!}{\includegraphics[]{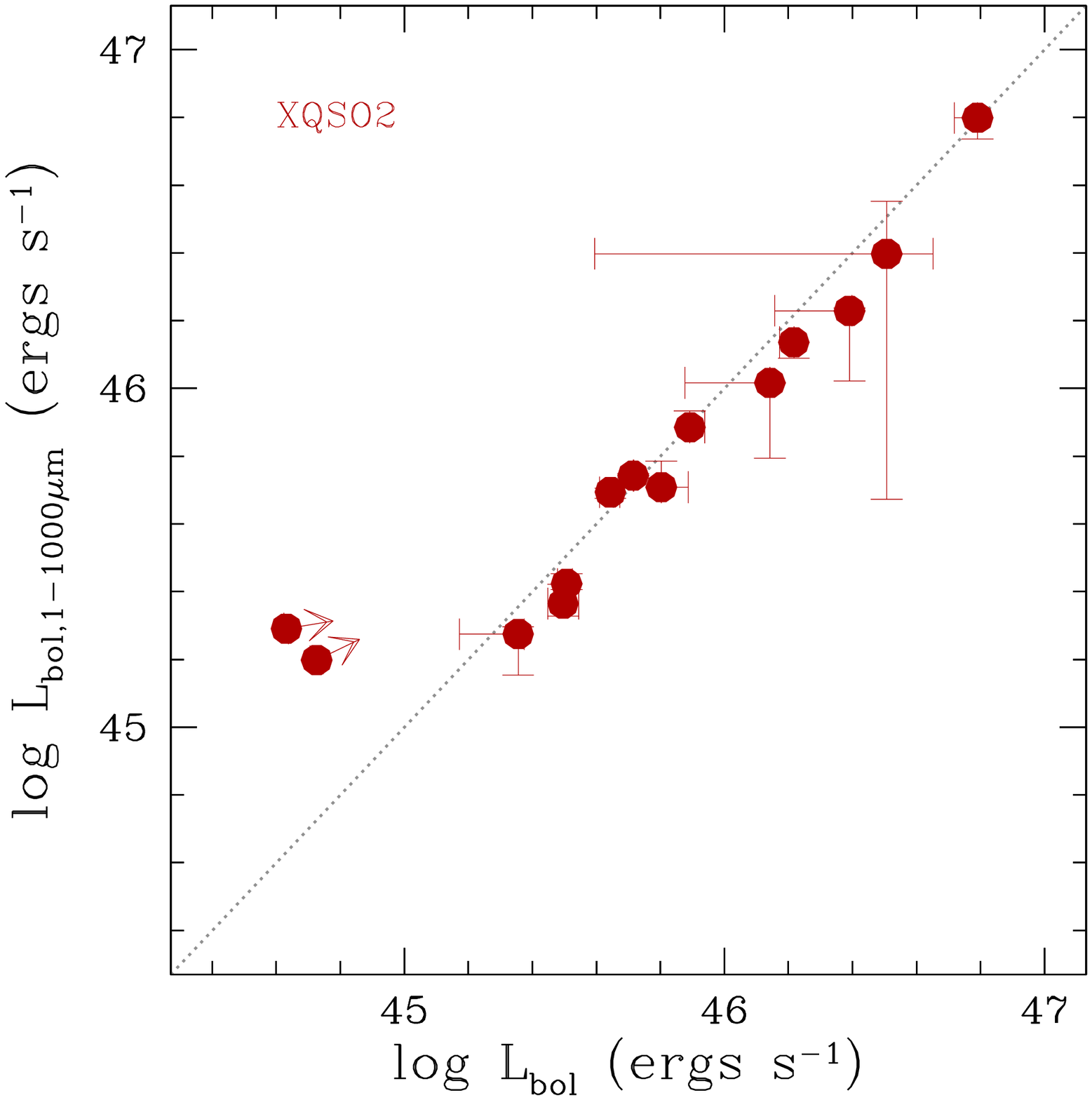}}
\caption{IR-based  \pedix{L}{bol} derived for the XQSO2s following the prescriptions 
by \citet{pier92} and \citet{pozzi07} vs. the WISE-based 
 \pedix{L}{bol} derived by applying the relation between the WISE W3 data and the bolometric luminosity
recovered from the analysis of the XQSO1s in the XBS; the bars identify the ranges of values obtained 
when all the possible modelling (with all the QSO templates and extinction curves) are considered.
The arrows mark the two objects for which we are able to put only limits to the nuclear emission 
(namely, \llqsoa\ and \llqsob).
The dotted line marks the one-to-one relation between the luminosities.
}
\label{fig:cflbollbolir}
\end{center}
\end{figure}
\addtocounter{numfig}{1}

In Sect.~\ref{sect:lboleddr}, we derived the bolometric luminosity of the XQSO2s using the properties of the XQSO1s.
The SED decomposition presented in Sect.~\ref{sect:sed} 
allows us to check these estimates.
The IR luminosity is considered an indirect probe of the accretion
disk optical/UV luminosity; using an approach similar to \citet{pozzi07,pozzi10} and \citet{vasudevan10},
the bolometric luminosity can be computed as the sum of the total IR
luminosity and the X-ray luminosity,  $\pedix{L}{bol}=\pedix{L}{IR}$+\pedix{L}{X}, 
where the former can be obtained from the nuclear luminosity between 
$1$ and $1000\,\mu$m.

We calculate the \pedix{L}{1-1000\,\mum} by integrating the QSO template between 
$1$ and $1000\,\mu$m after correcting for absorption.
Following the works of \citet{pier92} and \citet[see also \citealt{lusso11}]{pozzi07},  \pedix{L}{1-1000\,\mum} can be converted
into the nuclear accretion disk luminosity when corrected for geometrical effects related to the covering factor $f$ of the torus.
Following \citet{pozzi07,pozzi10}, we assumed $f\sim0.67$, corresponding to an angle $\theta\sim48\arcdeg$ between 
the axis of the disk
and the edge of the torus.
This is of course 
a mean value, and the derived luminosities must be assumed just as a first-order estimate.

The bolometric luminosities as derived from equation~\ref{eq:lirlbol1} are compared to the $1-1000\,\mu$m-based bolometric luminosities 
in Fig.~\ref{fig:cflbollbolir}.
The $12$ sources with well-constrained nuclear component in our SED modelling fall around the equal-luminosity relation, with a mean and a
maximum scatter of $0.07\,$dex and $0.16\,$dex, respectively.
The agreement of the two estimates of \pedix{L}{bol} implies that absorbed and unabsorbed QSOs in the XBS 
have the same direct-to-reprocessed luminosity ratio.
Moreover, this confirms that equation~\ref{eq:lirlbol1} can be confidently used to derive a first-order estimate of the bolometric luminosity,
at least for unabsorbed high-luminosity QSOs, where the host galaxy is not expected to contaminate the emission in the WISE W3 
band.

\subsection{The effective Eddington limit}\label{sect:nhlambda}

\begin{figure*}
\begin{center}
 \resizebox{0.5\hsize}{!}{\includegraphics[]{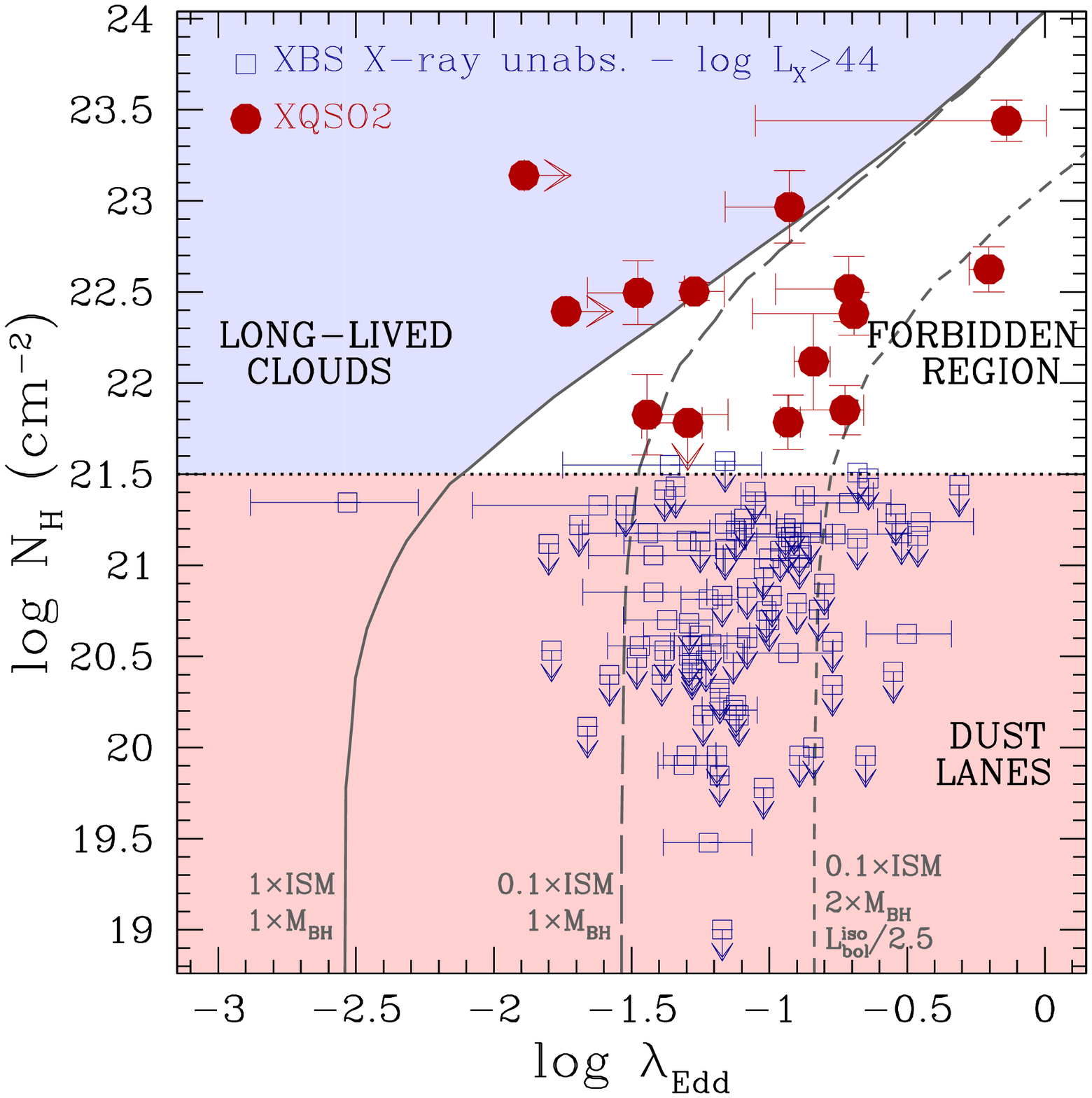}}
\caption{Hydrogen column density vs. Eddington ratio for the XQSO2s analysed in this work (red filed circles); 
the bars identify the ranges of values obtained 
when all the possible modelling (with all the QSO templates and extinction curves) are considered.
The horizontal arrows mark the two objects for which we are able to put only limits to the nuclear emission 
(namely, \llqsoa\ and \llqsob).
With the blue open symbols we show 
the whole sample (\ie, $\nhsym<4\times 10^{21}\,$\nh) of X-ray unabsorbed high-luminosity
($\pedix{L}{2-10\kev}>10^{44}\,$\lum) AGN 
in the XBS \citep[from ][]{marchese12}.
The lines and the different regions in the plot summarize the ``effective Eddington limit paradigm'' 
\citep[adapted from][see the details in the text]{fabian09}: 
the grey continuous line marks the effective Eddington limit for standard ISM and enclosed mass of \pedix{M}{BH}; 
the grey long-dashed line show how the effective Eddington limit moves when $1/10$ of standard ISM
grain abundance is assumed; if in addition the enclosed mass is twice the \pedix{M}{BH} and the incident radiation is reduced by a factor of $2.5$ (accounting for
an anisotropic emission from the disk; see the text for details), the limit can further increase (grey dashed line).
}
\label{fig:nheddr}
\end{center}
\end{figure*}
\addtocounter{numfig}{1}

The studies of \citet{fabian08} and \citet{fabian09} identify an ``effective Eddington limit''
for dusty gas in the \nhsym- \pedix{\lambda}{Edd} plane (see Fig.~\ref{fig:nheddr}).
Starting from the model presented in 
\citet{fabian06}, the authors suggest a connection between the radiation pressure exerted
by the AGN on its proximity (\ie, where the influence of the BH is dominant)
and the structure of the surrounding material.
Briefly, for a fixed nuclear luminosity, the incident radiation would result
on an easier sweeping away of the absorber when the medium is more dusty.
Assuming a dust-to-gas ratio constant during the process of ``cleaning'', this translates in a  ``forbidden region'' at high \nhsym\
and high ``classical''  \pedix{\lambda}{Edd} (\ie, derived by assuming a distribution of gas composed only of ionized hydrogen).
In this region, absorbing dusty gas clouds are unstable to radiation and absorption may be transient or variable.
For lower ``classical'' Eddington ratios, also quite dense absorbing clouds can survive to the
radiation pressure (``long-lived clouds region'').
Finally, sources with low \nhsym\ ($\lesssim 3\times10^{21}\,$\nh) can accrete at high \pedix{\lambda}{Edd}, being the absorption 
possibly associated with dust lanes, far away from the nucleus, so as
to be retained by a gravitational mass large enough to balance also an intense nuclear emission.

In Fig.~\ref{fig:nheddr} we plot the XQSO2s studied here (red filled circles) in the \nhsym-\pedix{\lambda}{Edd} diagram; 
the blue open squares mark the positions of the X-ray unabsorbed AGN in the XBS with $\pedix{L}{2-10\kev}>10^{44}\,$\lum.
Unlike in the previous sections, to properly fill the plane we do not restrict the comparison to AGN 
with $\nhsym<5\times10^{20}\,$\nh, but we consider the whole sample studied by \citet[$\nhsym<4\times 10^{21}\,$\nh]{marchese12}.
As evident from the figure, our XQSO2s populate the ``forbidden region'' proposed by Fabian and collaborators (the white part of the plot).
The agreement found in Sect.~\ref{sect:cflbol} between the different estimates of \pedix{L}{bol} guarantees that this result does not depend
on the method adopted to derive the bolometric luminosity.
In particular, the small scatter found between the bolometric luminosities derived from equation~\ref{eq:lirlbol1} and from \pedix{L}{1-1000\,\mum} is not enough to 
move any source outside the ``forbidden region''.

The ``effective Eddington limit'' drawn in Fig.~\ref{fig:nheddr} (grey continuous line)
has been calculated for a grain abundance typical of the interstellar medium (ISM), and under the hypothesis that the only inward gravitational 
force of importance is that from the central BH \citep[see][]{fabian09}.
Different assumptions can modify the shape of the ``forbidden region''.
The lower is \nhsym, the stronger are the effects of a decreasing of the grain abundance (\eg, long-dashed grey line in Fig.~\ref{fig:nheddr}).
Ratios of optical extinction \pedix{A}{V} to X-ray column density \nhsym\ lower than the Galactic dust-to-gas ratio
are not unusual in AGN \citep{maiolino01}; in $9$ out of the $12$ XQSO2s, for which we were able to properly model the nuclear properties, 
we find a dust-to-gas ratio lower than the Galactic one
(we stress again that in Sect.~\ref{sect:sed} the dust extinction from the SED modelling 
was derived independently of the value of \nhsym\ from the X-ray spectra).

As suggested by  \citet{fabian09}, the presence of stars inwards from the dusty gas clouds 
close to the nucleus
would increase the effective Eddington limit by an amount proportional to the total enclosed mass, regardless of the amount of obscuration.
Near-IR high-resolution data for $9$ Seyfert galaxies \citep{davies07} have revealed evidence
for recent, intense, short-lived starbursts on few pc scale distance from the BH,
leaving behind massive and vertically thick nuclear star clusters.

We suggest here a further effect that could concur in reducing the proposed locus of short life: 
the estimation of the radiation pressure boost factor 
presumes isotropic emission from the accretion disk, while emission from an optically thick disk is, in principle, anisotropic.
Following \eg\ \citet{kawaguchi10}, the dependence from the polar angle $\theta$ of the radiation flux from a unit surface area 
of the disk 
can be expressed as  $F(\theta) \sim \cos \theta (1 + 2 \cos \theta)$ \citep[see also ][]{laor89,sun89}.
Therefore, the assumption of isotropy, that adopts the radiation observed face-on as an estimate of the flux 
incident on the torus at an angle \pedix{\theta}{torus} with respect to the axis of the system, overestimates 
$F(\pedix{\theta}{torus})$ by a factor $\mathcal{C}$:
\begin{equation}\label{eq:disk}
 \mathcal{C}=\frac{\cos \pedix{\theta}{torus} (1 + 2 \cos \pedix{\theta}{torus} )}{\cos \pedix{\theta}{face-on} (1 + 2 \cos \pedix{\theta}{face-on} )}
\end{equation}
where  \pedix{\theta}{face-on} is the angle between the rotation axis of the disk and the line of sight.
As an example, for a $\pedix{\theta}{face-on}\sim 25\arcdeg$, typical of type~1 AGN,
the incident radiation on the torus surface at an angle $\pedix{\theta}{torus} \sim 60\arcdeg$
would be reduced by a factor of $\mathcal{C}\sim 0.4$.
This would reflect on an increasing in the ``effective Eddington limit'' in this direction by a factor of $1/\mathcal{C}\sim2.5$.

Qualitatively, a combination of low grain abundance (\eg, of the order of one tenth of the ISM abundance),
enclosed mass higher than \pedix{M}{BH} (\eg, twice the mass of the central BH),
and anisotropy of the disk emission, would 
reduce the proposed ``forbidden region''
in the \nhsym-\pedix{\lambda}{Edd} plane
as marked by the dashed grey line in Fig.~\ref{fig:nheddr}.


\section{Summary}\label{sect:summ}

In this paper, we presented a detailed analysis of the accretion properties of a complete sample of $14$ XQSO2s
(column density $\nhsym > 4\times10^{21}\,$\nh\ and X-ray luminosity 
$\pedix{L}{2-10\kev} > 10^{44}\,$\lum; $0.35<z<0.9$, plus one object at $z\sim 2$) drawn from the XBS.

The bright flux of these sources 
guarantees a high detection rate in a number of multiwavelength catalogues 
(\spitzer, WISE, \sdss, \galex), thus allowing us (in combination with proprietary observations) to construct photometric broad-band SED 
from the optical range up to
$\sim 25\,\mu$m (observed frame), or even $\sim 70\,\mu$m (observed frame).
By analysing the SED with a combination of empirical templates we were able to  
separate the nuclear and the host-galaxy emission.
For all the sources, we obtained the masses of the central
black hole from the luminosity of the host galaxy, using the well-known scaling relations linking the BH mass and the host properties.

To derive the nuclear bolometric luminosities, 
we used the properties of the 
X-ray unabsorbed high-luminosity AGN  ($\nhsym < 5\times10^{20}\,$\nh and $\pedix{L}{2-10\kev} > 10^{44}\,$\lum; XQSO1s)
in the XBS.
We compared their WISE luminosity at $12\,\mu$m (observed frame, a wavelength range where a powerful QSO 
is expected to dominate over the host galaxy emission) with the \pedix{L}{bol}, derived from an optical/UV SED modelling
\citep[an approach that does not include the IR emission;][]{marchese12}.
We found that in these high-luminosity unabsorbed QSOs the bolometric luminosity and the W3 luminosity are well-correlated over $\sim 2\,$dex.
The best-fitting relation has been then applied to the XQSO2s, adopting as W3 luminosity the value derived from the nuclear component 
after correcting for the estimated absorption.

The main results of our work can be summarized as follows.
   \begin{enumerate}
      \item Being the unabsorbed population in the XBS previously analysed \citep{marchese12,caccianiga13,fanali13}, 
      we could compare the two samples of XQSO1s and XQSO2s, looking for possible statistically significant differences.
      We found that 
      X-ray-selected absorbed and unabsorbed QSOs share the same intrinsic properties 
      (\pedix{L}{X}, \pedix{L}{bol}, \pedix{k}{X}, \pedix{M}{BH}  and \pedix{\lambda}{Edd}).
      \item The WISE-based \pedix{L}{bol} have been compared with the IR-based \pedix{L}{bol}, derived from the intrinsic 
      (\ie, absorption-corrected) \pedix{L}{1-1000\mum} as computed from the SED modelling.
      We found a good agreement between the two estimates.
      We conclude that the absorbed ad unabsorbed QSOs in the XBS sample have similar observed torus-reprocessed 
      luminosity-to-bolometric luminosity ratio.
      \item The XQSO2s analysed here populate the proposed ``forbidden region'' in the \pedix{N}{H} vs. \pedix{\lambda}{Edd} diagram.
      This result could imply a grain abundance much lower than in the local ISM or the presence of stars inwards of the obscuring material 
      with mass comparable to the mass of the BH.
      In addition, we propose that the anisotropy in the nuclear emission can also play a significant role. 
   \end{enumerate}
In conclusion, our work suggests that the hypothesis of different evolutionary states for absorbed and unabsorbed QSOs 
in addition to the basic Unified Model is not required, at least up to $z\sim 1$.

\section*{Acknowledgements}
We thank the referee for her/his constructive comments that improved the paper.
The research leading to these results has received funding from the European Commission Seventh Framework Programme 
(FP7/2007-2013) under grant agreement n.~267251 ``Astronomy Fellowships in Italy'' (AstroFIt).
The authors acknowledge financial support from the Italian Ministry of Education,
Universities and Research (PRIN2010-2011, grant n.~2010NHBSBE) and from ASI (grant n.~I/088/06/0).
SM and FJC acknowledge financial support by the Spanish Ministry of Economy and Competitiveness through grant AYA2012-31447.
SM acknowledges support from the ARCHES project ($7$th Framework of the European Union, n.~ 313146).

We thank F.~Pozzi, A.~Feltre and E.~Lusso for the useful discussions; T.~Maccacaro and L. Maraschi for their helpful 
suggestions; 
and R.~Paladini for her assistance with the WISE data and catalogue.

Based on observations obtained with \xmm\ (an ESA science mission with instruments and contributions directly funded by
ESA Member States and the USA, NASA).
Based on observations made under the proposals GTC18-10B and GTC44-11A with the Gran Telescopio Canarias (GTC), 
installed at the Spanish ``Observatorio del Roque de los Muchachos'' 
of the Instituto de Astrof\'\i{}sica de Canarias, in the island of La Palma.
This work is based in part on observations made with the \spitzer\ Space Telescope, which is operated by the Jet
Propulsion Laboratory, California Institute of Technology under a contract with NASA.

Funding for the \sdss\ and \sdss-II has been provided by the Alfred P. Sloan Foundation, the Participating Institutions,
the National Science Foundation, the U.S. Department of Energy, the National Aeronautics and Space Administration, the
Japanese Monbukagakusho, the Max Planck Society, and the Higher Education Funding Council for England.
The \sdss\ Web Site is http://www.sdss.org/.
The \sdss\ is managed by the Astrophysical Research Consortium for the Participating Institutions. 
This publication makes use of data products from the Wide-field Infrared Survey Explorer, which is a joint project of 
the University of California, Los Angeles, and the Jet Propulsion Laboratory/California Institute of Technology, funded 
by the National Aeronautics and Space Administration.
This research has made use of NASA's Astrophysics Data System.

\appendix


\section{New \xmm\ observations}\label{sect:xmm}

%
\begin{table*}
\begin{minipage}[h]{1\textwidth}
 \caption{EPIC {\xmm} observation details for the new data.}             
 \label{tab:xmmlog}      
 \begin{center}
 {
 \scriptsize
  \begin{tabular}{l@{\extracolsep{0.2cm}} l@{\extracolsep{0.2cm}} c@{\extracolsep{0.3cm}} c@{\extracolsep{0.3cm}} c@{\extracolsep{0.3cm}} c@{\extracolsep{0.3cm}} c@{\extracolsep{0.3cm}} c@{\extracolsep{0.3cm}} c@{\extracolsep{0.2cm}} c@{\extracolsep{0.2cm}} c@{\extracolsep{0.2cm}} c}
   \hline\hline       
    \multicolumn{1}{c}{Name} & \multicolumn{1}{c}{{\sc OBSID}} & Instr. & Filter & \multicolumn{2}{c}{Start Date \& Time} & \multicolumn{2}{c}{Stop Date \& Time}  & Tot. Exp. Time & Net Exp. Time  & Net Count Rates & $S/N$ \\
     \multicolumn{1}{c}{(1)} & \multicolumn{1}{c}{(2)} & (3) & (4) & (5) & (6) & (7) & (8) & (9) & (10) & (11) & (12)
    \vspace{0.1cm} \\
   \hline
    \vspace{-0.2cm} \\
            &		   & MOS1       & medium & 2008/12/16   & 18$:$55$:$40  & 2008/12/17   & 08$:$09$:$47  & $47.0$ & $34.03$ & $1.91\pm 0.08$ & 22.88 \\
     XBSJ013240.1-133307 & $0550960101$ & MOS2     & medium & 2008/12/16   & 18$:$55$:$40  & 2008/12/17   & 08$:$09$:$52  & $46.9$ & $36.05$ & $2.22\pm 0.08$ & 25.82 \\
             &  	   & pn       & thin   & 2008/12/16   & 19$:$18$:$25  & 2008/12/17   & 08$:$10$:$07  & $45.8$ & $22.97$ & $6.03\pm 0.17$ & 34.05	 
    \vspace{0.1cm} \\
   \hline   
    \vspace{-0.2cm} \\
            &		   & MOS1     & thin   & 2008/11/26   & 17$:$48$:$40  & 2008/11/27   & 00$:$24$:$27  & $23.6$ & $22.24$ & $0.45\pm 0.05$ &  8.13 \\
     XBSJ113148.7+311358 & $0550960301$ & MOS2     & thin   & 2008/11/26   & 17$:$48$:$40  & 2008/11/27   & 00$:$24$:$32 &  $23.6$ & $21.97$ & $0.53\pm 0.06$ &  8.85 \\
            &		   & pn       & thin   & 2008/11/26   & 18$:$11$:$25  & 2008/11/27   & 00$:$24$:$47  & $21.9$ & $17.92$ & $1.44\pm 0.10$ & 13.46  
    \vspace{0.1cm} \\
   \hline   
    \vspace{-0.2cm} \\
            &		   & MOS1     & thin   & 2009/02/01   & 18$:$09$:$06  & 2009/02/02   & 02$:$59$:$53  & $31.7$ & $27.57$ & $0.70\pm 0.08$ &  7.35 \\
     XBSJ160645.9+081525 & $0550960601$ & MOS2     & thin   & 2009/02/01   & 18$:$09$:$06  & 2009/02/02   & 02$:$59$:$58  & $31.7$ & $26.53$ & $0.63\pm 0.08$ &  7.10 \\
            &		   & pn       & thin   & 2009/02/01   & 18$:$31$:$51  & 2009/02/02   & 03$:$00$:$13& $30.0$ & $13.25$ & $2.22\pm 0.24$ &  7.77 \\
  \end{tabular}
 }
 \end{center}       
 {
 \footnotesize Column (1): source name in the XBS sample.
 \footnotesize Column (2): {\sc OBSID} of the new \xmm\ observation.
 \footnotesize Column (3): EPIC instrument.
 \footnotesize Column (4): EPIC filter.
 \footnotesize Columns (5) and (6): observation start date and time.
 \footnotesize Columns (7) and (8): observation end date and time.
 \footnotesize Column (9): performed duration, in units of ksec.
 \footnotesize Column (10): exposure time after removing high-background intervals, in units of ksec.
 \footnotesize Column (11): count rate after removing high-background intervals, in the energy range $0.3 - 10\,$keV, in units of $10^{-2}\,$counts~sec$^{-1}$.
 \footnotesize Column (12): signal-to-noise ratio after removing high-background intervals, in the energy range $0.3 - 10\,$keV.}
\end{minipage}
\end{table*}
\addtocounter{numtab}{1}
%

%
\begin{table*}
\begin{minipage}[h]{1\textwidth}
 \caption{X-ray spectral analysis of the new pn and MOS data. }
 \label{tab:xbf}
 \begin{center}
 {
  \begin{tabular}{r@{\extracolsep{0.1cm}}l@{\extracolsep{0.2cm}} l@{\extracolsep{0.2cm}} r@{\extracolsep{0.cm}.}l@{\extracolsep{0.2cm}} 
  r@{\extracolsep{0.cm}.}l@{\extracolsep{0.2cm}} r@{\extracolsep{0.cm}.}l@{\extracolsep{-0.1cm}} 
  r@{\extracolsep{0.cm}.}l@{\extracolsep{0.2cm}} r@{\extracolsep{0.cm}.}l@{\extracolsep{0.1cm}} 
  r@{\extracolsep{0cm}}l@{\extracolsep{0.2cm}} c@{\extracolsep{0.2cm}} c@{\extracolsep{0.2cm}} 
  r@{\extracolsep{0.cm}.}l@{\extracolsep{0.cm}/}l}
   \hline\hline       
    \multicolumn{2}{c}{Name} & \multicolumn{1}{c}{BF model} & \multicolumn{2}{c}{\nhsym} & \multicolumn{2}{c}{$\Gamma$} & 
    \multicolumn{2}{c}{\pedSM{N}{unabs}/\pedSM{N}{abs}} & \multicolumn{2}{c}{$R$} & 
    \multicolumn{2}{c}{\pedSM{E}{g}} & \multicolumn{2}{c}{EW} & \pedSM{F}{2-10\kev} & \pedapSM{L}{2-10\kev}{intr} & 
    \multicolumn{3}{c}{\chidof} \\

    \multicolumn{2}{c}{(1)} & \multicolumn{1}{c}{(2)} & \multicolumn{2}{c}{(3)} & \multicolumn{2}{c}{(4)} & 
    \multicolumn{2}{c}{(5)} & \multicolumn{2}{c}{(6)} & 
    \multicolumn{2}{c}{(7)} & \multicolumn{2}{c}{(8)} & (9) & (10) & 
    \multicolumn{3}{c}{(11)}
    \vspace{0.1cm} \\
   \hline
    \vspace{-0.2cm} \\
    XBSJ013240.1-133307 & & aPL+scatt+line & $ 3$&$22\errUD{0.40}{0.49}$ & $ 1$&$47\errUD{0.09}{0.13}$ &
     $ 0$&$05\pm 0.02$ & \multicolumn{2}{c}{-} & 
     $ 6$&$25\errUD{0.09}{0.08}$ & $108$&$\errUD{80}{68}$ & $3.81\errUD{0.28}{0.31}$ & $4.39\errUD{0.27}{0.29}$ &
     $135$&$8$&$128$ \\ 
    XBSJ113148.7+311358 & & aPL+scatt+refl & $ 4$&$05\errUD{2.39}{1.84}$ & $ 1$&$59\errUD{0.39}{0.40}$ &
    $ 0$&$07\errUD{0.06}{0.04}$ & $0$&$98\errUD{0.02}{0.66}$ & 
     \multicolumn{2}{c}{-} & \multicolumn{2}{c}{-} & $1.17\errUD{1.20}{0.60}$ & $1.01\errUD{1.83}{0.41}$ &
     $ 14$&$9$&$16$ \\ 
    XBSJ160645.9+081525 &  & aPL+scatt & $29$&$62\errUD{9.63}{8.01}$ & $ 2$&$02\errUD{0.56}{0.52}$ &
     \multicolumn{2}{c}{$<0.05$} & \multicolumn{2}{c}{-} & 
     \multicolumn{2}{c}{-} & \multicolumn{2}{c}{-} & $1.77\errUD{3.84}{1.20}$ & $5.87\errUD{12.96}{3.98}$ &
     $16$&$6$&$25$ \\ 
  \end{tabular}
 }
 \end{center}       
 
  {\footnotesize   {\sc Note:} Fluxes and luminosities refer to the MOS calibration. Errors are quote at $90$\% confidence level.
 \vspace{0.1cm}\\
 \footnotesize Column (1): source name in the XBS sample.
 \footnotesize Column (2): best-fitting model: aPL~=~absorbed power law; scatt~=~scattering component (additional unabsorbed power law
 with the same photon index as the primary one); refl~=~neutral Compton reflection (continuum + Fe and Ni lines; 
 the {\sc pexmon} model in XSPEC, with the same photon index as the primary power law); line~=~narrow Gaussian line. 
 \footnotesize Column (3): intrinsic column density; in units of $10^{22}\,$\nh.
 \footnotesize Column (4): source photon index.
 \footnotesize Column (5): scattering fraction, defined as the ratio of the unabsorbed and the direct power-law normalizations.
 \footnotesize Column (6): reflection scaling factor of the {\sc pexmon} component.
 \footnotesize Column (7): line energy.
 \footnotesize Column (8): line EW.
 \footnotesize Column (9): observed flux (deabsorbed by our Galaxy) in the $2-10\,$keV energy band, in units of $10^{-13}\,$\flux.
 \footnotesize Column (10): luminosity of the direct component (deabsorbed by intrinsic and Galactic \nhsym) in the $2-10\,$keV energy band,
  in units of $10^{44}\,$\lum.
 \footnotesize Column (11): \apix{\chi}{2} and number of degree of freedom.
}
\end{minipage}
\end{table*}
\addtocounter{numtab}{1}
%


\begin{figure*}
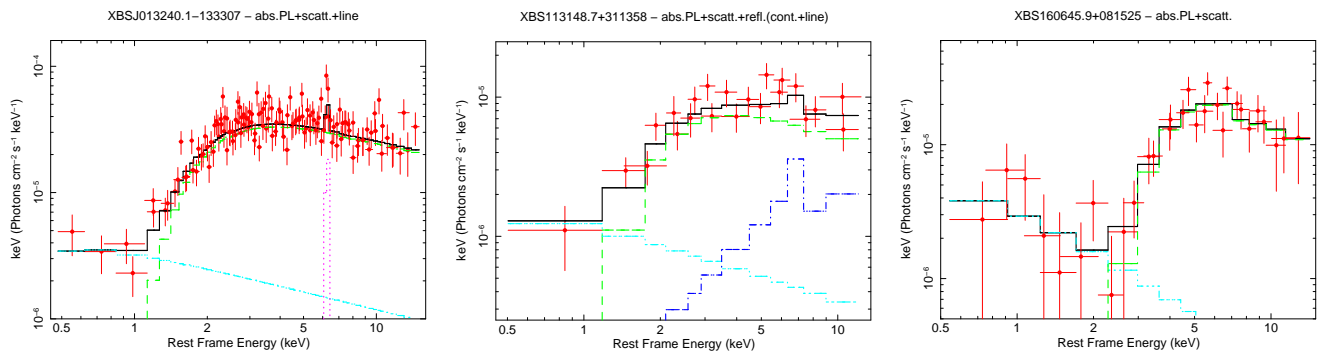

 \centering
 \resizebox{0.33\hsize}{!}{\includegraphics[angle=270]{figA1a.ps}}
 \resizebox{0.33\hsize}{!}{\includegraphics[angle=270]{figA1b.ps}}
 \resizebox{0.33\hsize}{!}{\includegraphics[angle=270]{figA1c.ps}}
 \caption{New \xmm\ EPIC spectra (in red). 
 The black continuous line represents the
 total best-fitting model; the underlying continuum is modelled as detailed: for XBSJ013240.1-133307 ({\it left-hand panel}),
 a dominant absorbed power-law component (dashed green line) plus a scattered power law (dash-dot-dot-dotted cyan line) and
 a \feka\ emission line (magenta dotted line).
 For XBSJ113148.7+311358 ({\it central panel}), an absorbed power law (dashed green line) plus a reflection component (continuum+line; dot-dashed blue line), 
 with the addition of a scattered power law (dash-dot-dot-dotted cyan line).
 Finally, for XBSJ160645.9+081525 ({\it right-hand panel}) the model is an absorbed (dashed green line) plus an unabsorbed (dash-dot-dot-dotted cyan line) 
 power-law components. 
 }
 \label{fig:xmmnew}%
\end{figure*}
\addtocounter{numfig}{1}

\subsection{Data reduction}\label{sect:xmmred}

The observations were performed with the European Photon Imaging Camera (EPIC),
the OM \citep{om} and the Reflection Grating Spectrometer (RGS);
the sources are not detected with the last instrument.
The three EPIC cameras \citep[pn, MOS1, and MOS2;][]{pn,mos} were operating in
full frame mode with the thin filter, except for XBSJ013240.1-133307 that was
observed by the MOS cameras with the medium filter.
The observation details are reported in Table~\ref{tab:xmmlog}.

The \xmm\ data have been processed using the Science Analysis Software (SAS
version~9.0) with the calibration from 2009~July; the tasks \emph{epproc}
and \emph{emproc} were run to produce calibrated and concatenated event lists
for the EPIC cameras.

EPIC event files have been filtered for high-background time intervals,
following the standard method consisting in rejecting periods of high count rate
at energies~$\,>10\,$keV.
Strong flares affect the last part of the observation of XBSJ013240.1-133307 and the
whole observation of XBSJ160645.9+081525;
in this case, we tried different filtering, to find a compromise
between cleaning and net exposure time.
After checking the consistency of the spectral results, we chose a rather relaxed filtering, 
that warrants the 
collection of a number of source counts high enough.

Events corresponding to patterns $0 - 12$  (MOS1\&2) and $0 - 4$ (pn) have been
used \citep[see][]{xmmhb}.
We have also generated the spectral response matrices at the sources position
using the SAS tasks \emph{arfgen} and \emph{rmfgen}.
We exclude event pile-up in the MOS and pn data.
The net exposure times at the
source positions, the net count rates and the
signal-to-noise ratios ($S/N$) in the energy range $0.3-10\;$keV are reported in
Table~\ref{tab:xmmlog}.
We tested short-time variability within each observation, generating source 
light curves in different energy ranges.
No bin shows significant deviation from the mean value.
To improve statistics, the MOS1 and MOS2 data have been combined together, and
the MOS and pn spectra have been fitted simultaneously, keeping the relative
normalization free.

The OM performed several observations for each source, with the $U$-filter 
($\lambda_{\mbox{\scriptsize eff}}=3440\,$\AA; XBSJ013240.1-133307, for a total of $\sim 43\,$ksec) 
and the $UVW1$-filter 
($\lambda_{\mbox{\scriptsize eff}}=2910\,$\AA; XBSJ113148.7+311358 and XBSJ160645.9+081525, for a 
total of $\sim 20$ and $\sim 28\,$ksec, respectively) in the optical light path.
All OM exposures were performed in the ``Science User Defined'' image mode (with
a window size of $5\arcmin \times 5\arcmin$).
Only XBSJ013240.1-133307 is detected in the OM 
images obtained running the standard data
reduction (performed with the \emph{omichain} routine of SAS).
In the OM observing window there is no evidence of variability of the source emission in the $U$-filter;
the averaged magnitude (in the AB system) and flux density, extracted from the
\emph{omichain} output, are $22.36\pm0.23$ and $(1.05\pm0.23)\times
10^{-17}\,$\fluxA, respectively.

\subsection{X-ray spectral analysis}\label{sect:xmman}

The MOS and pn spectra have been analysed over the energy range from $0.3$ to $10\,$keV
using standard software packages
\citep[FTOOLS version~6.9, XSPEC version~12.7.0;][]{xspec}.
All the models discussed in the following assume Galactic absorption (see
Table~\ref{tab:xbs}); Galactic and intrinsic absorptions have been parametrized with the
{\sc (z)phabs} model in XSPEC, adopting cross-sections
and abundances of \citet{wilms}.
When a reflection component has to be tested, we adopted the {\sc pexmon} model in XSPEC \citep{pexmon}. 
{\sc pexmon} is an additive component self-consistently incorporating the Compton-reflected continuum from
a neutral slab combined with emission from \feka, \fekb, \nika, and the \feka\ Compton shoulder.
In this model, the reflection scaling factor is defined so that $R = 1$ corresponds to reflection from a semi-infinite
slab subtending $2\pi\,$sr to the central source of X-rays.
During the fit, the inclination angle was fixed to $i \sim 60\arcdeg$ (an average value for Seyfert galaxies).
We assumed also a fixed energy cut-off, $\pedix{E}{c}=100\,$keV, since our data are not sensitive to this parameter; the 
adopted value of \pedix{E}{c} is typical for local Seyfert galaxies \citep[\eg,][]{dadina08,derosa08,derosa12,malizia14}.
Unless otherwise stated, the figures and fit parameters are in the rest-frame of the sources.

The data were grouped to ensure a minimum of $20$ net source counts per bin
in the pn and MOS spectra; the use of $\chi^2$ statistics is then allowed.
All uncertainties are quoted at the 90\% confidence level for one parameter of interest 
($\Delta\apix{\chi}{2}=2.71$).
The results of our analysis are reported in Table~\ref{tab:xbf}, while the unfolded spectra are shown in Fig.~\ref{fig:xmmnew}.

\begin{description}
 \item[]{\bf XBSJ013240.1-133307} \\
  Unlike the archival spectra analysed in \citet{corral11}, the quality of the proprietary data allows
us to investigate both the power-law
photon index 
and the intrinsic absorption at the same time. 
Using an absorbed (Galactic $+$ intrinsic \nhsym) power-law model we obtain a flat photon index, $\Gamma=1.34\pm0.10$, 
with an absorption $\nhsym=(2.33\errUD{0.35}{0.37}) \times 10^{22}\,$\nh\
($\chidof=160.9/131$, for $135$ bins). 
When a scattering component (modelled with a second, unabsorbed power law, with the same $\Gamma$ as the
direct one) is added to the model, the fit improves ($\dchidof=17.1/1$, $\pedap{\chi}{r}{2}=1.11$), but it still results in a quite low photon index, 
$\Gamma=1.47\errUD{0.09}{0.13}$.
A narrow ($\sigma<50\,$eV) \feka\ line is partially detected at $E=6.25\errUD{0.09}{0.08}\,$keV 
[$F=(1.38\pm0.82)\times 10^{-6}\,$\normGAUSS, equivalent width with respect to the observed continuum 
$\mbox{EW}=108\errUD{80}{66}\,$eV; $\dchidof=7.9/2$].
The best-fitting parameters for this model are reported in the first row in Table~\ref{tab:xbf}.

A flat observed continuum could suggest the presence of a reflection component; we then investigated this possibility.
A pure neutral reflection model is rejected by the data ($\chidof=423.6/135$), while assuming an absorbed power law plus a reflection component
strong residuals below $1\,$keV are evident.
When a model composed by an intrinsic absorbed power law, a scattering component and a reflection component is fitted to the data,
we obtain a photon index closer to typical values for unabsorbed AGN, 
$\Gamma=1.60\errUD{0.09}{0.16}$ \citep{piconcelli05,mateos10,corral11}; the column density is 
$\nhsym=(3.36\errUD{0.55}{0.50}) \times 10^{22}\,$\nh, while the scattering fraction is $\sim4$\%.
The magnitude of the reflection scaling factor is found to be $R = 0.44\errUD{0.47}{0.41}$.
Although this model provides an 
acceptable description for the $0.3-10\,$keV spectrum of XBSJ013240.1-133307 ($\chidof=140.1/129$), 
the statistics favours the unabsorbed power-law plus absorbed power-law plus Gaussian line model:  $\dchidof=4.3/1$, 
F-test probability $\pedix{P}{F}=95.4$\%.

A joint fit of the new and old datasets, with only the normalizations of the intrinsic power laws free to change between the two observations, 
shows that the best-fitting model (intrinsic power law $+$ scattering component $+$ emission line) 
provides a satisfactory description of all the available data: $\Gamma=1.47\errUD{0.06}{0.10}$, 
$\nhsym=(3.18\errUD{0.37}{0.54}) \times 10^{22}\,$\nh, scattering fraction (consistent within the errors) 
changing from $\sim8$\% (old dataset) to $\sim5$\% (new dataset); 
line energy $E=6.25\errUD{0.09}{0.07}\,$keV 
and equivalent width (consistent within the errors) 
changing from $\mbox{EW}\sim 193\,$eV (old dataset) to $\mbox{EW}\sim 117\,$eV (new dataset);
$\chidof=154.8/149$, for $157$ bins between $0.3$ and $10\,$keV.
The intrinsic $2-10\,$keV luminosity slightly increase, of a factor of $\sim 1.7$.
\\
 \item[]{\bf XBSJ113148.7+311358} \\
The photon index required by the data when an absorbed power law is assumed, is extremely flat,
$\Gamma=1.18\errUD{0.28}{0.22}$ [$\nhsym=(2.31\errUD{1.60}{1.13}) \times 10^{22}\,$\nh, consistent within the errors with the 
previous measurement; $\chidof=20.7/18$, for $22$ bins].
The addition of a scattering component 
does not change significantly the resulting value of the photon index ($\Gamma=1.34\errUD{0.31}{0.33}$).

As in the previous case, we tested the data for the presence of a reflection component.
A pure neutral reflection model can reproduce the observed spectra assuming a steep intrinsic continuum,
$\Gamma$ between $2.4$ and $2.5$ (the upper limit of the range allowed in {\sc pexmon} for the photon index), although the fit is poor
($\chidof=28.2/19$).
The fit improves when both the reflection component and the intrinsic, absorbed power
law are included in the model ($\dchidof=9.6/2$, F-test probability $\pedix{P}{F}=97.0$\% with respect to the pure reflection model).
The addition of a scattering component, with an intensity of $\sim 7$\% of the intrinsic emission, is significant at $93.7$\% ($\dchidof=3.7/1$).
The final spectral parameters are $\Gamma=1.59\errUD{0.39}{0.40}$, $\nhsym=(4.05\errUD{2.39}{1.84}) \times 10^{22}\,$\nh; 
the reflection component contributes significantly to the observed emission, with the reflection factor $R\sim 1$
(see Table~\ref{tab:xbf}, second row). 

A joint fit of the new and old datasets, with only the normalisations of the intrinsic power laws free to change between the two states, 
shows that this model provides a satisfactory description of all the available data: $\Gamma=1.52\pm{0.40}$, 
$\nhsym=(3.13\errUD{1.26}{1.38}) \times 10^{22}\,$\nh, $R=0.31\pm{0.27}$ 
($\chidof=49.5/51$ for $58$ bins).
Although the values are marginally consistent within the errors, our analysis suggests a decrease of a factor of $\sim 2.8$ 
in the intrinsic $2-10\,$keV luminosity.
\\
 \item[]{\bf XBSJ160645.9+081525} \\
The best-fitting model reported by \citet{corral11} can reproduce the new data above $\sim 1\,$keV (\ie, in the energy range where the source is detected
in the old spectra); the best-fitting values for the photon index and the column density are consistent within the (large) errors with
the previous values: $\Gamma=2.08\errUD{0.62}{0.55}$ and  $\nhsym=(25.35\errUD{10.10}{7.79}) \times 10^{22}\,$\nh\ 
($\chidof=12.5/19$, for $23$ bins).
However, strong residuals are clearly visible below $\sim 1\,$keV, suggesting that a more complex model must be tested.
With the addition of an unabsorbed power law, the observed residuals are completely accounted for 
($\chidof=16.6/25$, for $30$ bins between $0.3$ and $10\,$keV).
Although this scattering component is weak (scattering fraction $<5$\%), its inclusion is statistically significant 
($\dchidof=9.4/1$, F-test probability $\pedix{P}{F}=99.9$\%; see Table~\ref{tab:xbf}, third row).

A joint fit of the new and old datasets, with only the normalisations of the intrinsic power laws free to change between the two states, 
shows that this model provides a satisfactory description of all the available data: $\Gamma=2.07\errUD{0.44}{0.42}$, 
$\nhsym=(27.53\errUD{7.18}{6.32}) \times 10^{22}\,$\nh, scattering fraction $<5$\% for both datasets
($\chidof=30.1/38$, for $44$ bins).
Although the values are consistent within the errors, our analysis suggests a decrease of a factor of two in the intrinsic $2-10\,$keV luminosity.
\\
\end{description}


\section[]{Notes on individual objects}\label{sect:note}

\begin{description}
 \item[]{\bf XBSJ000100.2-250501} \\
The optical spectrum is characterized by broad emission lines ($\mbox{FWHM}>2000\,$km/s), typical of type~1 AGN.
By using only a type 1 QSO template to recover the SED, we are not able to simultaneously reproduce the optical and IR photometric points. 
The addition of an early-type component is needed to take into account the WISE data at lower wavelengths. 
Nevertheless, due to the scarcity of photometric points available for this source, the 
normalization of the host galaxy component remains poorly constrained.

 \item[]{\bf XBSJ013240.1-133307} \\
The optical spectrum is characterized by a galaxy continuum with a reduced Ca break, 
suggesting the
presence of a non-negligible contribution of the nuclear radiation in the bluer part of the spectrum. 
These spectral characteristics strongly constrain the level of dust absorption to be used in the SED modelling, \pedix{A}{V} between 
$1$ and $2\,$mag. 
For this low amount of absorption both the Galactic and the Galactic Centre extinctions curves produce a very similar shape above $\sim 0.5\,\mu$m, 
while below this wavelength they have different behaviour. 
The \sdss, \galex, and OM data seem to prefer the Galactic extinction curve, 
although also the Galactic Centre curve gives a reasonable reproduction of the SED.

 \item[]{\bf XBSJ021642.3-043553} \\
This is the high-redshift Extremely Red Object discussed in details by \citet{severgnini06}.
Its optical spectrum, characterized by narrow emission lines plus a host galaxy continuum, 
strongly suggests a nuclear absorption larger than $2\,$mag to be applied to our SED modelling.
As for the $K$-band flux, as noted by \citet{severgnini06}, two possible weak near-IR objects are visible in this band
within the same distance from the X-ray position, separated by $\sim 1\arcsec$ and with a similar 
magnitude.
The two sources cannot be resolved either by \spitzer\ or by WISE; to take into account the contamination due to the second 
source, expected at least at the lowest wavelengths, we associated with the observed fluxes a larger error than the quoted one
($50$\%).

 \item[]{\bf XBSJ022707.7-050819} \\
Although the IR SED data can be well modelled both by a nuclear dust extinction of $\pedix{A}{V,opt}\sim15\,$mag plus an early-type galaxy 
or by an $\pedix{A}{V,opt}\sim30\,$mag plus a late-type host, the optical spectrum is better reproduced by the former combination of template and \pedix{A}{V}.
As for the extinction curve, the spectral shape described by the \spitzer\ and WISE data is inconsistent with the deep absorption features observed when
the extinction curve of the Galactic Centre is assumed.
Therefore, we assumed as reference modelling the one with lowest \pedix{A}{V}; the ranges of physical parameters have been 
evaluated by considering all the best modelling obtained with the three nuclear templates quoted in the text with only the Galactic curve applied.

Finally, we note that in the optical image we can see several objects near the QSO. 
We therefore cannot exclude a contamination at some level of the IR fluxes.

 \item[]{\bf XBSJ050536.6-290050} \\
In the \spitzer\ images at $3.6$ and $4.5\,\mu$m, a second source is detected at a distance of $12\arcsec$; 
however, in the image at $8\,\mu$m this source disappears.
Therefore, we do not expect a contamination from this source in the WISE upper limit at $22\,\mu$m \citep[that has the 
largest  PSF, $\mbox{FWHM}=12.0\arcsec$;][]{wise}.
Moreover, the profile-fitting method adopted to recover the WISE magnitudes assumed here
strongly reduces the level of contamination expected.
With this method, carried out using the data from all bands simultaneously, the higher resolution data at the shorter wavelengths can 
guide the extraction at the longer wavelengths.

 The optical spectrum requires a dust obscuration of $\pedix{A}{V,opt}=3.6\,$mag, while the Ca break 
 is typical of a late-type galaxy.

 \item[]{\bf XBSJ051413.5+794345} \\
The morphological type of the host is strongly driven by the Ca break observed in the optical spectrum, 
suggesting an elliptical or an early spiral.
The spectrum provides only a lower limit to the dust extinction, $\pedix{A}{V,opt}>3\,$mag.

 \item[]{\bf XBSJ052128.9-253032} \\
This is the source with the poorest photometric coverage: only an $R$-band magnitude and two detections at $3.4\,\mu$m and 
$4.6\,\mu$m by WISE.
At longer wavelength, the WISE catalogue does not provide any flux, even a $2\sigma$ limit.
We therefore estimated upper limits to the fluxes at $12\,\mu$m and $22\,\mu$m using the weakest source detected in the field. 

The host galaxy morphology is well constrained by the Ca break measured from the optical spectrum.
Concerning the nuclear emission, we cannot constrain at the same time both its intensity and the absorption; 
as said in the text, the quoted physical parameters are limits, in the sense detailed below.
In modelling the SED, we first fixed the dust extinction to the lower limit obtained from the optical spectrum, 
$\pedix{A}{V,opt}>1.6\,$mag.
 An estimate of the way in which a different optical obscuration affects our determination of the physical parameters has been 
 obtained by increasing \pedix{A}{V} up to the value expected from the column density derived from the X-ray data 
 \citep[when a Galactic dust-to-gas ratio is assumed, $\pedix{A}{V}/\nhsym=5.27\times 10^{-22}\,$mag/cm$^{-2}$;][]{bohlin78}  
 and performing again the SED modelling.

The point-like appearance of the source in the WISE images at $3.4$ and $4.6\, \mu$m (dominated in 
our modelling by the host), is consistent with the WISE resolution being not good enough to resolve the galaxy at a redshift
of $z\sim 0.6$ (where $1\arcsec$ corresponds to about $7.2\,$kpc with the cosmology 
adopted here).
The result of our modelling is in agreement with the elusive nature of this QSO; in particular, taking into account the observed 
X-ray luminosity, the expected $\pedix{L}{\mbox{\ion{O}{iii}}}$ is too low to be detected out from the host galaxy light.

 \item[]{\bf XBSJ080411.3+650906} \\
For this source, both the morphological type and the intensity of the host galaxy is well constrained by the availability of the 
\sdss\ data and by the measurement of the Ca break obtained by the optical spectrum.
Concerning the nuclear emission, the quality of the mid-IR coverage does not allow us to 
constrain at the same time both its intensity and the absorption; 
as done for XBSJ052128.9-253032, in the text we quote limits to the physical parameters.
A first SED modelling has been done with the dust extinction fixed to the lower limit obtained from the optical spectrum, 
$\pedix{A}{V,opt}>3.1\,$mag.
To investigate the effects of a different optical obscuration on our determination of the physical parameters, we 
increased \pedix{A}{V} up to the value expected for a Galactic dust-to-gas ratio
\citep[$\pedix{A}{V}/\nhsym=5.27\times 10^{-22}\,$mag/cm$^{-2}$;][]{bohlin78}, assuming the \nhsym\
derived from the X-ray data; then we performed again the SED modelling.

Again, taking into account the WISE resolution, not good enough to resolve the galaxy at 
$z\sim 0.6$ (where $1\arcsec$ corresponds to about $7.2\,$kpc with the cosmology 
adopted here), we expect a point-like appearance of the source in the WISE images also in the bands 
dominated in our modelling by the host ($3.4$ and $4.6\, \mu$m).

The result of our modelling is in agreement with the elusive nature of this QSO.
  
 \item[]{\bf XBSJ113148.7+311358} \\
A second source is clearly detected in the optical at a distance of $\sim 12\arcsec$; although it is still present in the \spitzer\ 
images at low wavelength, in the image at $8\,\mu$m it disappears.
Therefore, we do not expect a contamination due to this source either in the WISE upper limit at $22\,\mu$m \citep[that has the 
largest  PSF, $\mbox{FWHM}=12.0\arcsec$;][]{wise} or in the \spitzer\ flux at $24\,\mu$m (that we recovered from 
IRSA\footnote{The NASA/IPAC Infrared Science Archive, 
http://irsa.ipac.caltech.edu/cgi-bin/Gator/nph-scan?submit=Select\&projshort=SPITZER.}).
Moreover, the profile-fitting method adopted to recover the WISE magnitudes assumed here
strongly reduces the level of contamination expected.
With this method, carried out using the data from all bands simultaneously, the higher resolution data at the shorter wavelengths can 
guide the extraction at the longer wavelengths.

The optical spectrum suggests a high absorption, $\pedix{A}{V,opt}\sim 17\,$mag, and a late-type host galaxy. 
The absorption feature due to silicates present in the Galactic Centre extinction curve is in better agreement with the $12\,\mu$m 
upper limit than the shape obtained by applying the Galactic curve, although both curves provide an acceptable representation of the SED.

 \item[]{\bf XBSJ122656.5+013126} \\
 The optical spectrum is characterized by a galaxy continuum with a strongly reduced Ca break, 
 suggesting the
presence of a significant contribution of the nuclear radiation in the bluer part of the spectrum. 
For the low amount of absorption implied by the optical spectrum,  $\pedix{A}{V,opt}\sim 1.6\,$mag, 
both the Galactic and the Galactic Centre extinctions curves produce a very similar shape above $\sim 0.5\,\mu$m, 
while below this wavelength they have different behaviour. 
The \galex\ data in the near-UV and far-UV allowed us to prefer the Galactic extinction curve; 
therefore, the ranges for the physical parameters have been evaluated by considering only the three best modelling with the 
three nuclear templates with only this extinction curve applied.

 \item[]{\bf XBSJ134656.7+580315} \\
The Ca break observed in the optical spectrum
suggests an
elliptical or an early spiral host galaxy.
The spectrum provides only a lower limit to the dust extinction, $\pedix{A}{V,opt}>3.1\,$mag.

The spectral shape described by the \spitzer\ and WISE data can not be reproduced if the Galactic extinction curve is assumed; 
therefore, the ranges for the physical parameters have been evaluated by considering only the three best modelling with the 
three nuclear templates with only the curve of the Galactic Centre applied.

 \item[]{\bf XBSJ144021.0+642144} \\
The optical spectrum of this source is characterized by broad emission lines ($\mbox{FWHM}>2000\,$km/s), typical of type~1 AGN, and
the amount of intrinsic absorption is constrained to be low, $\pedix{A}{V,opt}\sim 0.15$.
By using only a type 1 QSO template to recover the SED, we are not able to simultaneously reproduce the optical and IR photometric points. 
The addition of an early-type component is needed to well reproduce the data between $0.4$ and $3\,\mu$m. 

 \item[]{\bf XBSJ160645.9+081525} \\
 In the optical domain, both the spectrum 
 and the photometric points strongly constrain the host
 to be an early-type galaxy. 
 The spectrum provides only a lower limit to the dust extinction , $\pedix{A}{V,opt}>2\,$mag.
At wavelength longer than $\sim 5\,\mu$m (rest frame), the SED is better reproduced when a shape like that of the deep
due to the absorption from silicates is assumed.
We note that the feature obtained when the former is applied with the high \pedix{A}{V} required by the modelling 
is probably too deep to be realistic.
However, the exact shape of this feature does 
not affect our estimate of the physical parameters.

 \item[]{\bf XBSJ204043.4-004548} \\
The \spitzer\ images reveal the presence of a source at less than $12\arcsec$ from this object.
While the emission of the two objects can be resolved in the \spitzer/MIPS data at $24\,\mu$m, due to the larger PSF 
\citep[$\mbox{FWHM}=12.0\arcsec$;][]{wise} the flux measured  by WISE at $22\,\mu$m is probably contaminated by the 
second source.
According to the WISE catalogue, no deblending has been applied\footnote{Number of PSF components used simultaneously in 
the profile-fitting for the source, including the source itself, $nb=1$; active deblending flag, indicating if a single detection was split 
into multiple sources in the process of profile-fitting, $na=0$, \ie\ the source is not actively deblended; see 
http://wise2.ipac.caltech.edu/docs/release/allsky/expsup/sec2\_2a.html.}.
To estimate a correction factor for $\pedix{F}{22\,\mum}$ from the \spitzer\ image at $24\,\mu$m, we compared {\it (i)} the 
aperture-corrected fluxes from \spitzer\ (radius of $16.5\arcsec$) with and without deblending applied (to estimate how the 
second object affects the observed emission); and {\it (ii)} the aperture-corrected fluxes from \spitzer\ and WISE (radius of 
$16.5\arcsec$), both not corrected for blending (to take into account the different background of the two cameras).
The corrected flux is marked as an asterisk in the SED of XBSJ204043.4-004548 in Fig.~\ref{fig:sed} (bottom panel).

Concerning the SED modelling, the small decrease in the flux implied by the WISE data at $12\,\mu$m, followed by a new 
increase at longer wavelength, suggesting the presence of a deep due to the presence of silicates, seems to prefer the extinction 
curve of the centre of the Galaxy.
However, both the tested curves provide an acceptable modelling of the SED.
\end{description}


\label{lastpage}


\renewcommand{\thetable}{\arabic{numtabdata}}

%
\begin{landscape}
\begin{table}
\begin{center}
 \caption{IR data.}             
 \label{tab:ir}     
{
\scriptsize
  \begin{tabular}{l r@{\extracolsep{0.05cm}$\pm$}l r@{$\pm$}l r@{$\pm$}l r@{$\pm$}l r@{$\pm$}l r@{$\pm$}l l r@{$\pm$}l r@{$\pm$}l r@{$\pm$}l r@{$\pm$}l}
   \hline\hline
    & \multicolumn{12}{c}{\spitzer} & & \multicolumn{8}{c}{WISE} \\
    \cline{2-13}\cline{15-22}
    \multicolumn{1}{c}{Name} & \multicolumn{2}{c}{$3.6\,\mu\mbox{m}$} & \multicolumn{2}{c}{$4.5\,\mu\mbox{m}$} & \multicolumn{2}{c}{$5.8\,\mu\mbox{m}$} & \multicolumn{2}{c}{$8.0\,\mu\mbox{m}$} & \multicolumn{2}{c}{$24\,\mu\mbox{m}$} & \multicolumn{2}{c}{$70\,\mu\mbox{m}$} & Note & \multicolumn{2}{c}{$3.4\,\mu\mbox{m}$} & \multicolumn{2}{c}{$4.6\,\mu\mbox{m}$} & \multicolumn{2}{c}{$12\,\mu\mbox{m}$} & \multicolumn{2}{c}{$22\,\mu\mbox{m}$} \\
    \multicolumn{1}{c}{(1)} & \multicolumn{2}{c}{(2)} & \multicolumn{2}{c}{(3)} & \multicolumn{2}{c}{(4)} & \multicolumn{2}{c}{(5)} & \multicolumn{2}{c}{(6)} & \multicolumn{2}{c}{(7)} & (8) & \multicolumn{2}{c}{(9)} & \multicolumn{2}{c}{(10)} & \multicolumn{2}{c}{(11)} & \multicolumn{2}{c}{(12)}
    \vspace{0.1cm} \\
  \hline   
   \vspace{-0.2cm} \\
  XBSJ000100.2-250501 & \multicolumn{2}{c}{$-$} & \multicolumn{2}{c}{$-$} & \multicolumn{2}{c}{$-$} & \multicolumn{2}{c}{$-$} & \multicolumn{2}{c}{$-$} & \multicolumn{2}{c}{$-$}         & $-$         & $106.6 $&$7.8 $ & $142.5 $&$14.3$ & $568.1 $&$134.7$	       & \multicolumn{2}{c}{$<2807.2$} \\ 
  XBSJ013240.1-133307 & $1111.1$&$204.7$      & $1400.0$&$229.5$      & $1674.6$&$253.0$	 & $1899.7$&$280.0$	   & $4507.6$&$863.6 $     & $10974.0 $&$3292.2   $	       & own obs & $749.8 $&$20.2$ & $1134.0$&$32.6$ & $1903.7$&$156.6$  	       & \multicolumn{2}{c}{$<3390.5$}   \\ 
  XBSJ021642.3-043553 & $104.4 $&$20.9 $      & $132.6 $&$26.5 $      & $276.3 $&$55.3 $	 & $421.8 $&$84.4$	   & $1303.6$&$260.7$      & \multicolumn{2}{c}{$-$}         & SWIRE       & $54.0  $&$5.7 $ & $113.8 $&$11.4$ & \multicolumn{2}{c}{$<246.8$} & \multicolumn{2}{c}{$<1998.3$}   \\ 
  XBSJ022707.7-050819 & $404.6 $&$80.9 $      & $557.6 $&$111.5$      & $742.5 $&$148.5$	 & $1166.4$&$233.3$	   & $5893.4$&$1178.7$     & \multicolumn{2}{c}{$-$}         & SWIRE       & $373.4 $&$11.1$ & $524.6 $&$18.0$ & $1811.3$&$120.0$	       & $5930.2$&$864.6 $		\\ 
  XBSJ050536.6-290050 & $212.5 $&$87.9 $      & $265.7 $&$114.2$      & $351.4 $&$137.2$	 & $617.3 $&$200.6$	   & $2325.5$&$900.0  $     & \multicolumn{2}{c}{$-$}         & own obs     & $167.2 $&$7.6 $ & $194.0 $&$12.8$ & $606.5 $&$124.4$	       & \multicolumn{2}{c}{$<2283.8$} \\ 
  XBSJ051413.5+794345 & \multicolumn{2}{c}{$-$} & \multicolumn{2}{c}{$-$} & \multicolumn{2}{c}{$-$} & \multicolumn{2}{c}{$-$} & \multicolumn{2}{c}{$-$} & \multicolumn{2}{c}{$-$}         & $-$         & $172.1 $&$7.5 $ & $217.1 $&$12.9$ & $434.5 $&$105.4$	       & \multicolumn{2}{c}{$<1904.9$} \\ 
  XBSJ052128.9-253032 & \multicolumn{2}{c}{$-$} & \multicolumn{2}{c}{$-$} & \multicolumn{2}{c}{$-$} & \multicolumn{2}{c}{$-$} & \multicolumn{2}{c}{$-$} & \multicolumn{2}{c}{$-$}         & $-$         & $174.5 $&$8.4 $ & $178.7 $&$14.6$ & \multicolumn{2}{c}{$-$}       & \multicolumn{2}{c}{$-$}          \\ 
  XBSJ080411.3+650906 & \multicolumn{2}{c}{$-$} & \multicolumn{2}{c}{$-$} & \multicolumn{2}{c}{$-$} & \multicolumn{2}{c}{$-$} & \multicolumn{2}{c}{$-$} & \multicolumn{2}{c}{$-$}         & $-$         & $76.3  $&$6.7 $ & $69.9  $&$12.4$ & \multicolumn{2}{c}{$<255.9$} & \multicolumn{2}{c}{$<2893.8$}   \\ 
   XBSJ113148.7+311358 & $169.5 $&$78.9 $      & $189.2 $&$83.6 $      & $202.2 $&$88.3 $	 & $362.7 $&$115.9$	   & $1050.0$&$210.0 $     & \multicolumn{2}{c}{$<5200.0$}   & own obs & $127.8 $&$7.2 $ & $161.4 $&$12.7$ & \multicolumn{2}{c}{$<208.9$} & \multicolumn{2}{c}{$<2411.4$} \\ 
 XBSJ122656.5+013126 & $484.7 $&$131.4$      & $685.0 $&$156.9$      & $1085.4$&$200.3$	 & $1816.0$&$263.5$	   & $5600.0$&$1120.0$     & \multicolumn{2}{c}{$-$}         & archive     & $458.9 $&$14.9$ & $714.2 $&$25.2$ & $2015.6$&$219.7$	       & \multicolumn{2}{c}{$<4194.4$} \\ 
  XBSJ134656.7+580315 & $1429.8$&$228.3$      & $2054.3$&$270.6$      & $3292.8$&$344.6$	 & $4411.8$&$882.4$	   & $7236.9$&$1059.8$     & $38305.0 $&$11491.5  $	       & own obs     & $1193.8$&$26.6$ & $1978.0$&$42.2$ & $3407.2$&$131.1$	       & $7824.7$&$821.7 $		\\ 
  XBSJ144021.0+642144 & \multicolumn{2}{c}{$-$} & \multicolumn{2}{c}{$-$} & \multicolumn{2}{c}{$-$} & \multicolumn{2}{c}{$-$} & $1999.5$&$435.2 $     & \multicolumn{2}{c}{$-$}         & archive & $418.2 $&$11.3$ & $517.4 $&$14.9$ & $1125.1$&$82.9 $ 	       & $2980.3$&$568.2 $	        \\ 
  XBSJ160645.9+081525 & $251.3 $&$97.2 $      & $215.3 $&$90.0 $      & $255.6 $&$98.6 $	 & $432.0 $&$133.3$	   & $1125.0$&$705.6 $     & $8180.0  $&$2454.0   $	       & own obs & $270.5 $&$10.3$ & $175.5 $&$13.8$ & $431.7 $&$128.7$	       & \multicolumn{2}{c}{$<1842.8$}   \\
  XBSJ204043.4-004548 & $270.1 $&$98.0 $      & $401.0 $&$119.3$      & $584.6 $&$145.4$	 & $1021.3$&$192.8$	   & $3290.1$&$870.2 $     & $8507.3 $&$2552.2 $	       & own obs     & $176.9 $&$9.4 $ & $369.7 $&$18.8$ & $1320.7$&$163.7$ 	       & $5358.8$&$1060.0$	        \\ 
  \end{tabular}
}
\end{center}
 {
 \footnotesize Column (1): source name in the XBS sample.
 \footnotesize Columns (2)-(7): \spitzer\ imaging photometric fluxes in $\mu$Jy, not corrected for Galactic extinction.
 \footnotesize Column (8): notes to the \spitzer\ data.
 \footnotesize Columns (9)-(12): WISE fluxes in $\mu$Jy, not corrected for Galactic extinction.}
\end{table}
\end{landscape}
\addtocounter{numtabdata}{1}
%

%
\begin{landscape}
\begin{table}
\begin{center}
 \caption{UV/Optical data.}             
 \label{tab:optir}     
{
 \scriptsize
  \begin{tabular}{l c c@{\extracolsep{0.2cm}} c c c c c c c@{\extracolsep{0.2cm}} c l}
   \hline\hline
    & \multicolumn{2}{c}{\galex} & & & \multicolumn{5}{c}{\sdss} & & \\
    \cline{2-3}\cline{6-10}
    \multicolumn{1}{c}{Name} & FUV & NUV & OM & Other UV & $u$ & $g$ & $r$ & $i$ & $z$ & $R$ & \multicolumn{1}{c}{Note} \\
    \multicolumn{1}{c}{(1)} & (2) & (3) & (4) & (5) & (6) & (7) & (8) & (9) & (10) & (11) & \multicolumn{1}{c}{(12)}
   \vspace{0.1cm} \\
  \hline   
   \vspace{-0.2cm} \\
  XBSJ000100.2-250501 &	 $	   -		   $	  & $	      - 		$      &     $- 		$      &     $- 		 $	&     $ 	-    $     &	 $	   -	$      &     $         -    $	   &	 $	   -	$      &     $         -    $ &     $21.86\pm6.57      $      &     (2)        \\
  XBSJ013240.1-133307           &	 $	   -		   $	  & $(5.1\pm1.3)\times10^{-29}  $      &     $22.21\pm0.23	$      &     $21.32\pm6.40	 $	&     $ 	-    $     &	 $	   -	$      &     $         -    $	   &	 $	   -	$      &     $         -    $ &     $19.96\pm6.00      $      &     (A,a,1)    \\
  XBSJ021642.3-043553 &	 $	   -		   $	  & $	      - 		$      &     $- 		$      &     $- 		 $	&     $ 	-    $     &	 $	   -	$      &     $         -    $	   &	 $	   -	$      &     $         -    $ &     $24.46\pm7.35      $      &     (3)        \\
  XBSJ022707.7-050819 &	 $	   -		   $	  & $	      - 		$      &     $- 		$      &     $21.24\pm6.39	 $	&     $ 	-    $     &	 $	   -	$      &     $         -    $	   &	 $	   -	$      &     $         -    $ &     $18.84\pm5.67      $      &     (b,3)    \\
  XBSJ050536.6-290050 &	 $	   -		   $	  & $	      - 		$      &     $- 		$      &     $- 		 $	&     $ 	-    $     &	 $	   -	$      &     $         -    $	   &	 $	   -	$      &     $         -    $ &     $21.17\pm6.36      $      &      (3)        \\
  XBSJ051413.5+794345 &	 $	   -		   $	  & $	      - 		$      &     $- 		$      &     $- 		 $	&     $ 	-    $     &	 $	   -	$      &     $         -    $	   &	 $	   -	$      &     $         -    $ &     $	    -	       $      &         -	       \\
  XBSJ052128.9-253032 &	 $	   -		   $	  & $	      - 		$      &     $- 		$      &     $- 		 $	&     $ 	-    $     &	 $	   -	$      &     $         -    $	   &	 $	   -	$      &     $         -    $ &     $20.69\pm6.23      $      &      (3)        \\
  XBSJ080411.3+650906 &	 $	   -		   $	  & $	      - 		$      &     $- 		$      &     $- 		 $	&     $23.767\pm1.351$     &	 $23.932\pm0.764$      &     $21.288\pm0.105$	   &	 $20.627\pm0.096$      &     $20.291\pm0.245$ &     $	    -	       $      &       -	       \\
  XBSJ113148.7+311358           &	 $	   -		   $	  & $	      - 		$      &     $- 		$      &     $- 		 $	&     $21.867\pm0.484$     &	 $21.174\pm0.081$      &     $20.379\pm0.058$	   &	 $19.559\pm0.041$      &     $19.430\pm0.145$ &     $	    -	       $      &       -	       \\
  XBSJ122656.5+013126 &	 $(1.5\pm0.6)\times10^{-29}$ & $(11.0\pm4.8)\times10^{-29} $  &     $20.71\pm0.23$      &     $- 		 $	&     $21.852\pm0.252$     &	 $20.907\pm0.046$      &     $20.114\pm0.029$	   &	 $19.805\pm0.032$      &     $18.882\pm0.055$ &     $	    -	       $      &        (B)	       \\
  XBSJ134656.7+580315 &	 $	   -		   $	  & $	      - 		$      &     $- 		$      &     $	    -	       $	&     $21.547\pm0.534$     &	 $20.084\pm0.067$      &     $18.279\pm0.019$	   &	 $17.651\pm0.017$      &     $17.254\pm0.045$ &     $	    -	       $      &         -        \\
  XBSJ144021.0+642144 &	 $	   -		   $	  & $	      - 		$      &     $- 		$      &     $- 		 $	&     $20.326\pm0.074$     &	 $19.704\pm0.015$      &     $19.482\pm0.016$	   &	 $19.309\pm0.023$      &     $19.066\pm0.051$ &     $	    -	       $      &        -	       \\
  XBSJ160645.9+081525	   &	 $	   -		   $	  & $	      - 		$      &     $- 		$      &     $- 		 $	&     $23.384\pm1.064$     &	 $22.535\pm0.210$      &     $21.028\pm0.080$	   &	 $20.059\pm0.049$      &     $19.407\pm0.102$ &     $19.99\pm6.03      $      &         (3)       \\
  XBSJ204043.4-004548 &	 $	   -		   $	  & $(5.0\pm1.0)\times10^{-29}  $      &     $- 		$      &     $- 		 $	&     $23.910\pm2.315$     &	 $22.543\pm0.298$      &     $20.956\pm0.097$	   &	 $20.065\pm0.059$      &     $19.802\pm0.193$ &     $21.07\pm6.36      $      &        (3)        \\
  \end{tabular}
}
\end{center}       
 {
 \footnotesize Column (1): source name in the XBS sample.
 \footnotesize Columns (2)-(3): \galex\ fluxes at $1539\,$\AA\ (FUV) and $2310\,$\AA\ (NUV) corrected for Galactic absorption, in units of \fluxHz.
 \footnotesize Column (4): OM AB magnitudes, corrected for Galactic extinction; for details, see column (13).
 \footnotesize Column (5): UV magnitudes found in the literature, corrected for Galactic extinction; for details, see column (13).
 \footnotesize Columns (6)-(10): \sdss\ DR7 magnitudes in the $u$ ($\lambda=3551\,$\AA), $g$ ($\lambda=4686\,$\AA), $r$ ($\lambda=6165\,$\AA), $i$ ($\lambda=7481\,$\AA), and $z$ ($\lambda=8931\,$\AA) filters, corrected for Galactic extinction.
 \footnotesize Column (11): $R$-band magnitudes found in literature, corrected for Galactic extinction; for details, see column (13).
 \footnotesize Column (12): notes to the OM, UV and optical data in cols. (4), (5) and (11): (A)=$U$ filter ($\lambda=3440\,$\AA); (B)=$B$ filter ($\lambda=4500\,$\AA); (a)=DSS magnitude $@4400\,$\AA; (b)=$g'$-band magnitude $@4872\,$\AA\ \citep{stalin10}; (1)=DSS $@6450\,$\AA\ \citep{caccianiga08}; (2)=Cousin $R$ $@6410\,$\AA\ \citep{fiore03}; (3)=Cousin $R$ $@6410\,$\AA\ \citep{caccianiga08}.
}
\end{table}
\end{landscape}
\addtocounter{numtabdata}{1}
%


\renewcommand{\thefigure}{\arabic{numfigsed}}

\begin{figure*}
 \centering
 \includegraphics[angle=0,height=6.5cm]{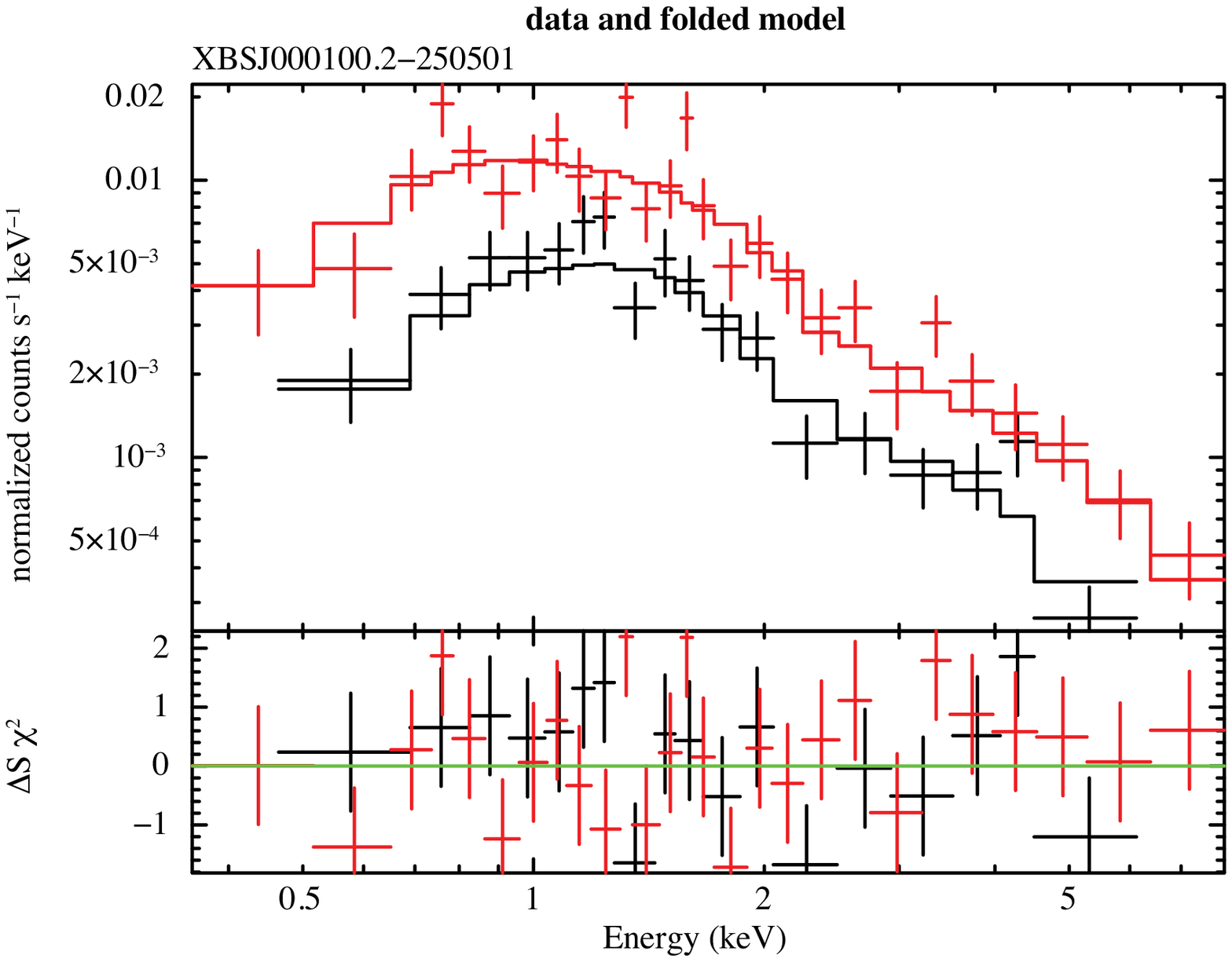}\\
 \vskip -0.5truecm
 \includegraphics[angle=-90,width=8.6cm]{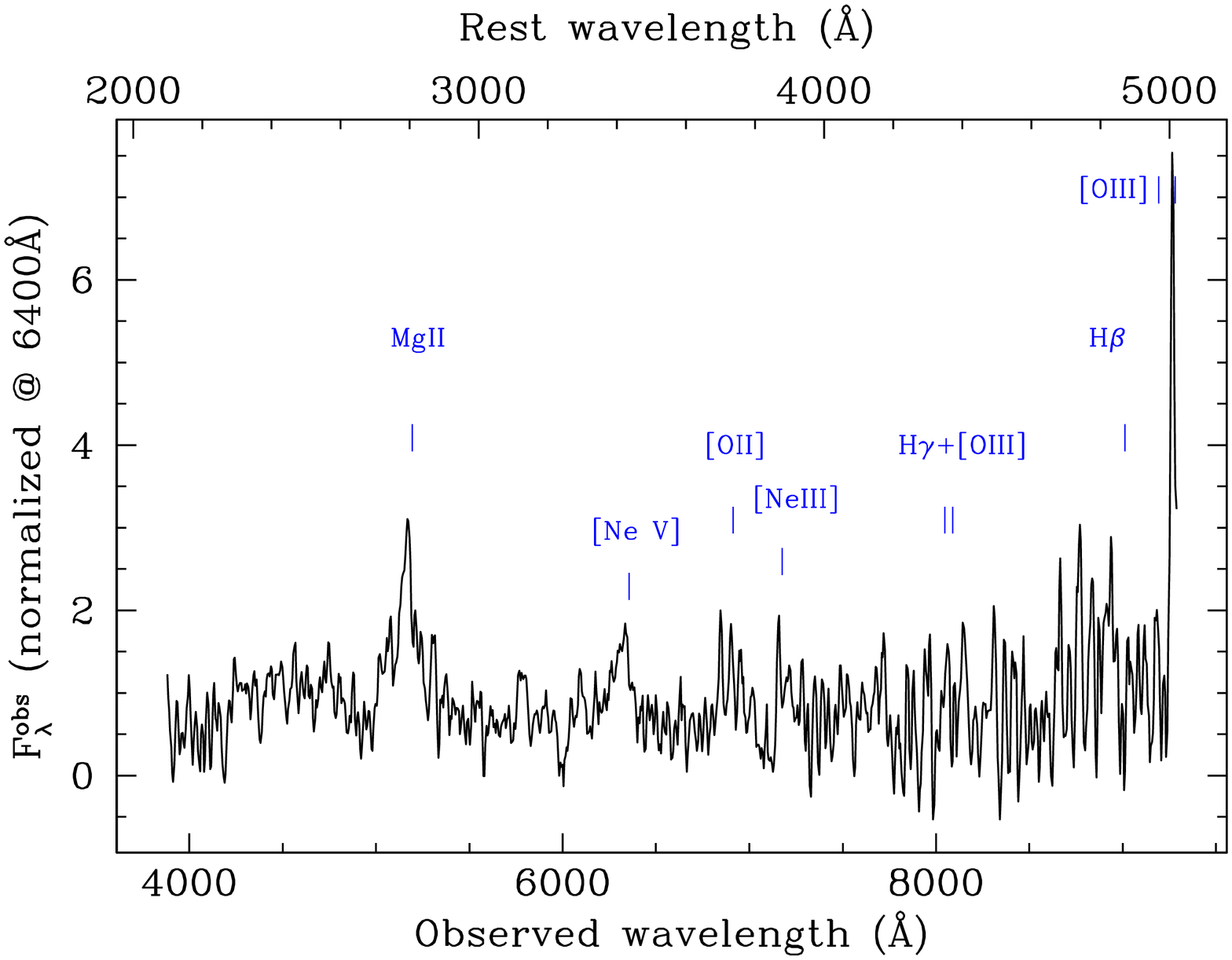}
 \vskip -0.5truecm
 \includegraphics[height=9cm]{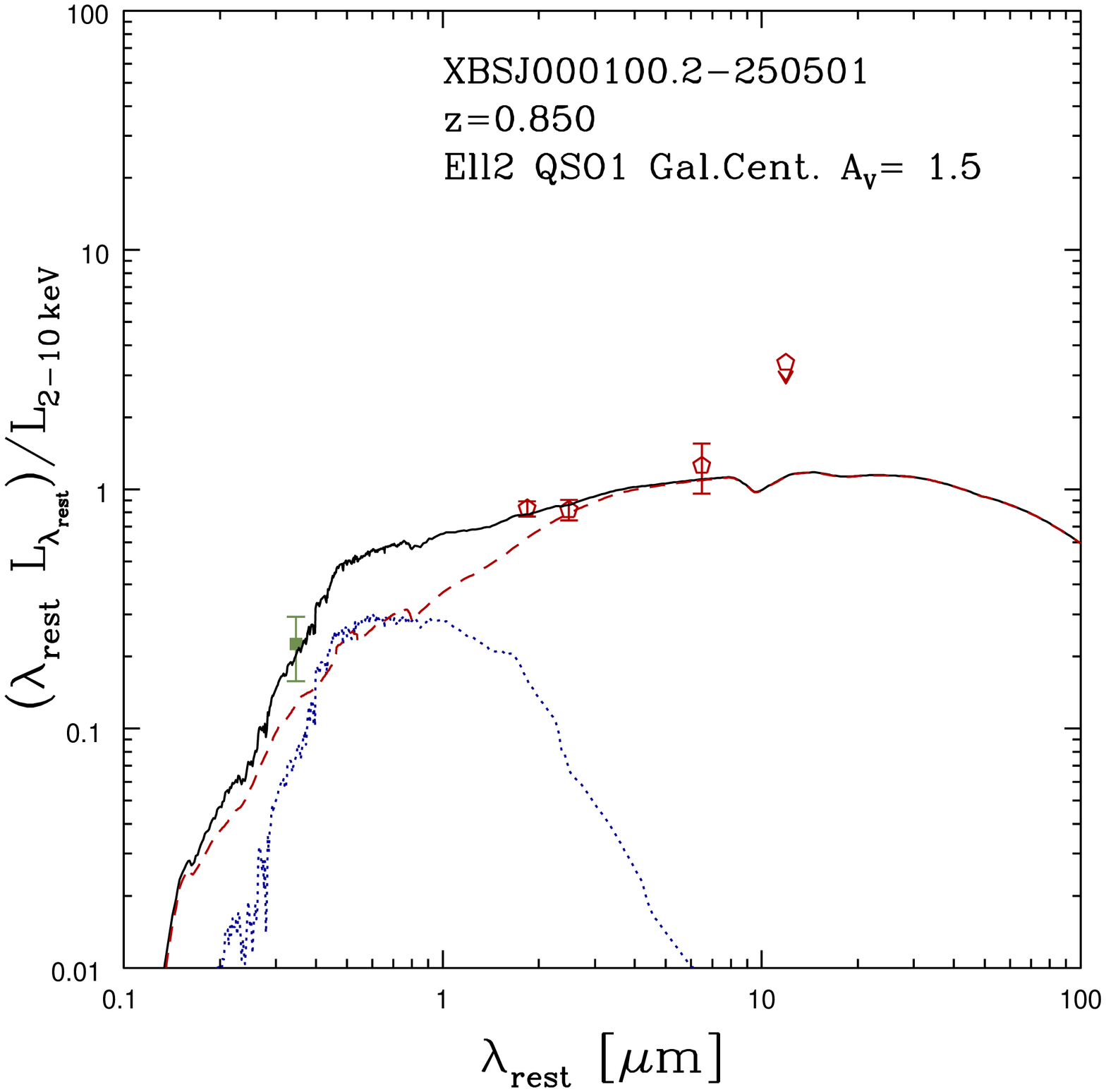}
 \caption{
  {\it Top panel:} \xmm\ data and residuals (pn $+$ combined MOS; red and black crosses and lines, respectively), from \citet{corral11}. 
 {\it Central panel:} optical spectrum (from HELLAS2XMM); the strongest emission lines are marked.
 {\it Bottom panel:} rest-frame SED fits: the data (red open pentagons, WISE; green filled
 square, data in the $R$-band from the literature), 
 plotted as luminosity ($\lambda
 \pedix{L}{$\lambda$}$, normalized to the X-ray luminosity as in Table~\ref{tab:xbs}) vs. the rest-frame wavelength, 
 are superposed on the corresponding best-fitting template SEDs (blue dotted line, host galaxy; red dashed line, AGN; black
 solid line, total).
  In addition to the name (first row) and the redshift (second row) of the source, in the last row of the legend we summarize the main parameters of the modelling:
 the morphological type of the host, the QSO template, the adopted extinction curve (see Fig.~\ref{fig:templ}), and the dust extinction.
 See Sect.~\ref{sect:sed} for details.}%
 \label{fig:sed}%
\end{figure*}

\renewcommand{\thefigure}{\arabic{numfigsed}}

\begin{figure*}
 \centering
 \includegraphics[angle=0,height=6.5cm]{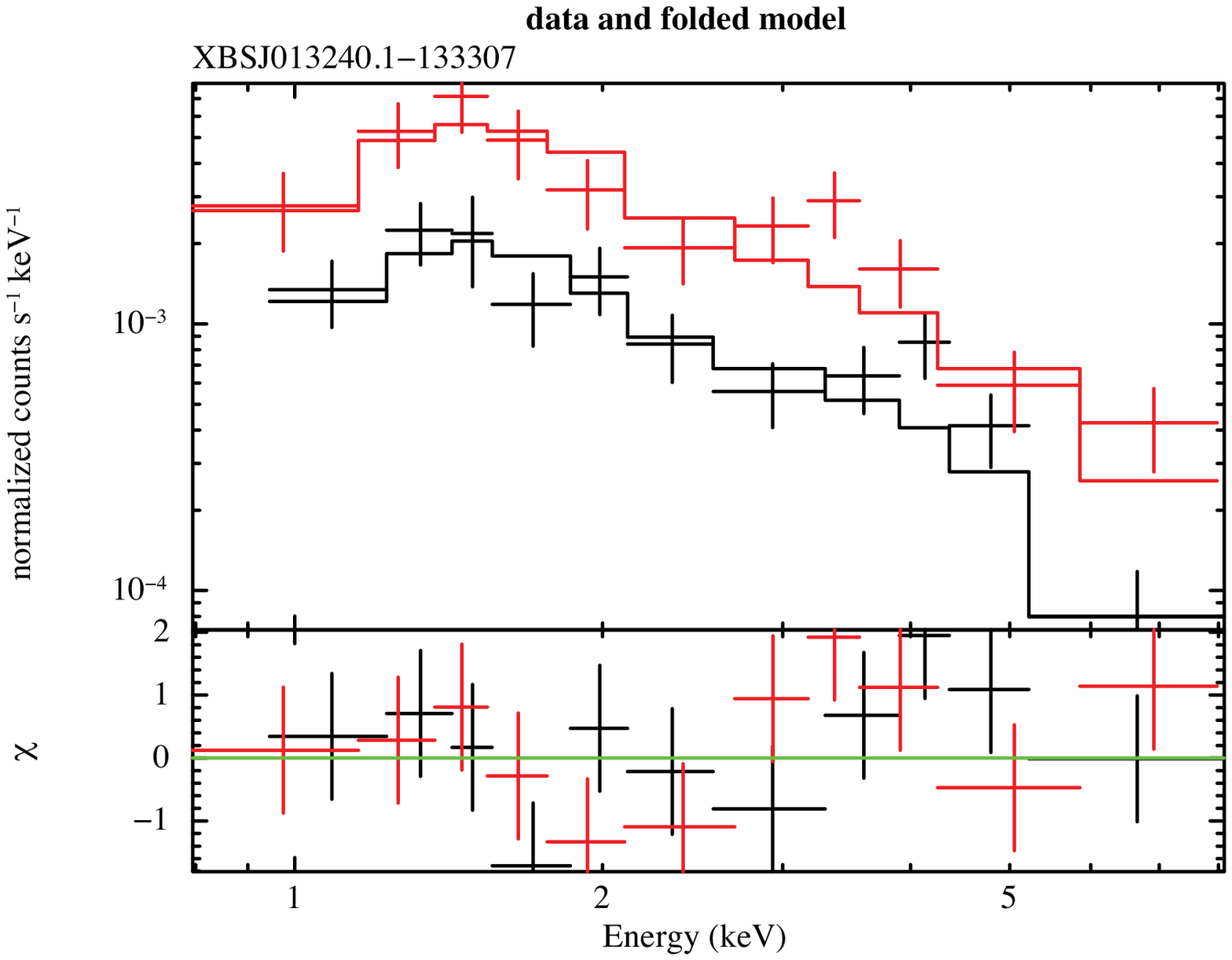}\\
 \vskip -0.5truecm
 \includegraphics[angle=-90,width=8.6cm]{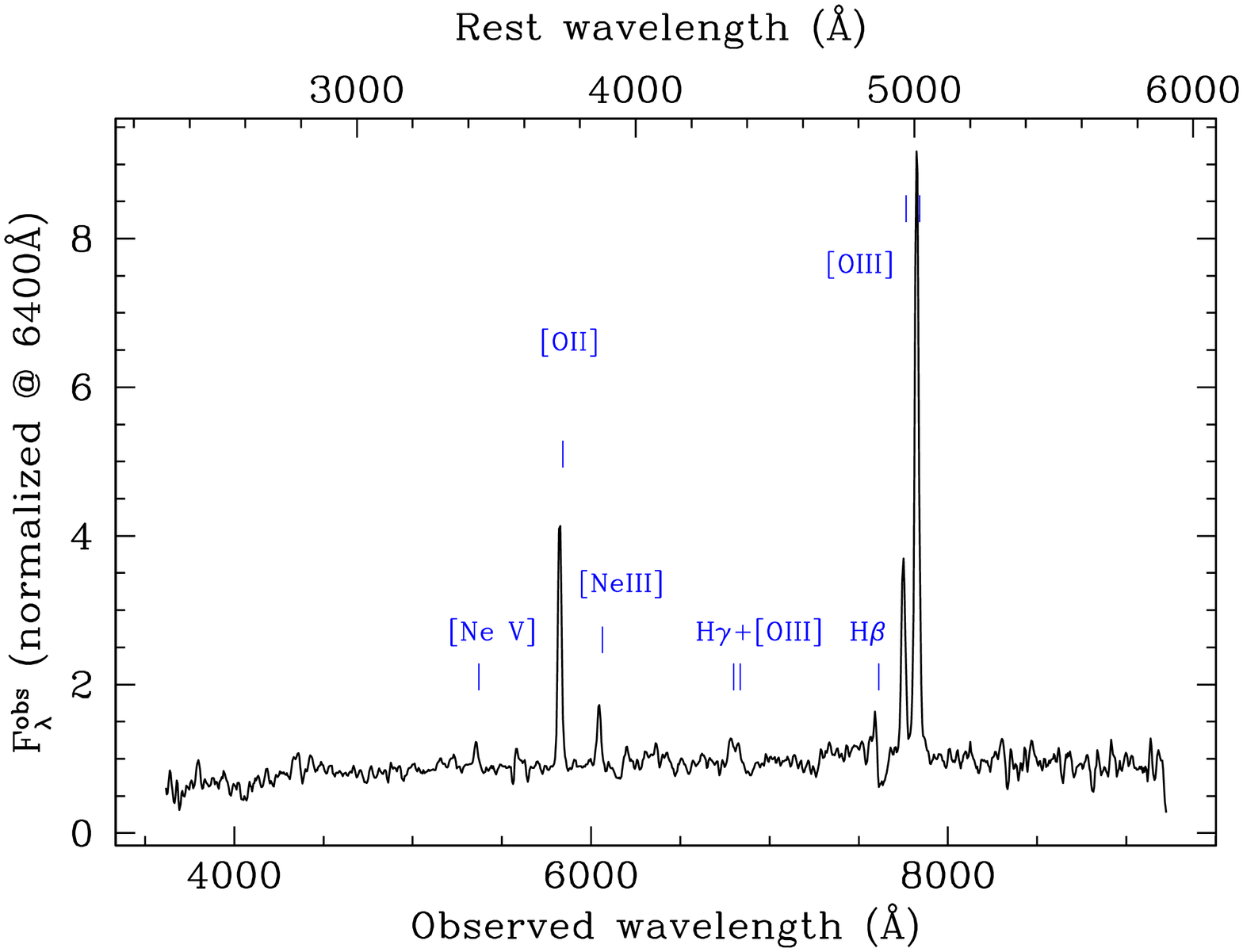}
 \vskip -0.5truecm
  \includegraphics[height=9cm]{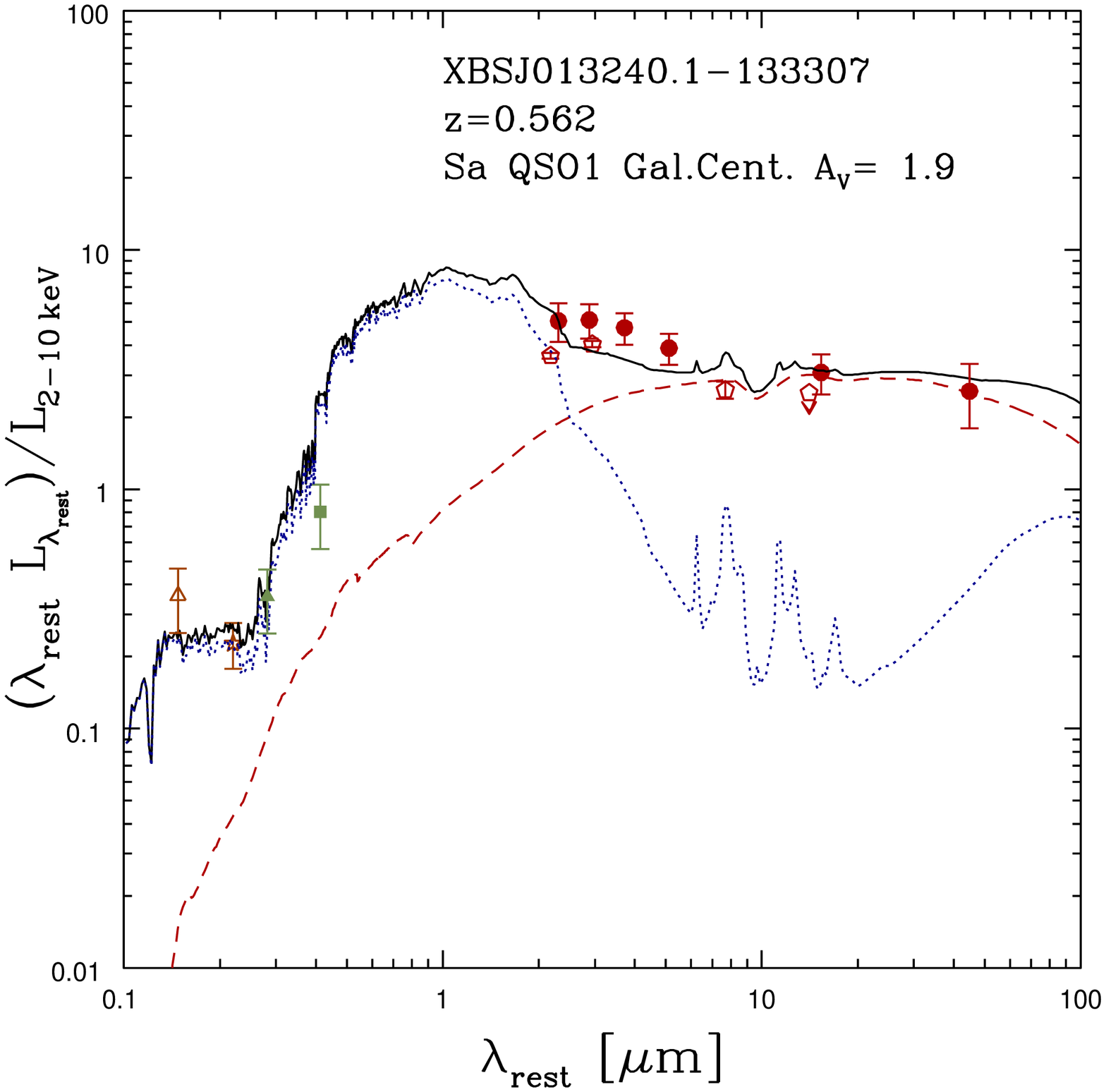}
\caption{ (cont.) {\it Top panel:} \xmm\ data and residuals (pn $+$ combined MOS; red and black crosses and lines, respectively), from \citet{corral11}. 
 {\it Central panel:} optical spectrum (from Caccianiga et al. 2004); the strongest emission lines are marked.
 {\it Bottom panel:} rest-frame SED fits: the data (red filled circles, \spitzer; red open pentagons, WISE; green filled
 square and triangle, data in the $R$-band and at $4400\,$\AA\ from the literature; brown open star and triangle, OM magnitude and \galex\ NUV flux), 
 plotted as luminosity ($\lambda
 \pedix{L}{$\lambda$}$, normalized to the X-ray luminosity as in Table~\ref{tab:xbs}) vs. the rest-frame wavelength, 
 are superposed on the corresponding best-fitting template SEDs (blue dotted line, host galaxy; red dashed line, AGN; black
 solid line, total).
  In addition to the name (first row) and the redshift (second row) of the source, in the last row of the legend we summarize the main parameters of the modelling:
 the morphological type of the host, the QSO template, the adopted extinction curve (see Fig.~\ref{fig:templ}), and the dust extinction.
 See Sect.~\ref{sect:sed} for details.}%
\end{figure*}

\renewcommand{\thefigure}{\arabic{numfigsed}}

\begin{figure*}
 \centering
 \includegraphics[angle=0,height=6.5cm]{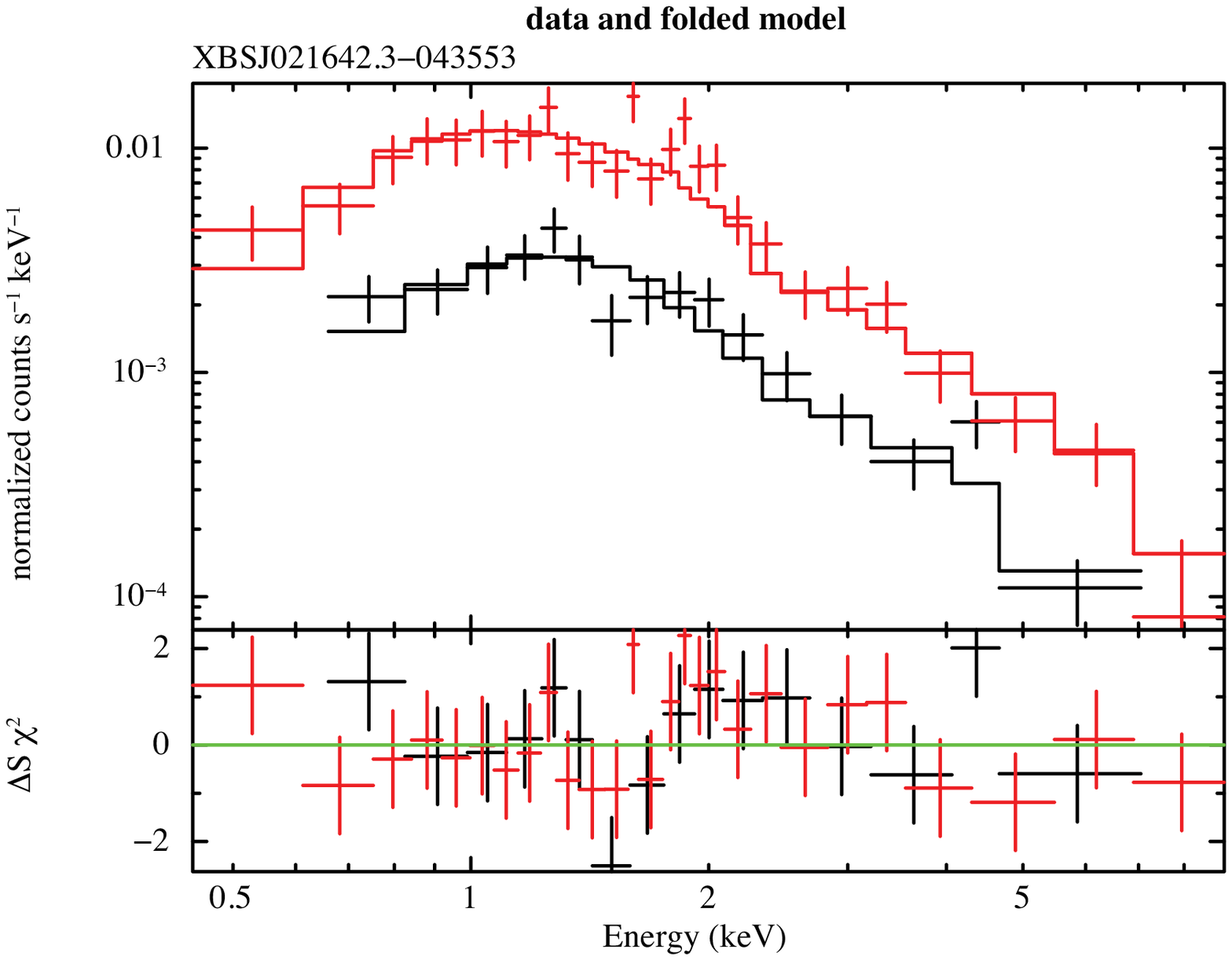}\\
 \vskip -0.5truecm
 \includegraphics[angle=-90,width=8.6cm]{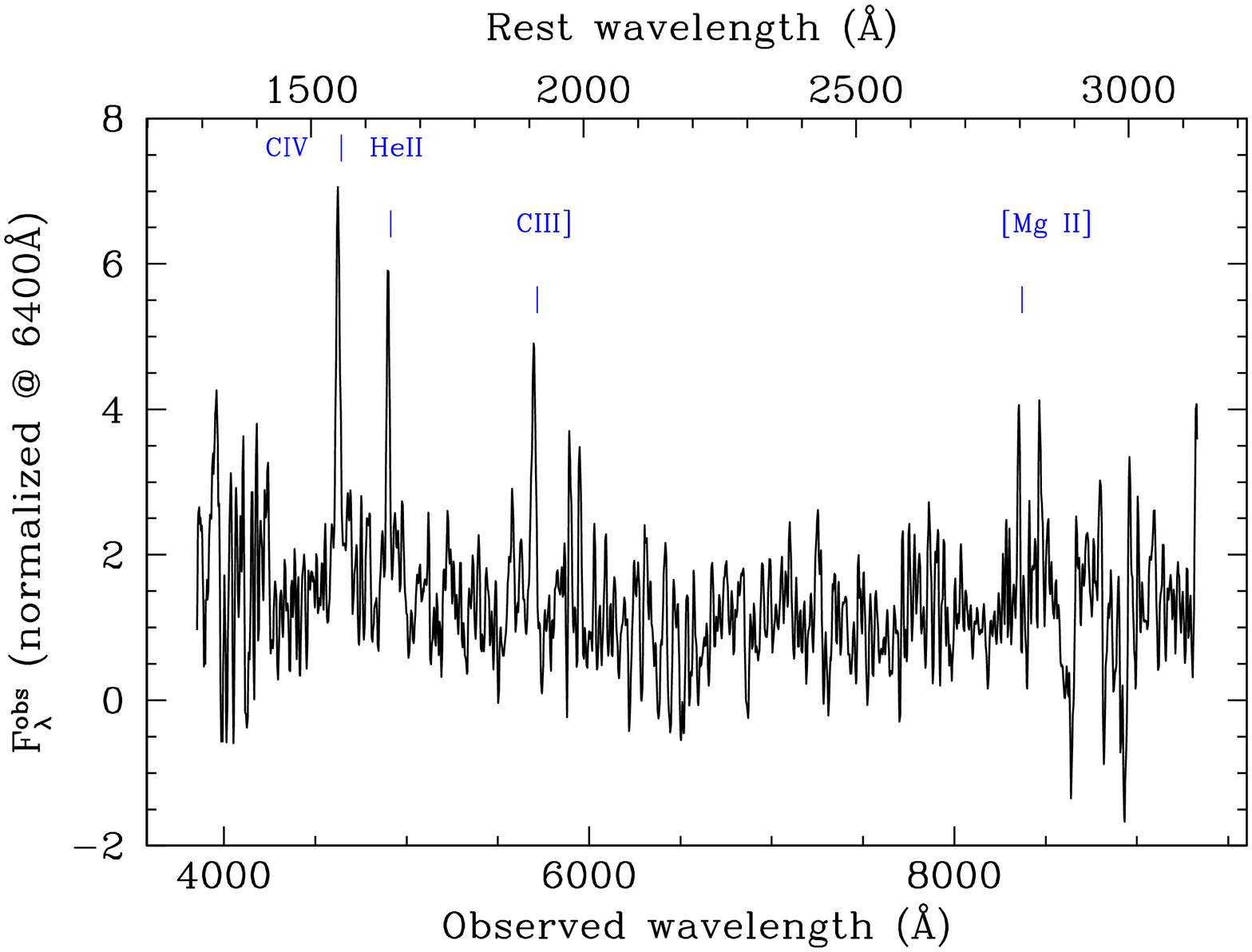}
 \vskip -0.5truecm
 \includegraphics[height=9cm]{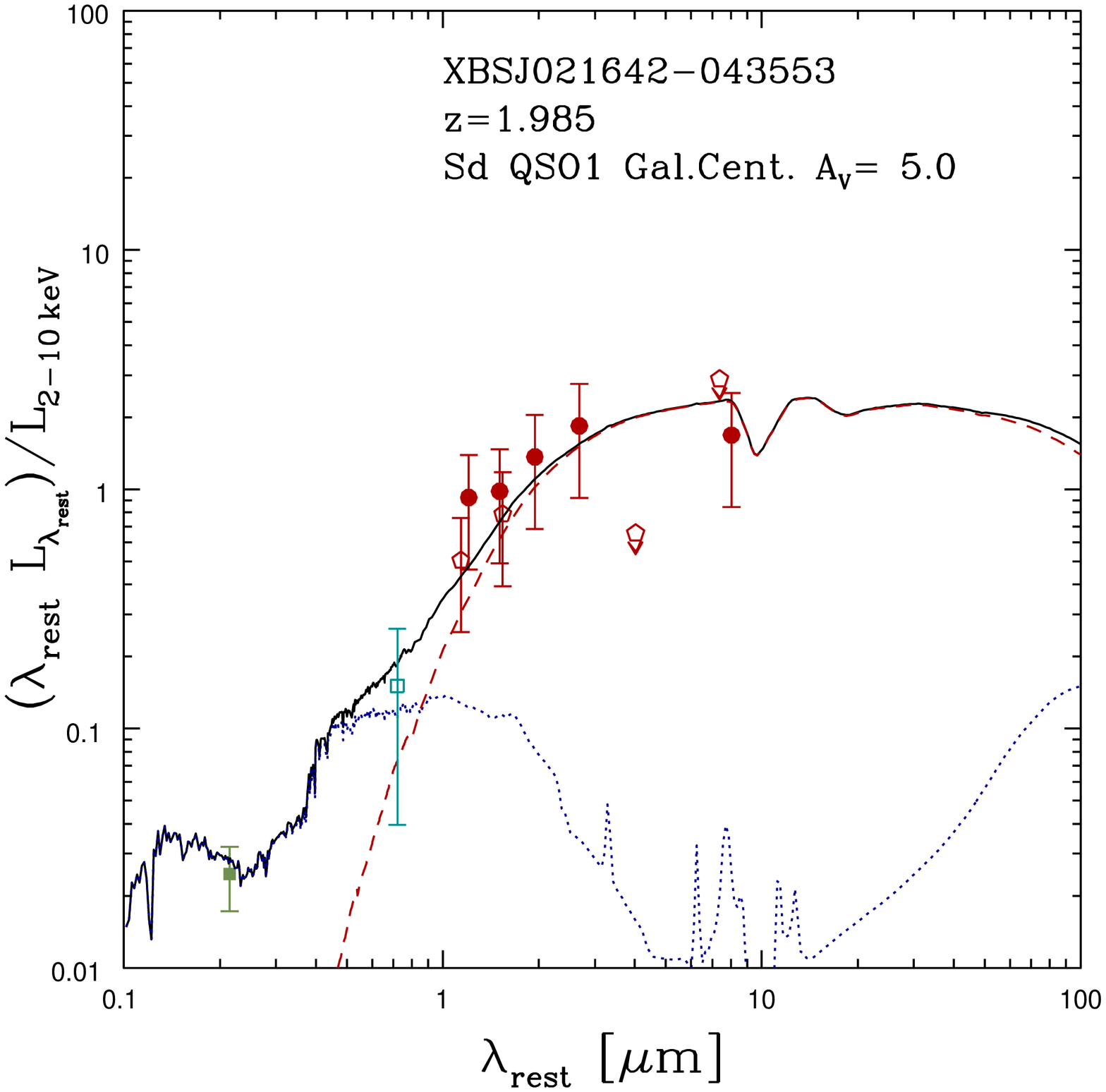}
\caption{ (cont.) {\it Top panel:} \xmm\ data and residuals (pn $+$ combined MOS; red and black crosses and lines, respectively), from \citet{corral11}. 
 {\it Central panel:} optical spectrum (from Severgnini et al. 2006); the strongest emission lines are marked.
 {\it Bottom panel:} rest-frame SED fits: the data (red filled circles, \spitzer; red open pentagons, WISE; green filled
 square data in the $R$-band from the literature; cyan open square, $K$-band flux from Severgnini et al. 2006), 
 plotted as luminosity ($\lambda
 \pedix{L}{$\lambda$}$, normalized to the X-ray luminosity as in Table~\ref{tab:xbs}) vs. the rest-frame wavelength, 
 are superposed on the corresponding best-fitting template SEDs (blue dotted line, host galaxy; red dashed line, AGN; black
 solid line, total).
  In addition to the name (first row) and the redshift (second row) of the source, in the last row of the legend we summarize the main parameters of the modelling:
 the morphological type of the host, the QSO template, the adopted extinction curve (see Fig.~\ref{fig:templ}), and the dust extinction.
 See Sect.~\ref{sect:sed} for details.}%
\end{figure*}

\renewcommand{\thefigure}{\arabic{numfigsed}}

\begin{figure*}
 \centering
 \includegraphics[angle=0,height=6.5cm]{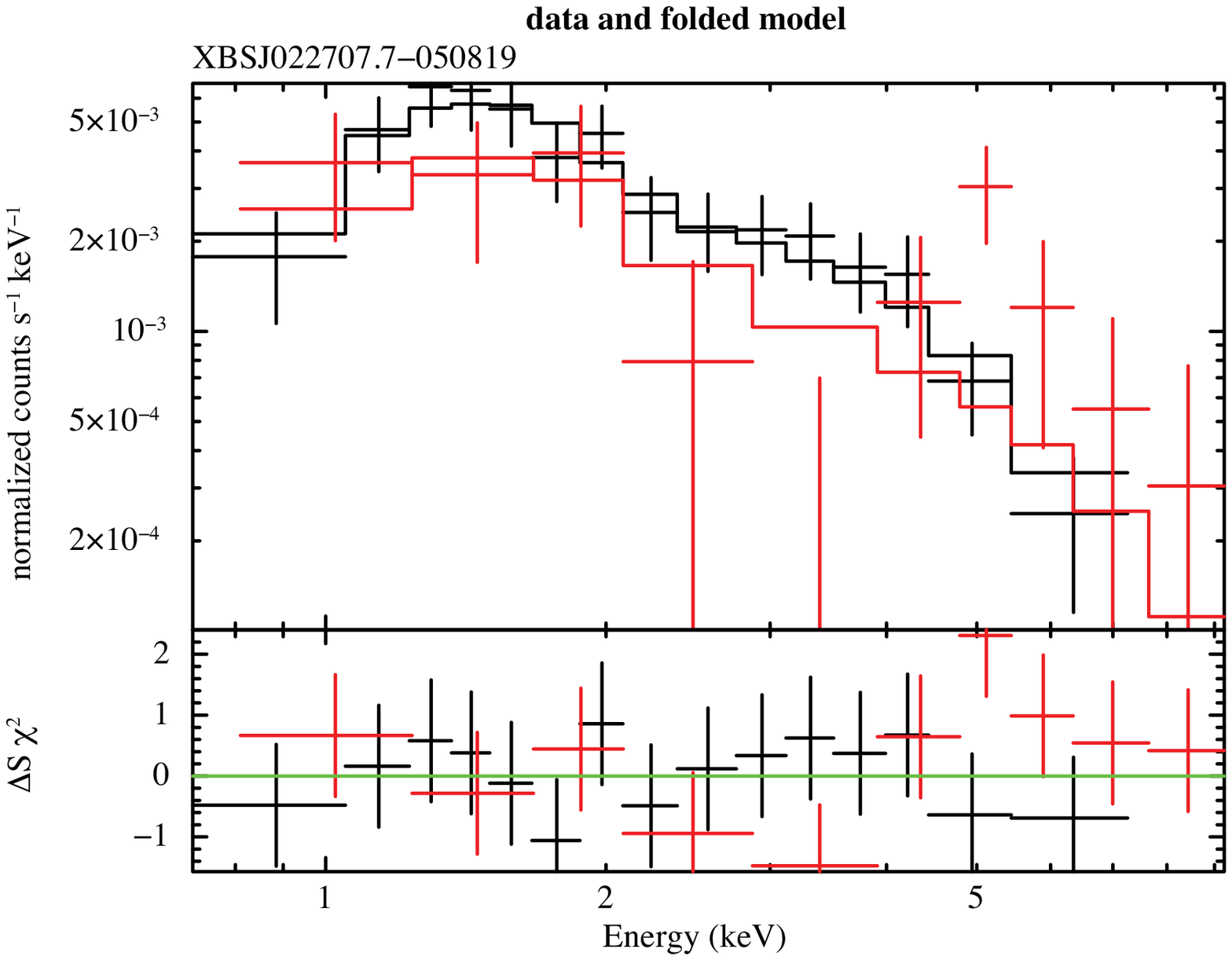}\\
 \vskip -0.5truecm
 \includegraphics[angle=-90,width=8.6cm]{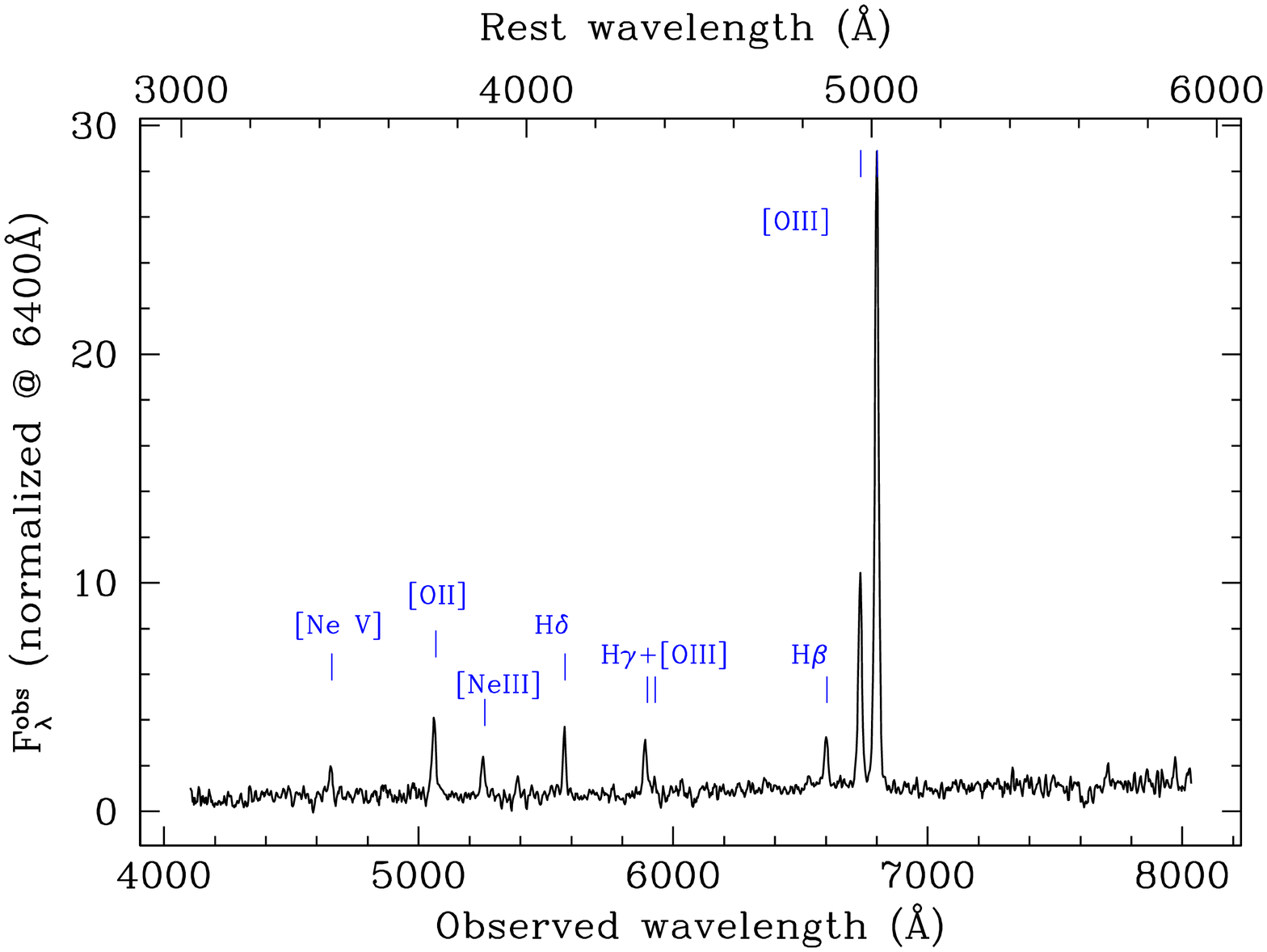}
 \vskip -0.5truecm
 \includegraphics[height=9cm]{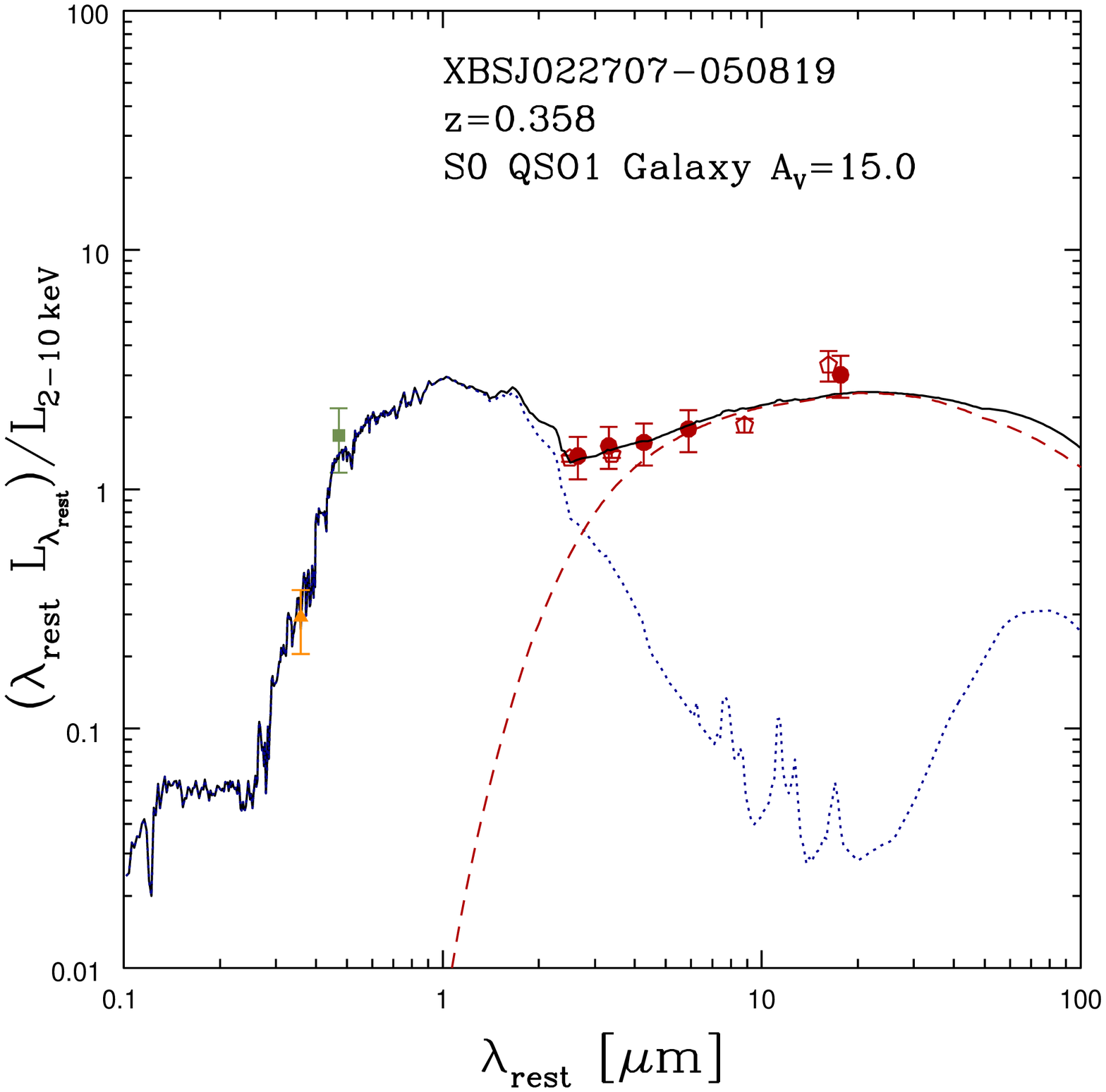}
\caption{ (cont.) {\it Top panel:} \xmm\ data and residuals (pn $+$ combined MOS; red and black crosses and lines, respectively), from \citet{corral11}. 
 {\it Central panel:} optical spectrum; the strongest emission lines are marked.
 {\it Bottom panel:} rest-frame SED fits: the data (red filled circles, \spitzer; red open pentagons, WISE; green filled
 square and orange filled triangle, data in the $R$-band and at $4872\,$\AA\ from the literature), 
 plotted as luminosity ($\lambda
 \pedix{L}{$\lambda$}$, normalized to the X-ray luminosity as in Table~\ref{tab:xbs}) vs. the rest-frame wavelength, 
 are superposed on the corresponding best-fitting template SEDs (blue dotted line, host galaxy; red dashed line, AGN; black
 solid line, total).
  In addition to the name (first row) and the redshift (second row) of the source, in the last row of the legend we summarize the main parameters of the modelling:
 the morphological type of the host, the QSO template, the adopted extinction curve (see Fig.~\ref{fig:templ}), and the dust extinction.
 See Sect.~\ref{sect:sed} for details.}%
\end{figure*}

\renewcommand{\thefigure}{\arabic{numfigsed}}

\begin{figure*}
 \centering
 \includegraphics[angle=0,height=6.5cm]{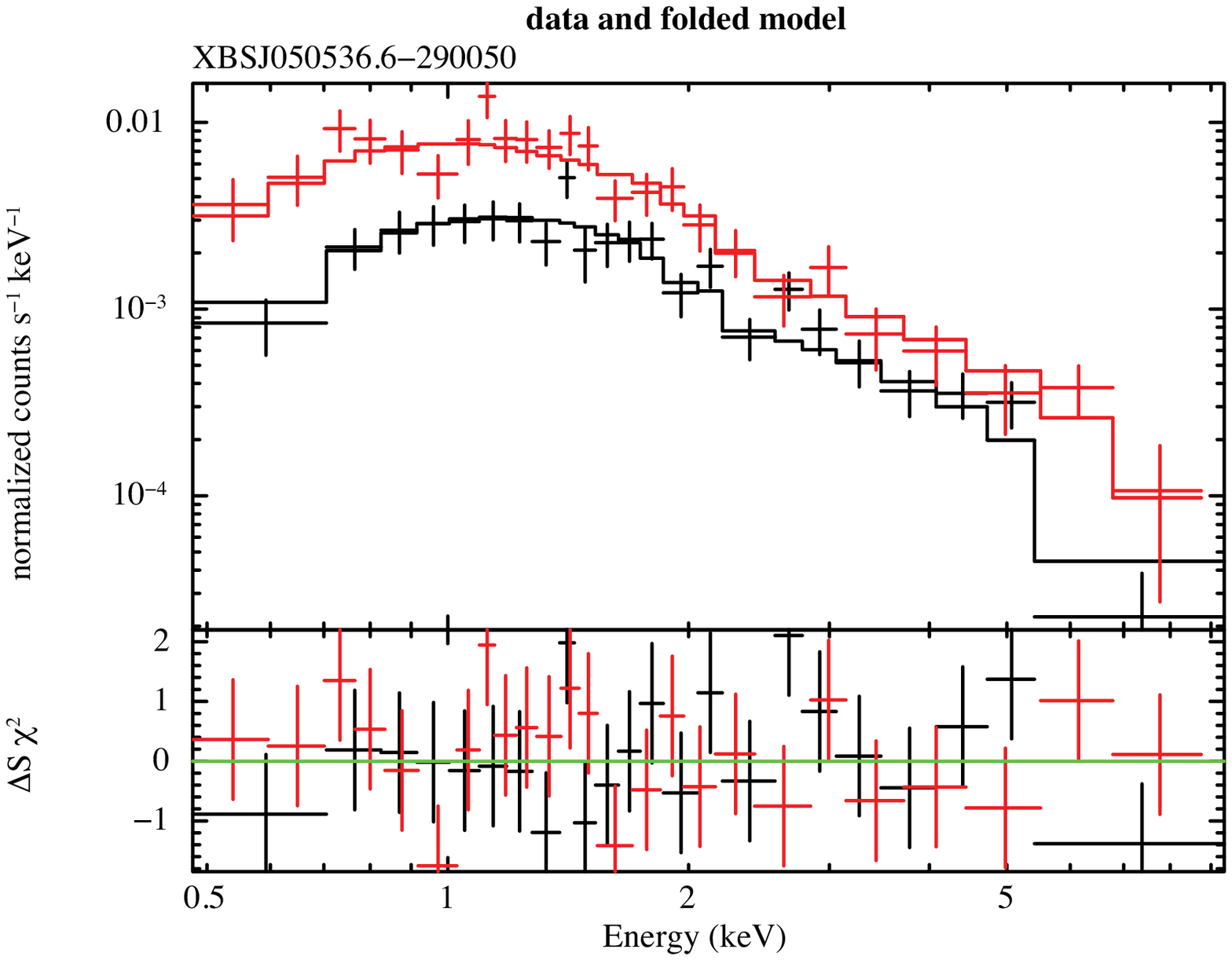}\\
 \vskip -0.5truecm
 \includegraphics[angle=-90,width=8.6cm]{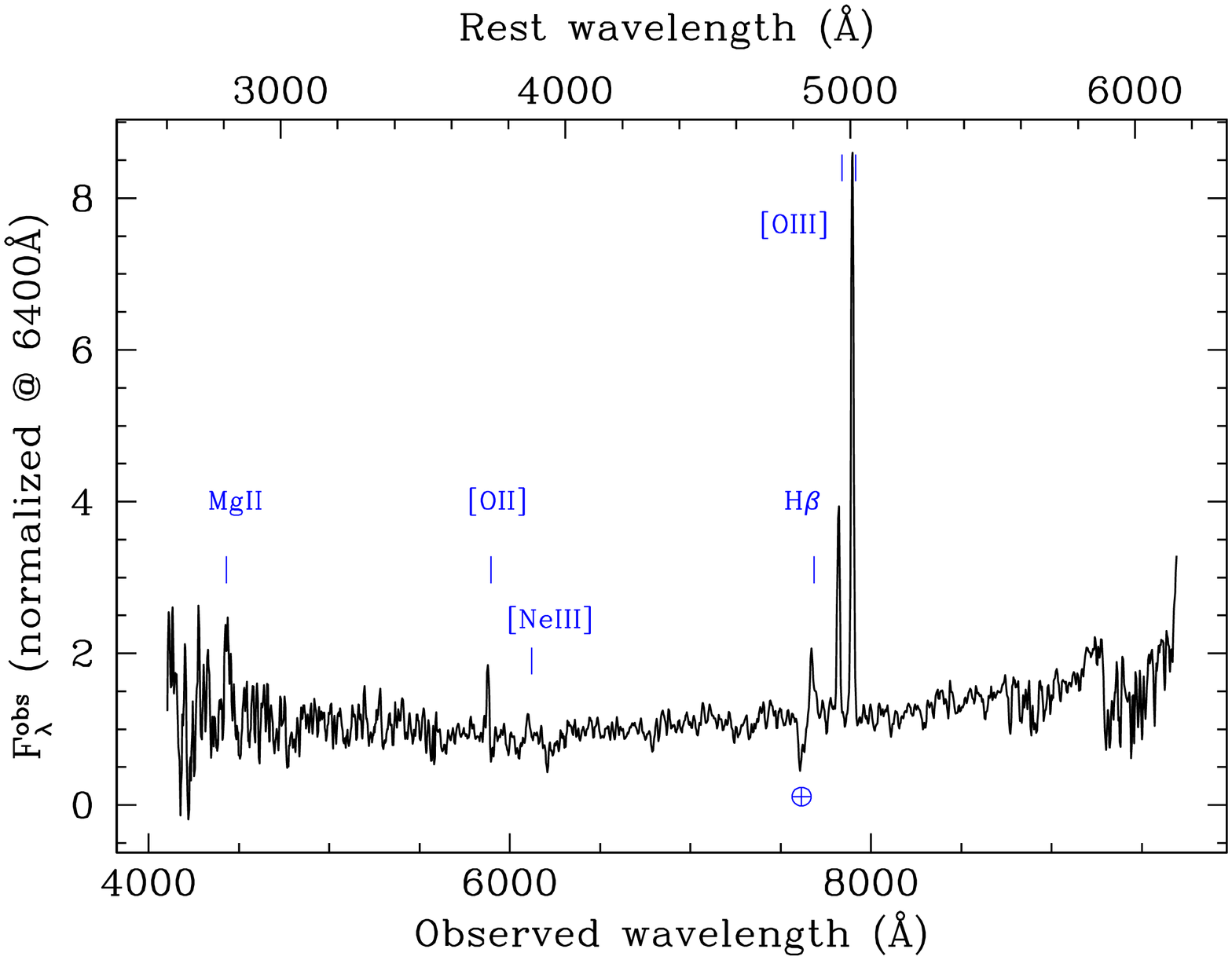}
 \vskip -0.5truecm
 \includegraphics[height=9cm]{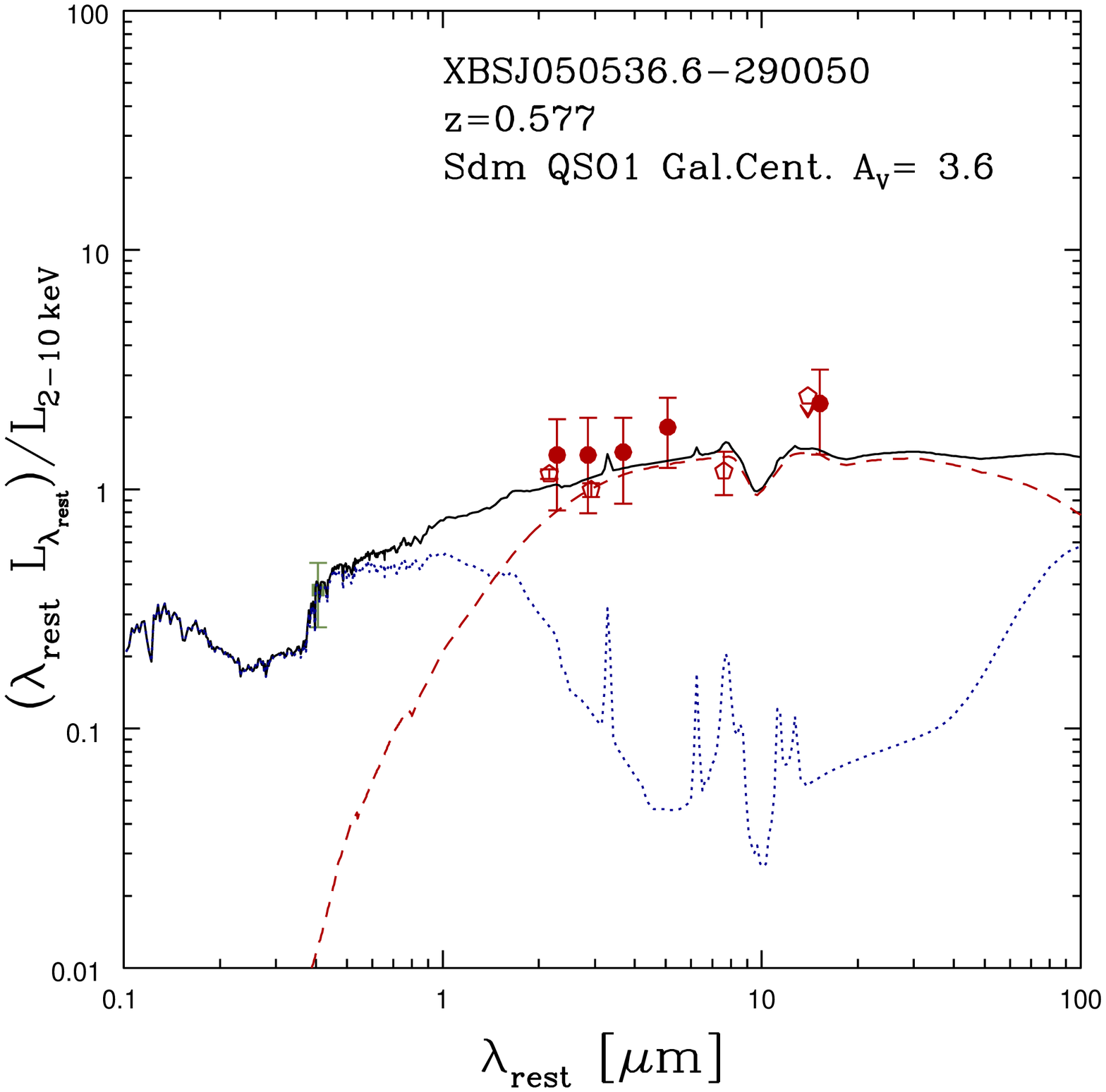}
\caption{ (cont.) {\it Top panel:} \xmm\ data and residuals (pn $+$ combined MOS; red and black crosses and lines, respectively), from \citet{corral11}. 
 {\it Central panel:} optical spectrum; the strongest emission lines are marked.
 {\it Bottom panel:} rest-frame SED fits: the data (red filled circles, \spitzer; red open pentagons, WISE; green filled
 square, data in the $R$-band from the literature), 
 plotted as luminosity ($\lambda
 \pedix{L}{$\lambda$}$, normalized to the X-ray luminosity as in Table~\ref{tab:xbs}) vs. the rest-frame wavelength, 
 are superposed on the corresponding best-fitting template SEDs (blue dotted line, host galaxy; red dashed line, AGN; black
 solid line, total).
  In addition to the name (first row) and the redshift (second row) of the source, in the last row of the legend we summarize the main parameters of the modelling:
 the morphological type of the host, the QSO template, the adopted extinction curve (see Fig.~\ref{fig:templ}), and the dust extinction.
 See Sect.~\ref{sect:sed} for details.}%
\end{figure*}

\renewcommand{\thefigure}{\arabic{numfigsed}}

\begin{figure*}
 \centering
 \includegraphics[angle=0,height=6.5cm]{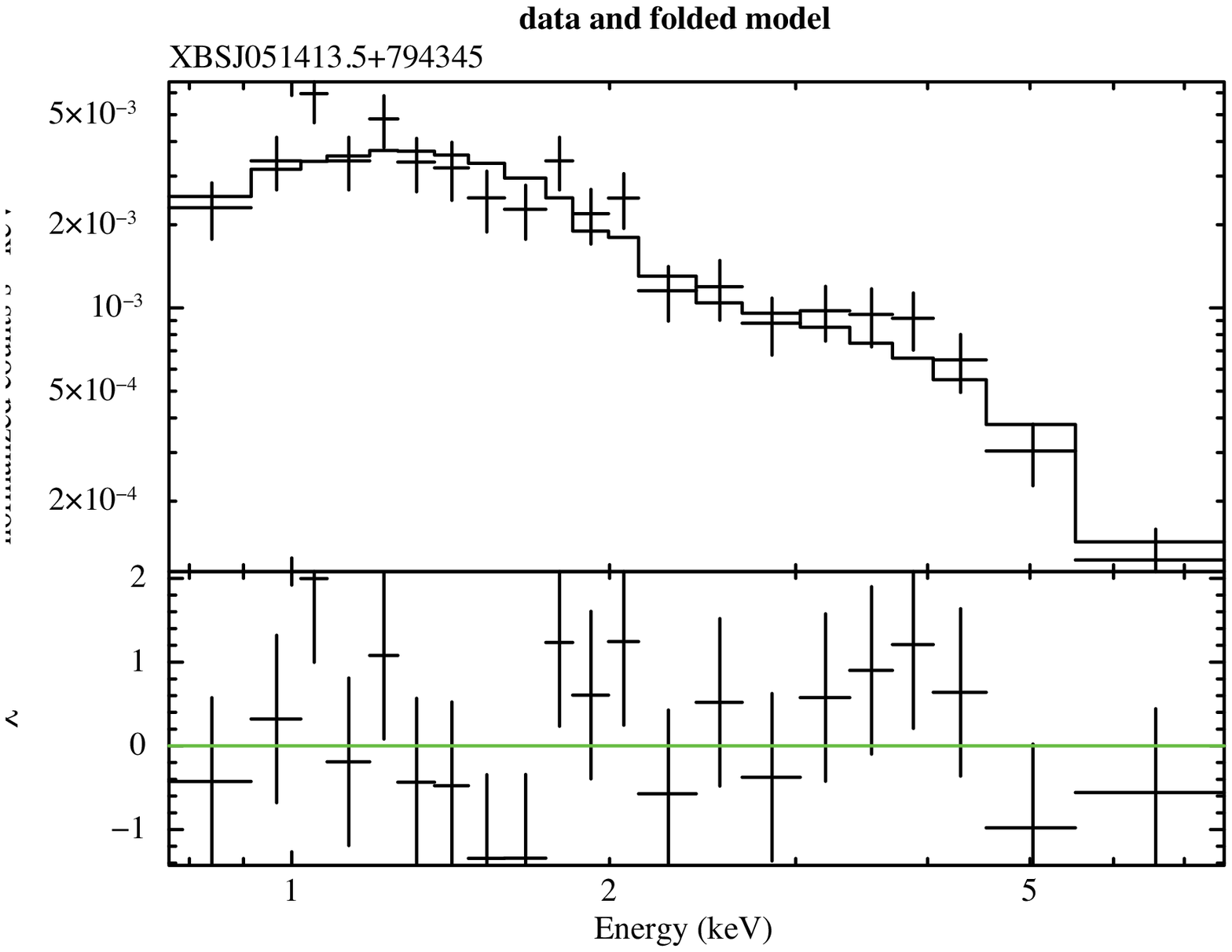}\\
 \vskip -0.5truecm
 \includegraphics[angle=-90,width=8.6cm]{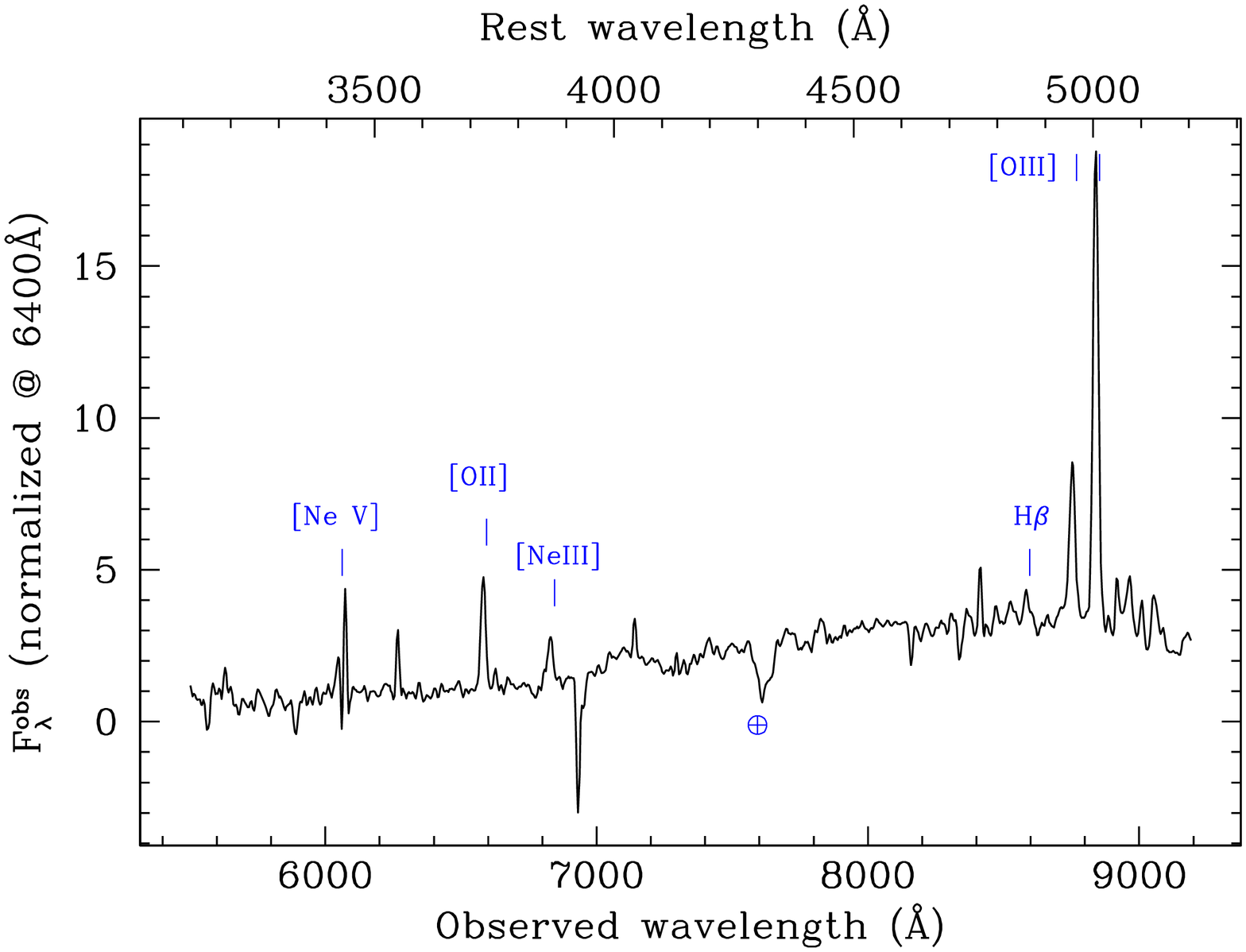}
 \vskip -0.5truecm
 \includegraphics[height=9cm]{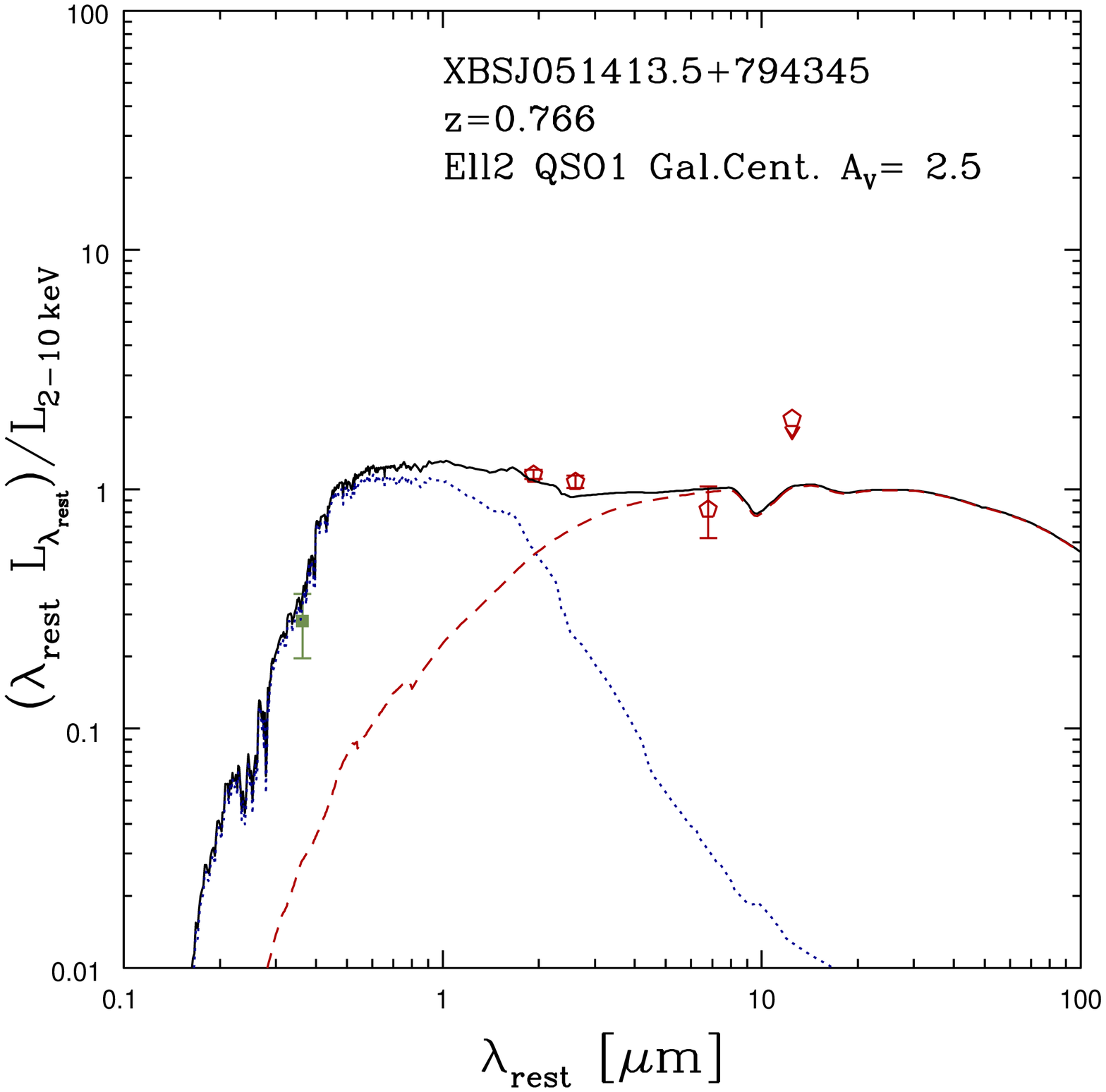}
\caption{ (cont.) {\it Top panel:} \xmm\ data and residuals (pn $+$ combined MOS; red and black crosses and lines, respectively). 
 {\it Central panel:} optical spectrum; the strongest emission lines are marked.
 {\it Bottom panel:} rest-frame SED fits: the data (red open pentagons, WISE; green filled
 square, data in the $R$-band from the literature), 
 plotted as luminosity ($\lambda
 \pedix{L}{$\lambda$}$, normalized to the X-ray luminosity as in Table~\ref{tab:xbs}) vs. the rest-frame wavelength, 
 are superposed on the corresponding best-fitting template SEDs (blue dotted line, host galaxy; red dashed line, AGN; black
 solid line, total).
  In addition to the name (first row) and the redshift (second row) of the source, in the last row of the legend we summarize the main parameters of the modelling:
 the morphological type of the host, the QSO template, the adopted extinction curve (see Fig.~\ref{fig:templ}), and the dust extinction.
 See Sect.~\ref{sect:sed} for details.}%
\end{figure*}

\renewcommand{\thefigure}{\arabic{numfigsed}}

\begin{figure*}
 \centering
 \includegraphics[angle=0,height=6.5cm]{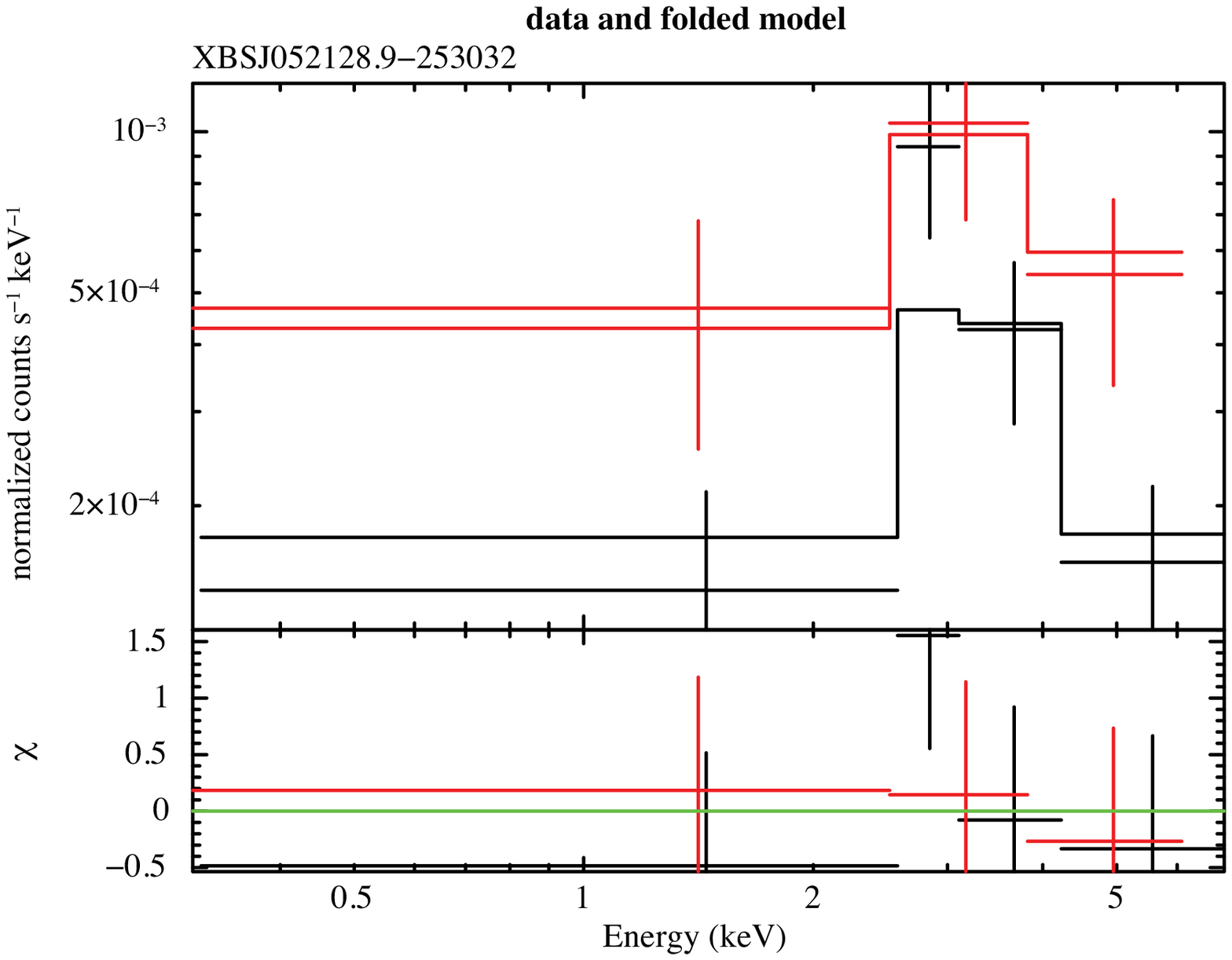}\\
 \vskip -0.5truecm
 \includegraphics[angle=-90,width=8.6cm]{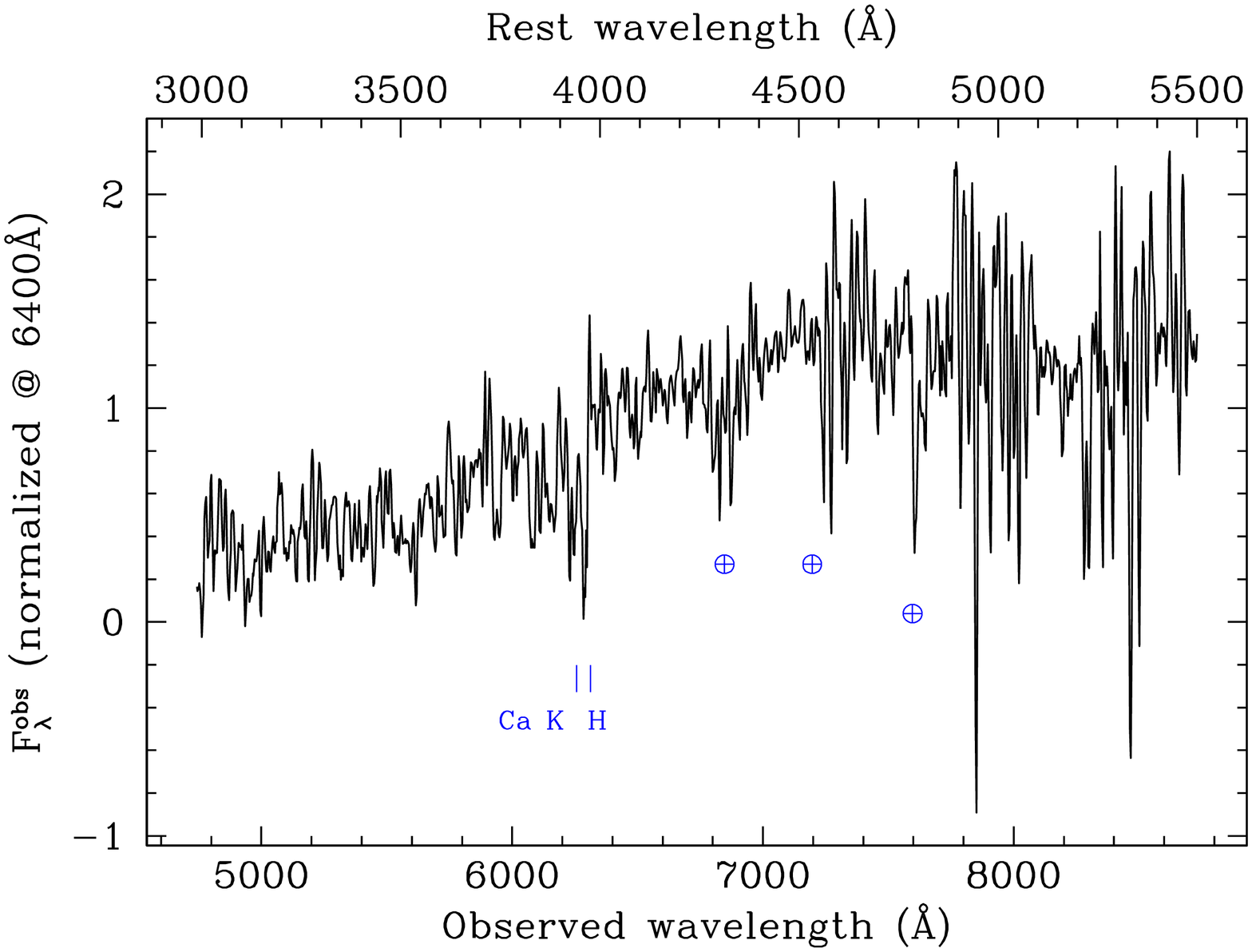}
 \vskip -0.5truecm
 \includegraphics[height=9cm]{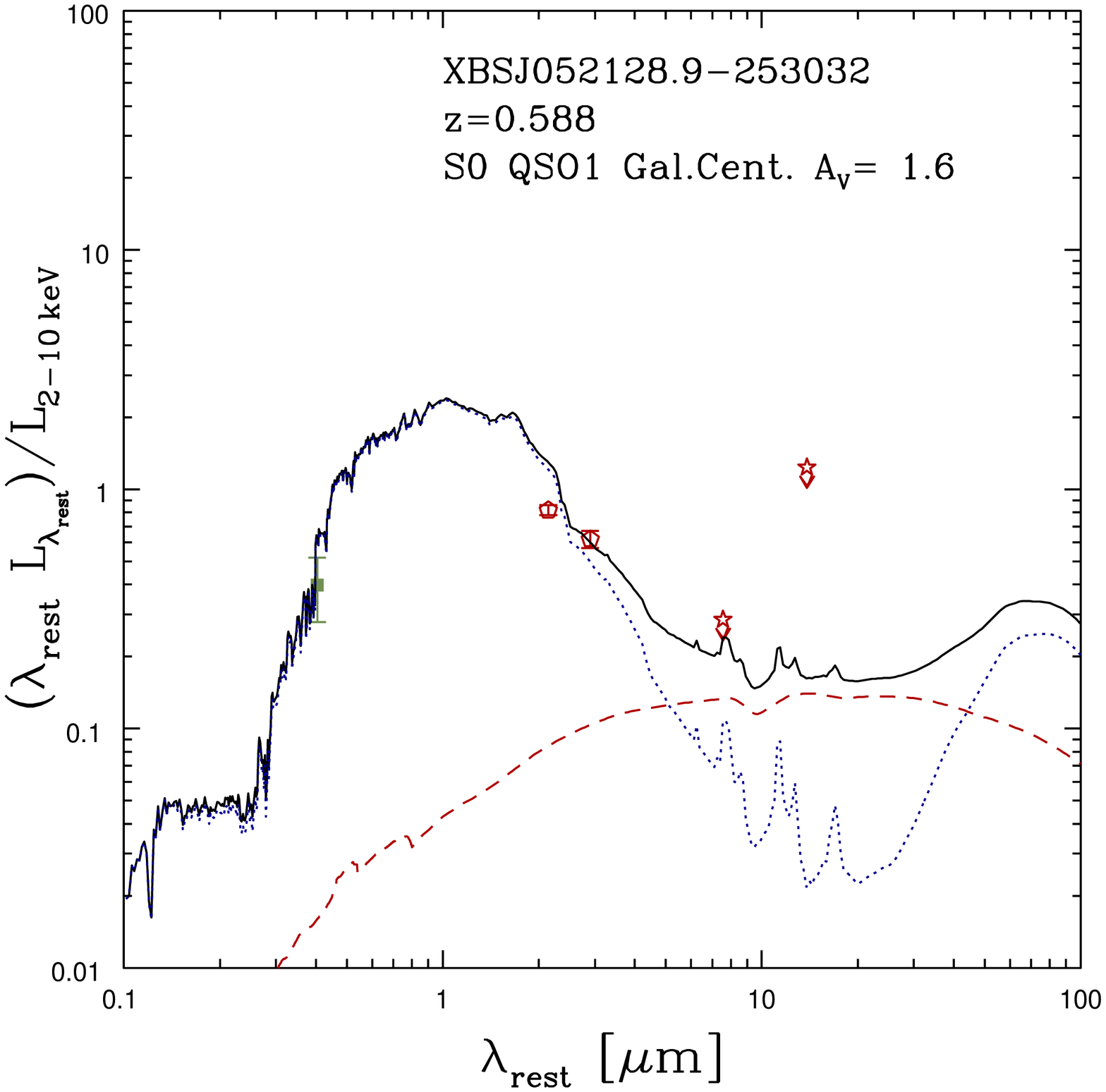}
\caption{ (cont.) {\it Top panel:} \xmm\ data and residuals (pn $+$ combined MOS; red and black crosses and lines, respectively), from \citet{corral11}. 
 {\it Central panel:} optical spectrum (from Caccianiga et al. 2004); the strongest emission lines are marked.
 {\it Bottom panel:} rest-frame SED fits: the data (red open pentagons, WISE; green filled
 square, data in the $R$-band from the literature), 
 plotted as luminosity ($\lambda
 \pedix{L}{$\lambda$}$, normalized to the X-ray luminosity as in Table~\ref{tab:xbs}) vs. the rest-frame wavelength, 
 are superposed on the corresponding best-fitting template SEDs (blue dotted line, host galaxy; red dashed line, AGN; black
 solid line, total).
  In addition to the name (first row) and the redshift (second row) of the source, in the last row of the legend we summarize the main parameters of the modelling:
 the morphological type of the host, the QSO template, the adopted extinction curve (see Fig.~\ref{fig:templ}), and the dust extinction.
 See Sect.~\ref{sect:sed} for details.}%
\end{figure*}

\renewcommand{\thefigure}{\arabic{numfigsed}}

\begin{figure*}
 \centering
 \includegraphics[angle=0,height=6.5cm]{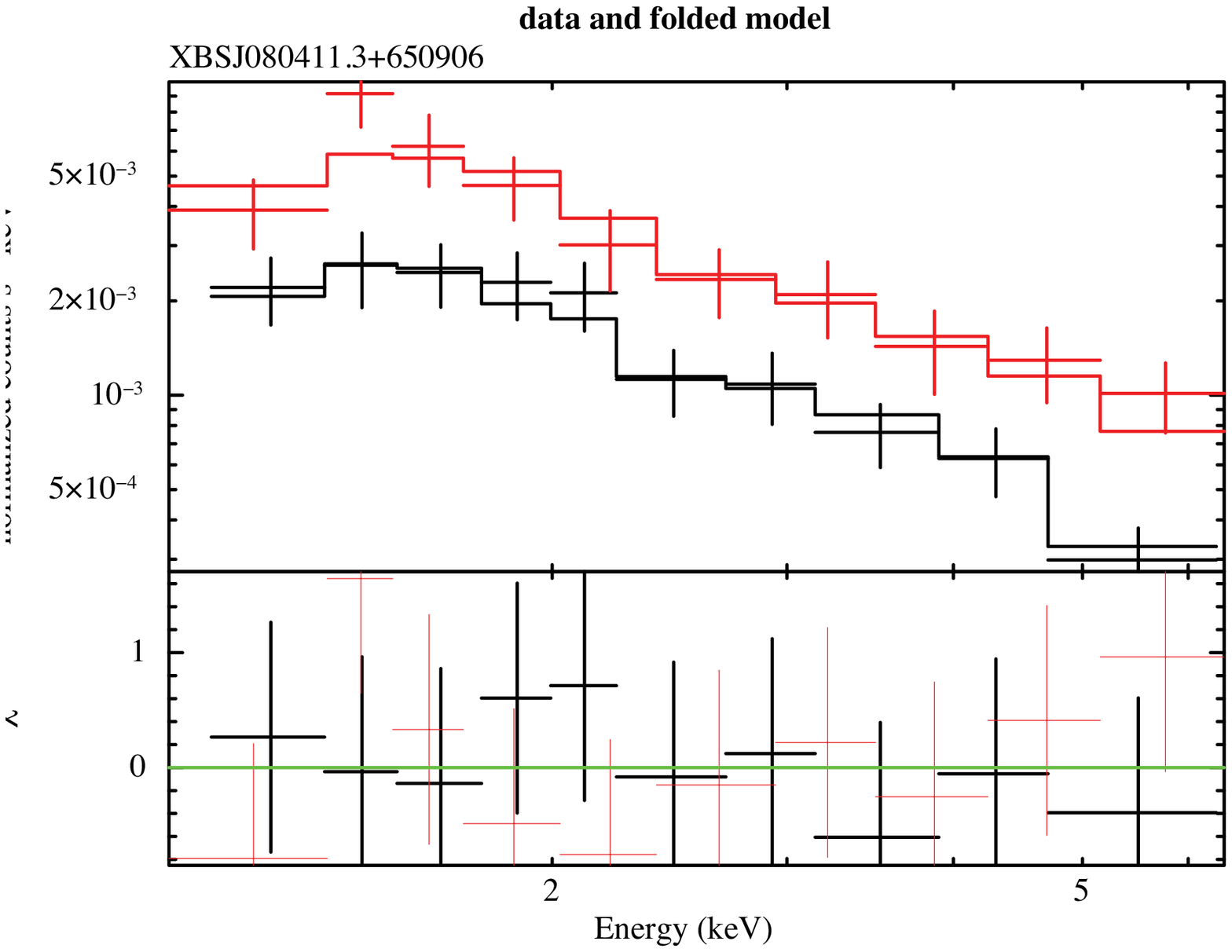}\\
 \vskip -0.5truecm
 \includegraphics[angle=-90,width=8.6cm]{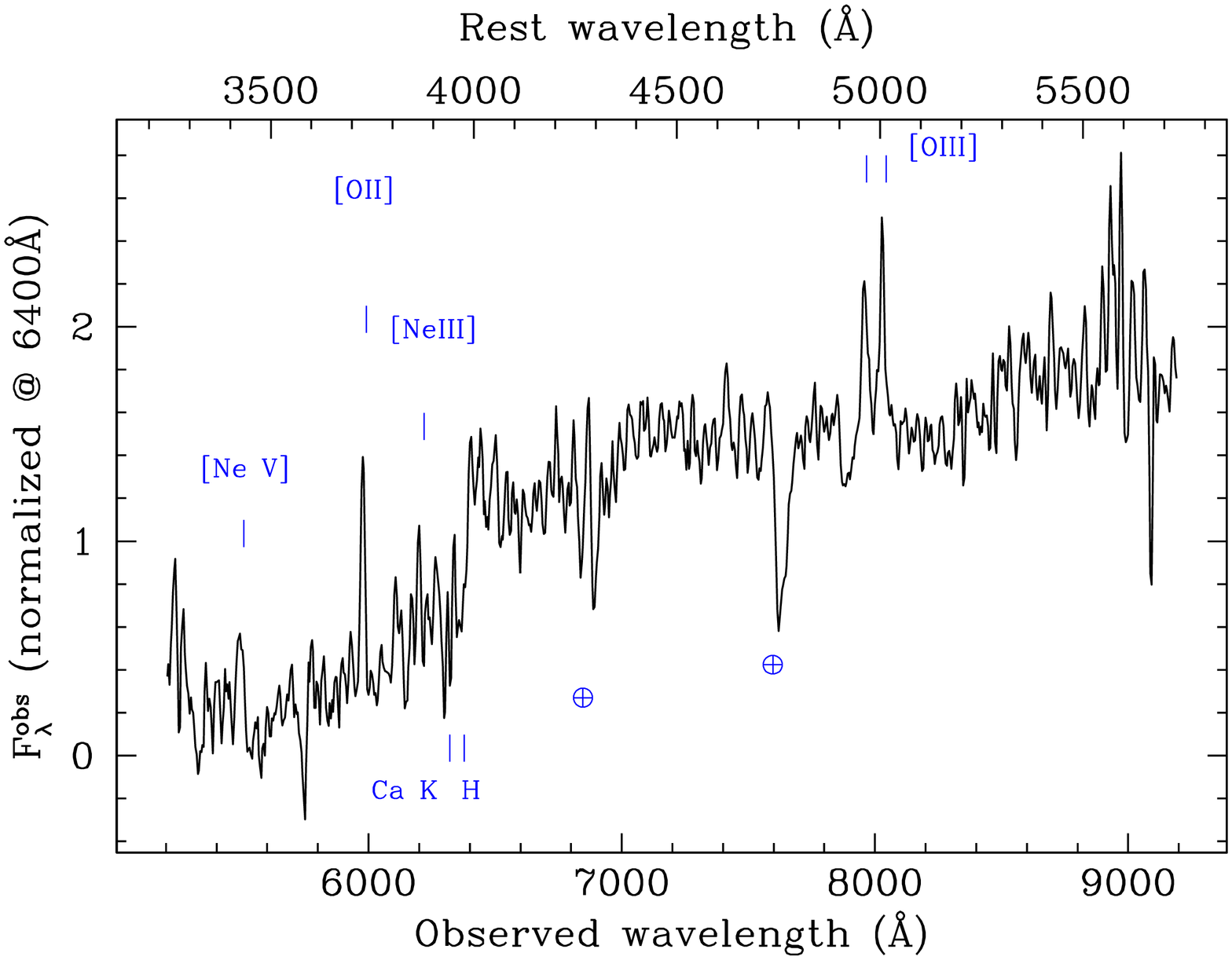}
 \vskip -0.5truecm
 \includegraphics[height=9cm]{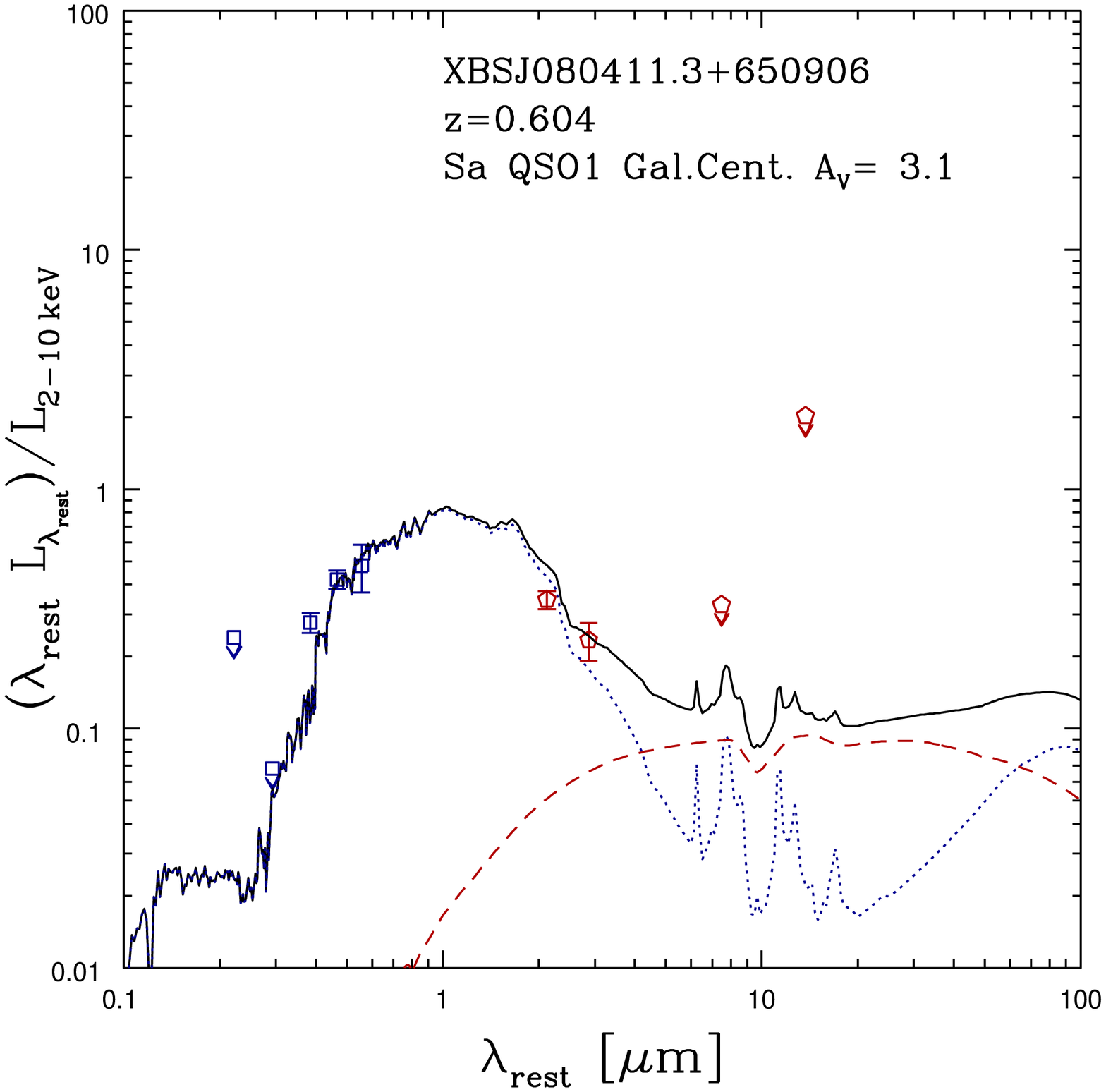}
\caption{ (cont.) {\it Top panel:} \xmm\ data and residuals (pn $+$ combined MOS; red and black crosses and lines, respectively). 
 {\it Central panel:} optical spectrum; the strongest emission lines are marked.
 {\it Bottom panel:} rest-frame SED fits: the data (red open pentagons, WISE; blue open
 squares, \sdss), 
 plotted as luminosity ($\lambda
 \pedix{L}{$\lambda$}$, normalized to the X-ray luminosity as in Table~\ref{tab:xbs}) vs. the rest-frame wavelength, 
 are superposed on the corresponding best-fitting template SEDs (blue dotted line, host galaxy; red dashed line, AGN; black
 solid line, total).
  In addition to the name (first row) and the redshift (second row) of the source, in the last row of the legend we summarize the main parameters of the modelling:
 the morphological type of the host, the QSO template, the adopted extinction curve (see Fig.~\ref{fig:templ}), and the dust extinction.
 See Sect.~\ref{sect:sed} for details.}%
\end{figure*}

\renewcommand{\thefigure}{\arabic{numfigsed}}

\begin{figure*}
 \centering
 \includegraphics[angle=0,height=6.5cm]{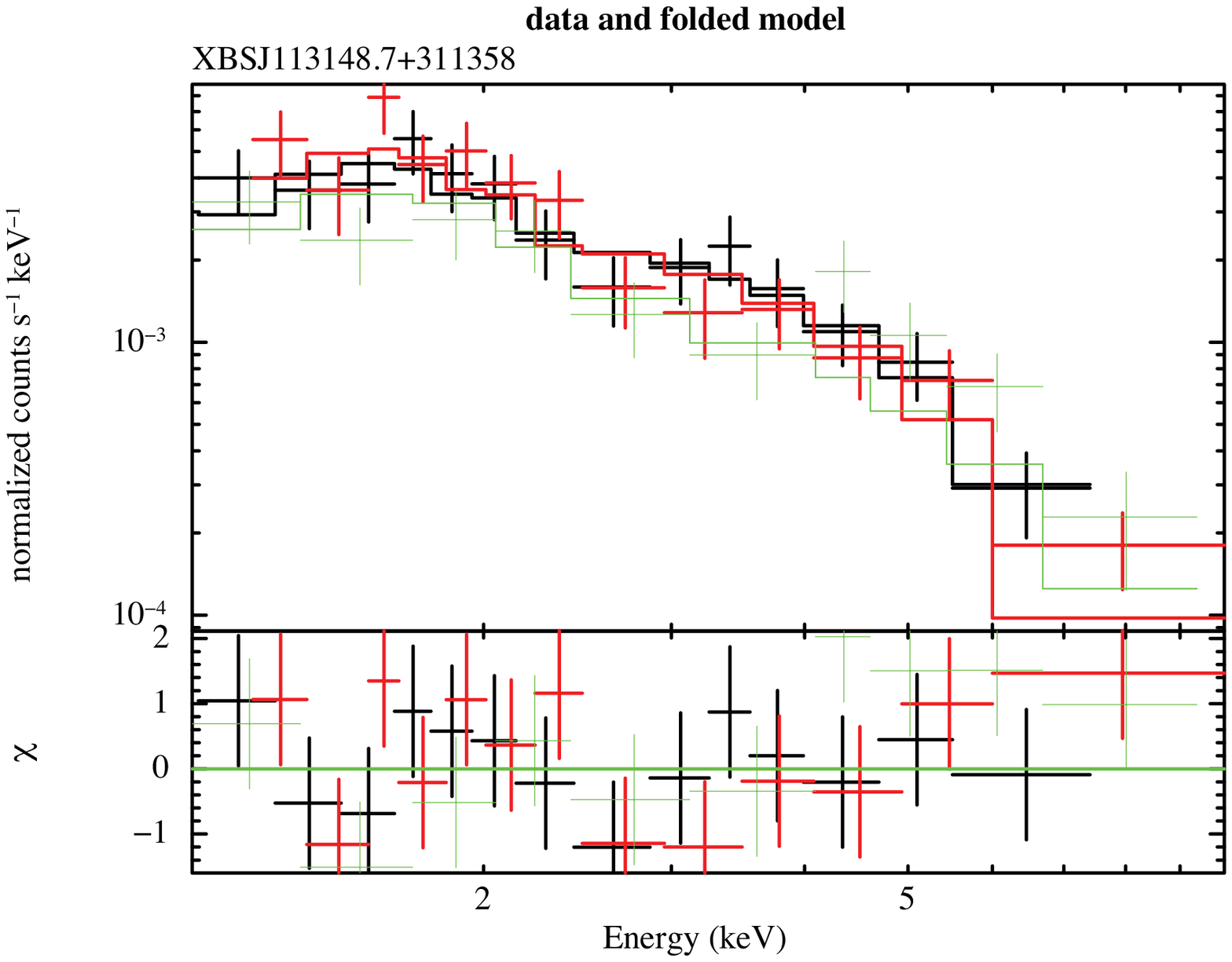}\\
 \vskip -0.5truecm
 \includegraphics[angle=-90,width=8.6cm]{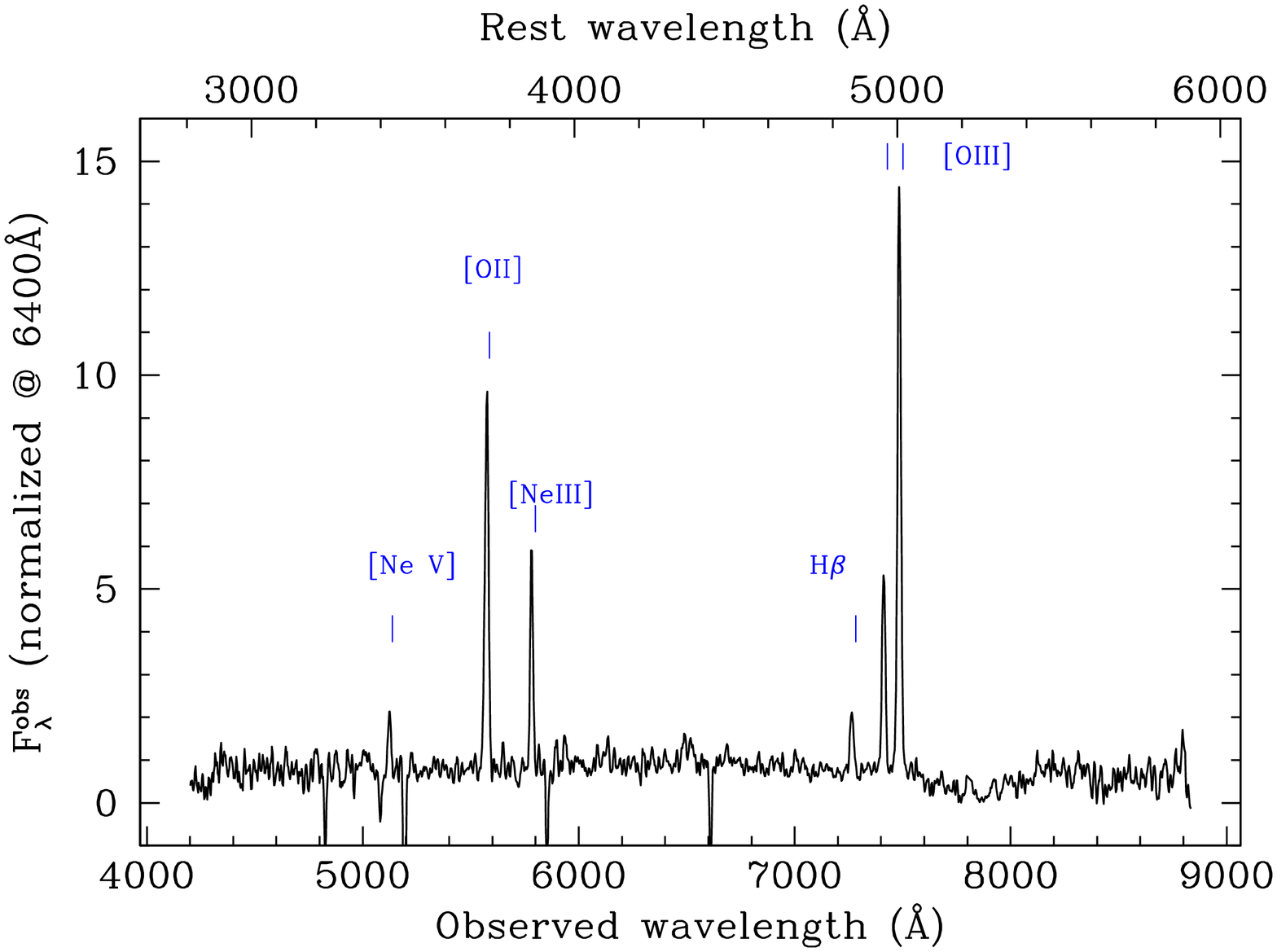}
 \vskip -0.5truecm
 \includegraphics[height=9cm]{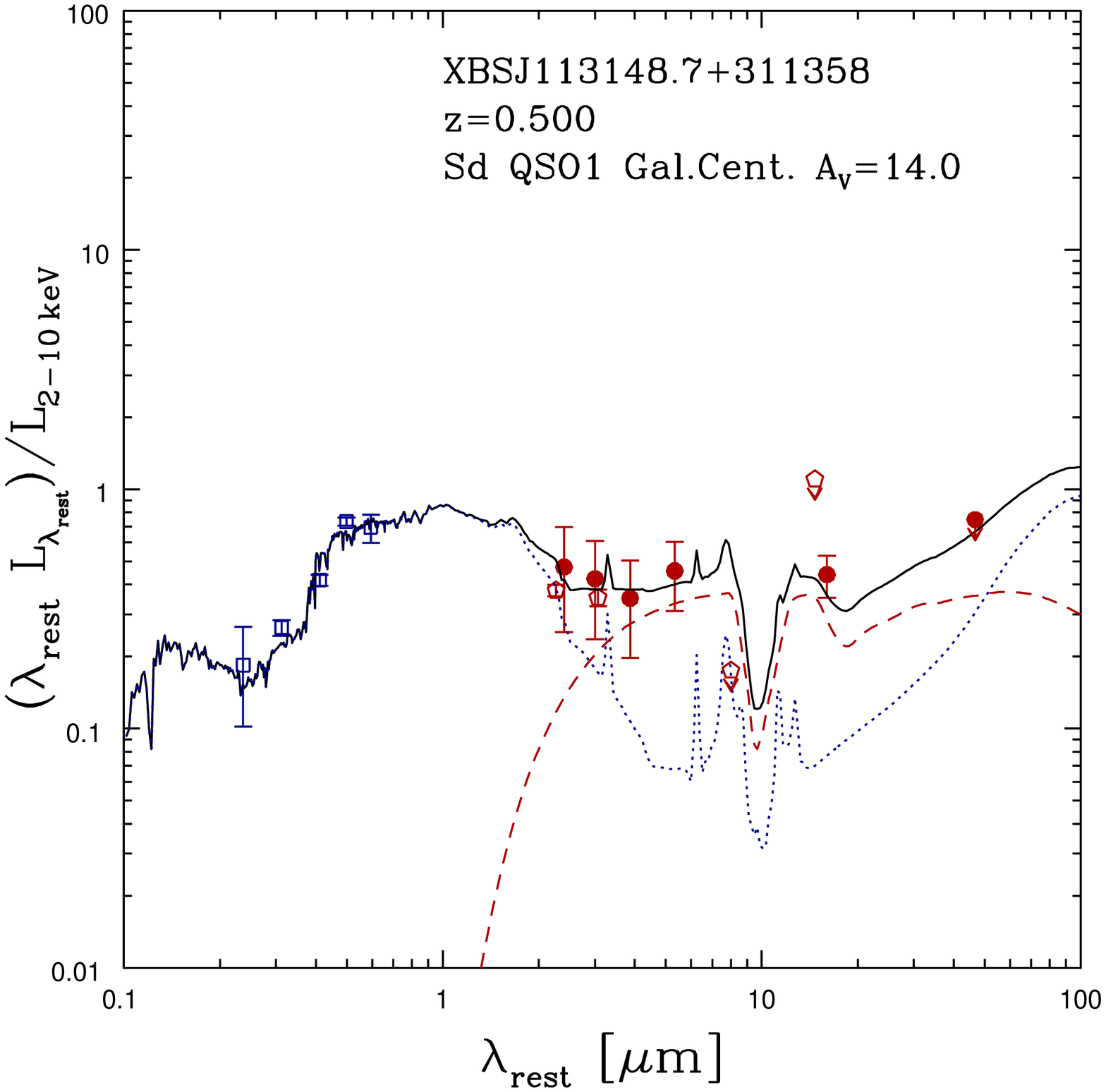}
\caption{ (cont.) {\it Top panel:} \xmm\ data and residuals (pn $+$ combined MOS; red and black crosses and lines, respectively), from \citet{corral11}. 
 {\it Central panel:} optical spectrum (from AXIS); the strongest emission lines are marked.
 {\it Bottom panel:} rest-frame SED fits: the data (red filled circles, \spitzer; red open pentagons, WISE; blue open
 squares, \sdss), 
 plotted as luminosity ($\lambda
 \pedix{L}{$\lambda$}$, normalized to the X-ray luminosity as in Table~\ref{tab:xbs}) vs. the rest-frame wavelength, 
 are superposed on the corresponding best-fitting template SEDs (blue dotted line, host galaxy; red dashed line, AGN; black
 solid line, total).
  In addition to the name (first row) and the redshift (second row) of the source, in the last row of the legend we summarize the main parameters of the modelling:
 the morphological type of the host, the QSO template, the adopted extinction curve (see Fig.~\ref{fig:templ}), and the dust extinction.
 See Sect.~\ref{sect:sed} for details.}%
\end{figure*}

\renewcommand{\thefigure}{\arabic{numfigsed}}

\begin{figure*}
 \centering
 \includegraphics[angle=0,height=6.5cm]{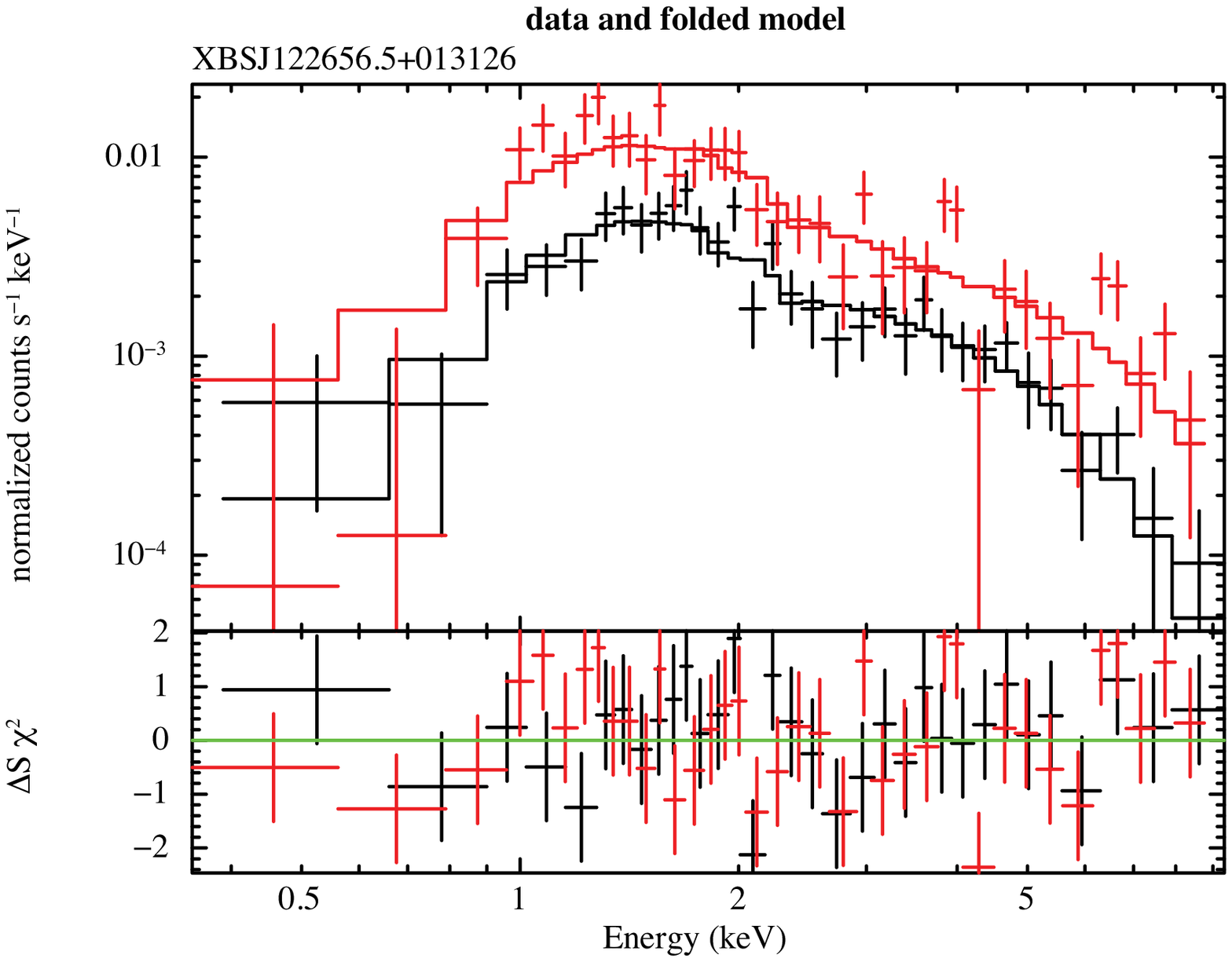}\\
 \vskip -0.5truecm
 \includegraphics[angle=-90,width=8.6cm]{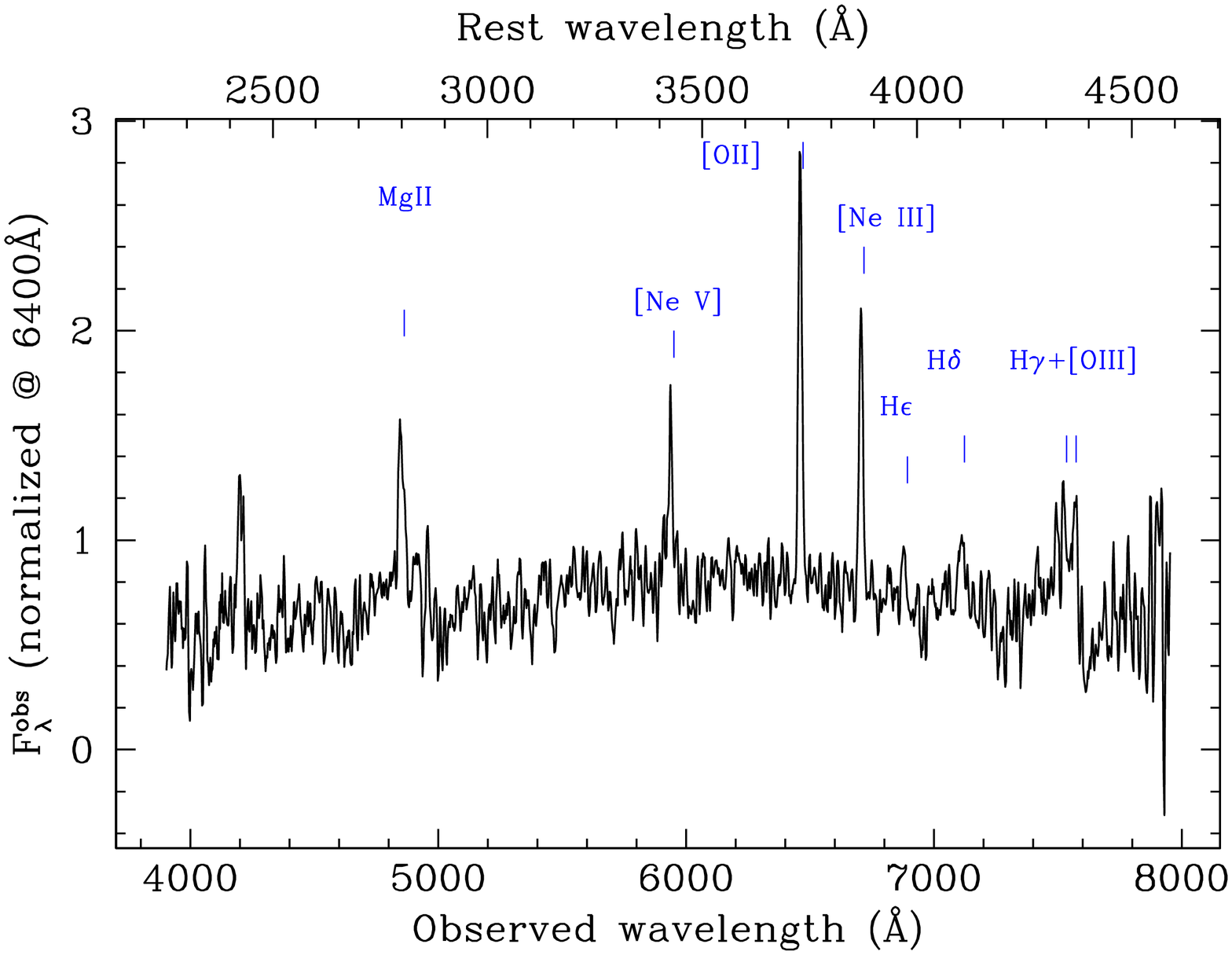}
 \vskip -0.5truecm
 \includegraphics[height=9cm]{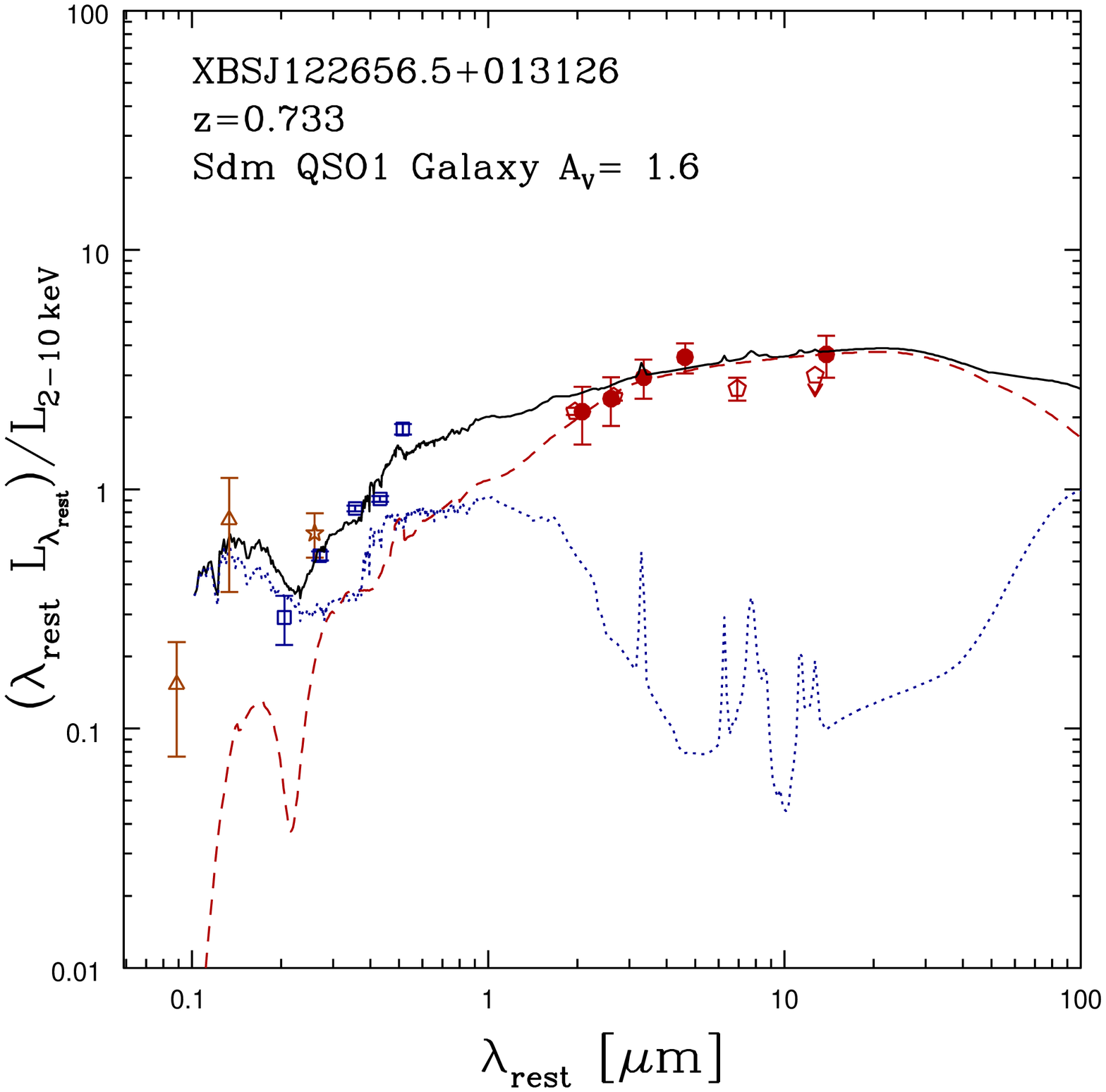}
\caption{ (cont.) {\it Top panel:} \xmm\ data and residuals (pn $+$ combined MOS; red and black crosses and lines, respectively), from \citet{corral11}. 
 {\it Central panel:} optical spectrum; the strongest emission lines are marked.
 {\it Bottom panel:} rest-frame SED fits: the data (red filled circles, \spitzer; red open pentagons, WISE; blue open
 squares, \sdss; brown open triangles, \galex\ NUV and FUV fluxes), 
 plotted as luminosity ($\lambda
 \pedix{L}{$\lambda$}$, normalized to the X-ray luminosity as in Table~\ref{tab:xbs}) vs. the rest-frame wavelength, 
 are superposed on the corresponding best-fitting template SEDs (blue dotted line, host galaxy; red dashed line, AGN; black
 solid line, total).
  In addition to the name (first row) and the redshift (second row) of the source, in the last row of the legend we summarize the main parameters of the modelling:
 the morphological type of the host, the QSO template, the adopted extinction curve (see Fig.~\ref{fig:templ}), and the dust extinction.
 See Sect.~\ref{sect:sed} for details.}%
\end{figure*}

\renewcommand{\thefigure}{\arabic{numfigsed}}

\begin{figure*}
 \centering
 \includegraphics[angle=0,height=6.5cm]{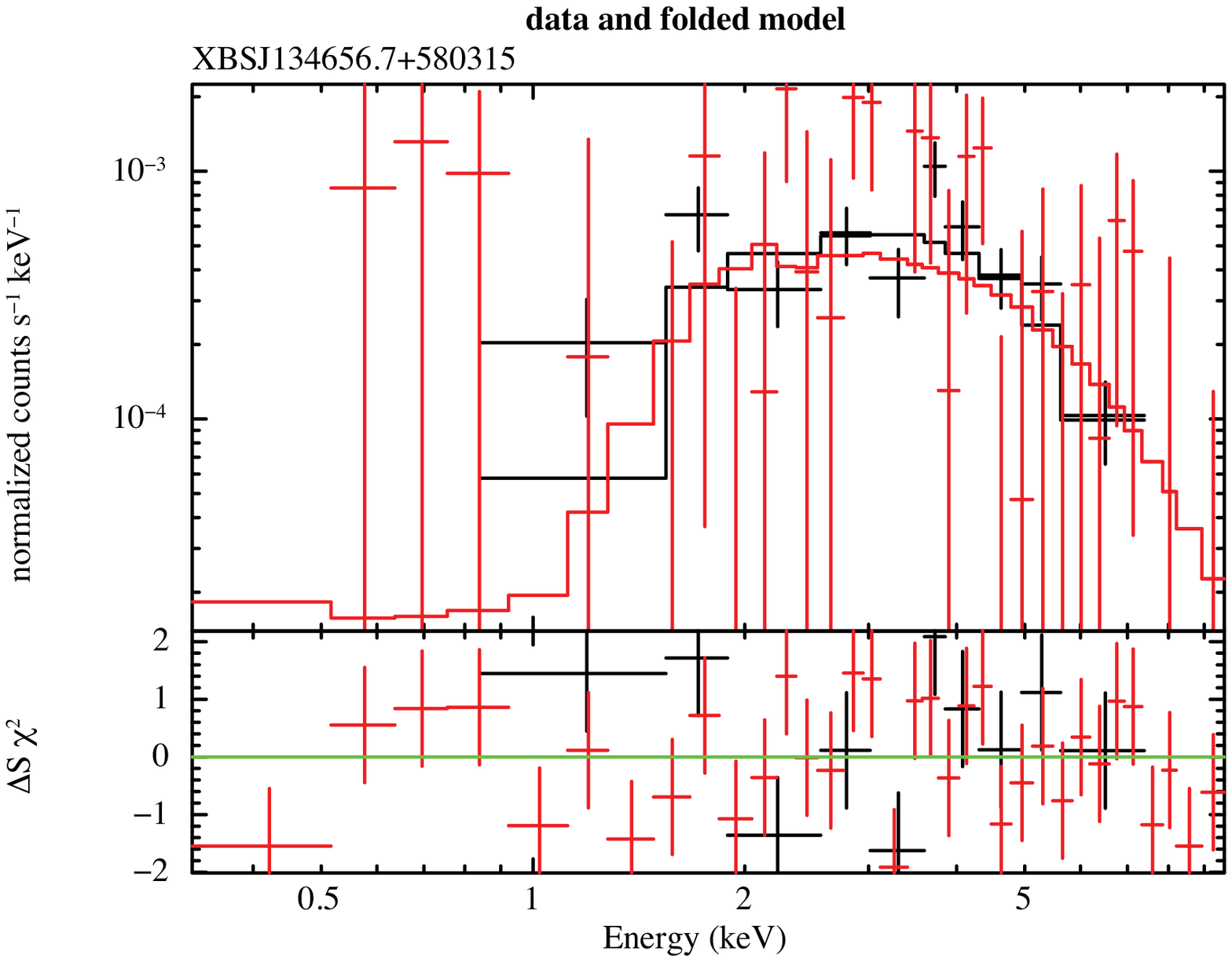}\\
 \vskip -0.5truecm
 \includegraphics[angle=-90,width=8.6cm]{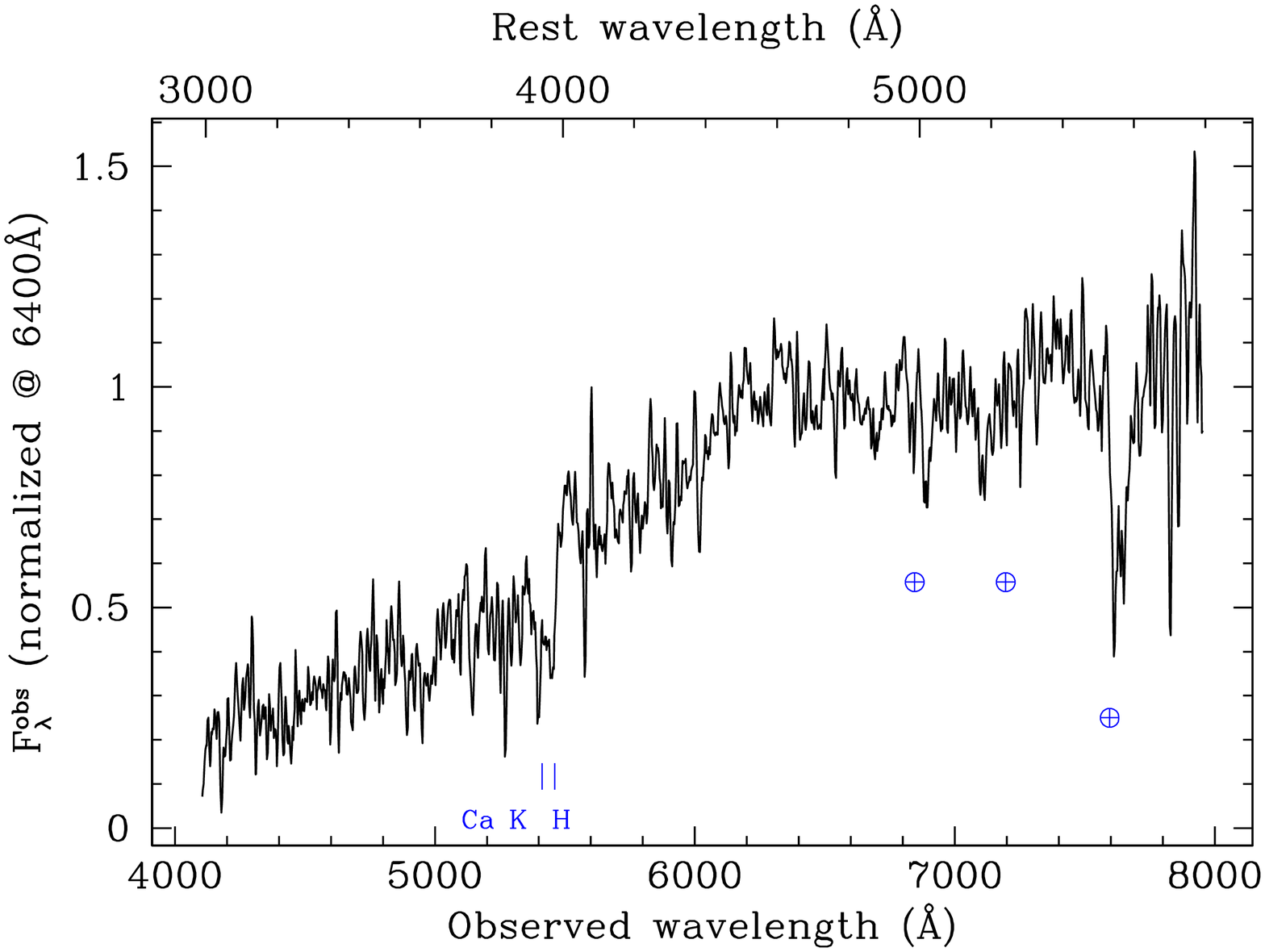}
 \vskip -0.5truecm
 \includegraphics[height=9cm]{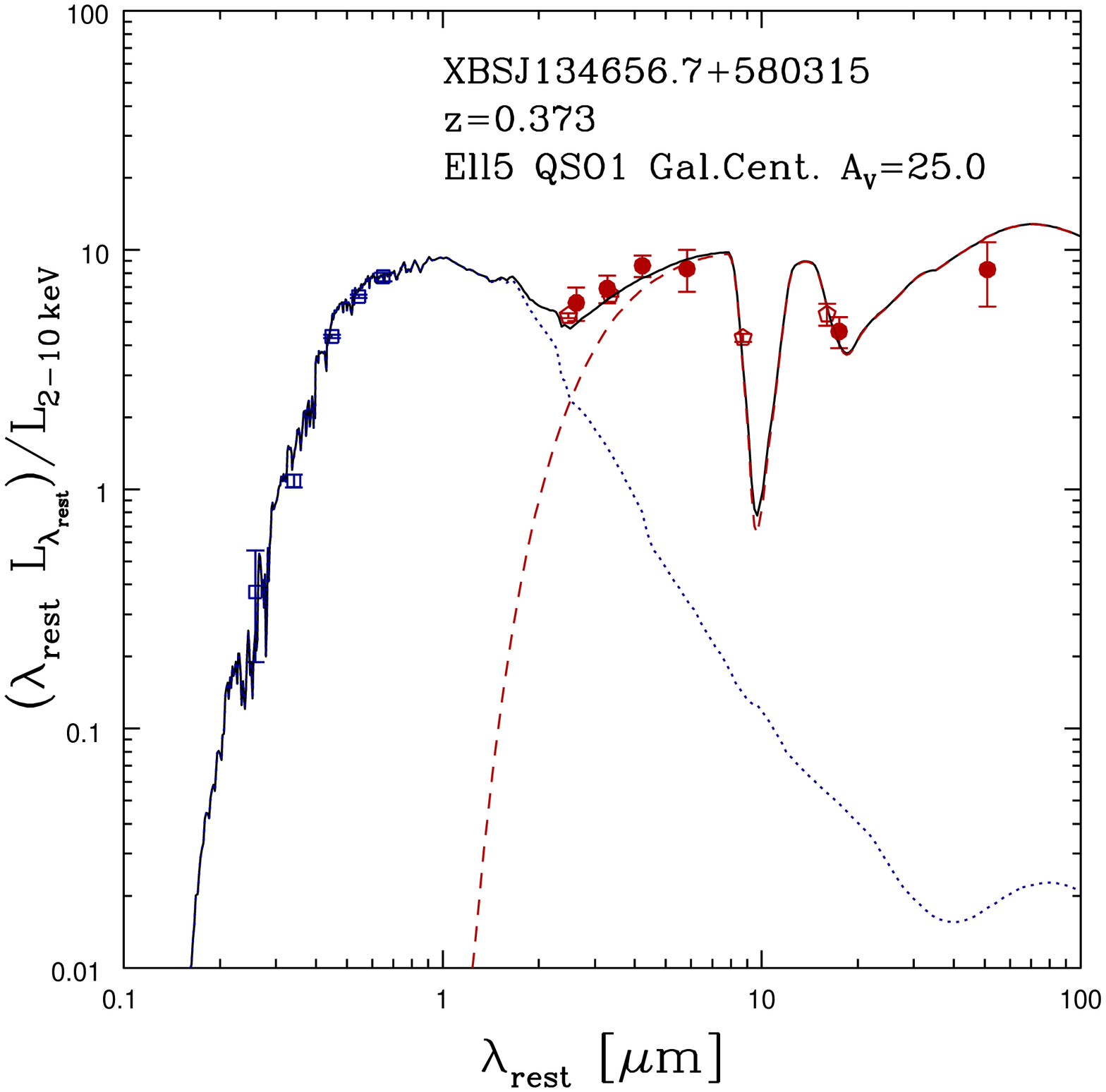}
\caption{ (cont.) {\it Top panel:} \xmm\ data and residuals (pn $+$ combined MOS; red and black crosses and lines, respectively), from \citet{corral11}. 
 {\it Central panel:} optical spectrum (from Caccianiga et al. 2007); the strongest emission lines are marked.
 {\it Bottom panel:} rest-frame SED fits: the data (red filled circles, \spitzer; red open pentagons, WISE; blue open
 squares, \sdss), 
 plotted as luminosity ($\lambda
 \pedix{L}{$\lambda$}$, normalized to the X-ray luminosity as in Table~\ref{tab:xbs}) vs. the rest-frame wavelength, 
 are superposed on the corresponding best-fitting template SEDs (blue dotted line, host galaxy; red dashed line, AGN; black
 solid line, total).
  In addition to the name (first row) and the redshift (second row) of the source, in the last row of the legend we summarize the main parameters of the modelling:
 the morphological type of the host, the QSO template, the adopted extinction curve (see Fig.~\ref{fig:templ}), and the dust extinction.
 See Sect.~\ref{sect:sed} for details.}%
\end{figure*}

\renewcommand{\thefigure}{\arabic{numfigsed}}

\begin{figure*}
 \centering
 \includegraphics[angle=0,height=6.5cm]{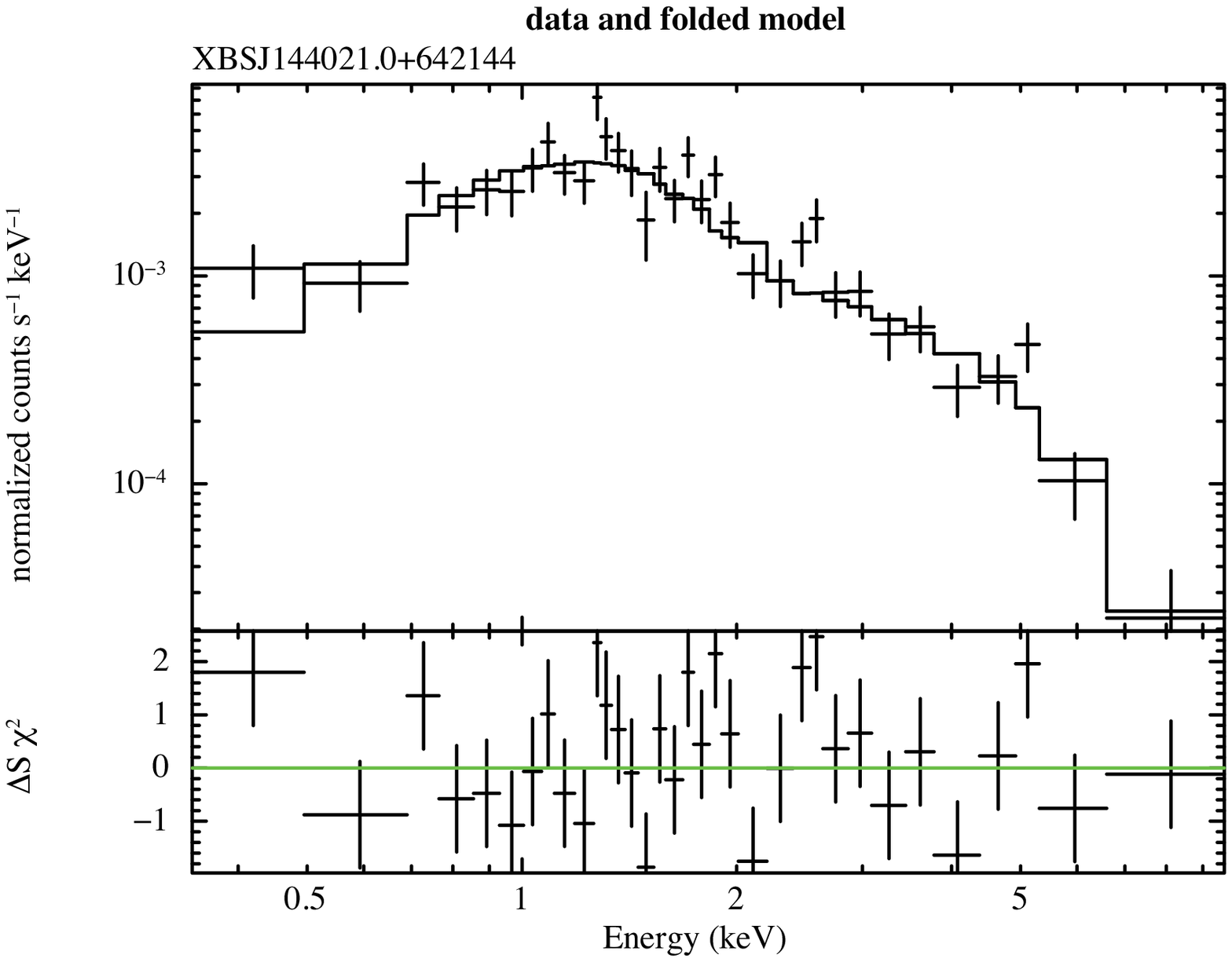}\\
 \vskip -0.5truecm
 \includegraphics[angle=-90,width=8.6cm]{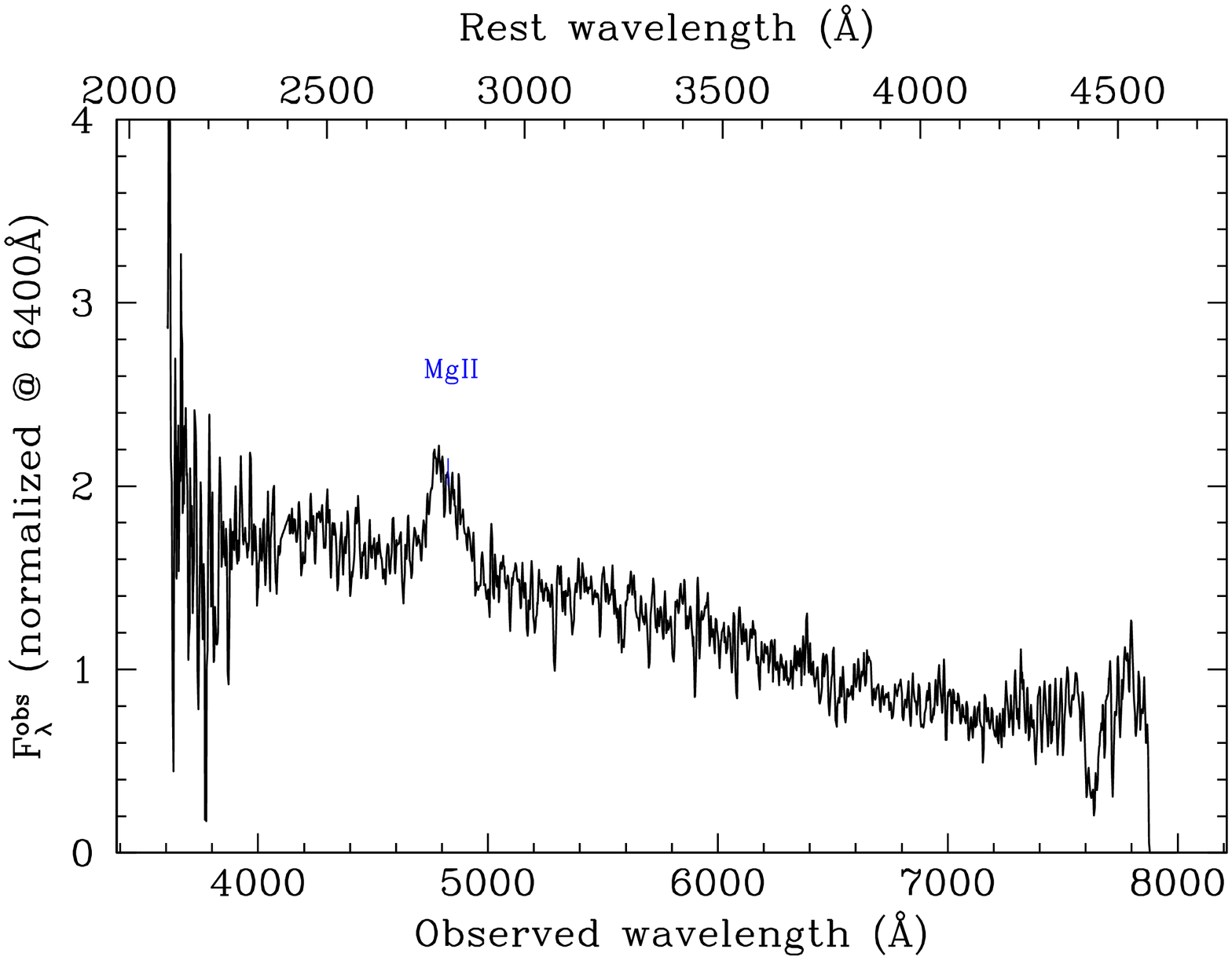}
 \vskip -0.5truecm
 \includegraphics[height=9cm]{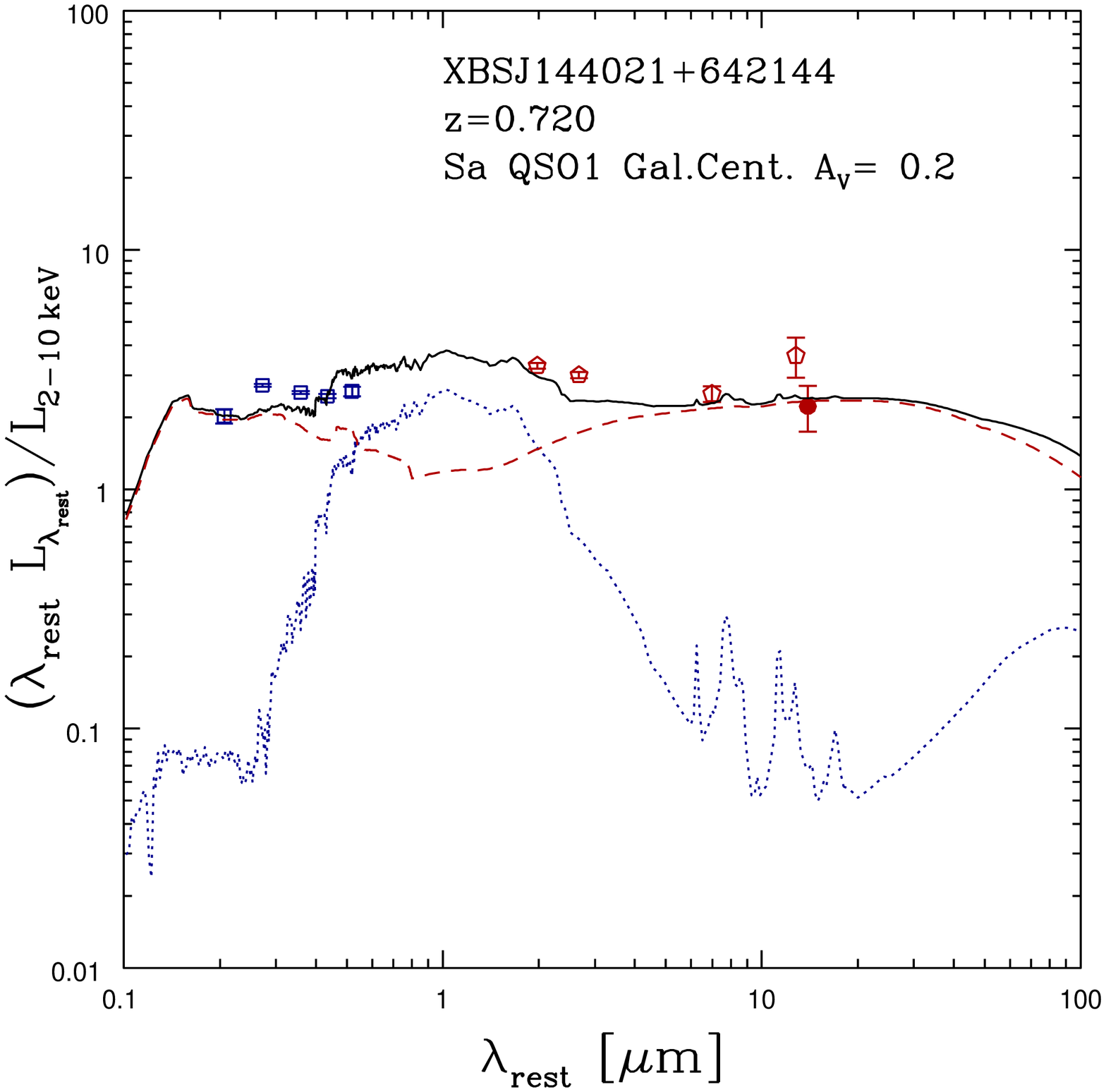}
\caption{ (cont.) {\it Top panel:} \xmm\ data and residuals (pn $+$ combined MOS; red and black crosses and lines, respectively), from \citet{corral11}. 
 {\it Central panel:} optical spectrum; the strongest emission lines are marked.
 {\it Bottom panel:} rest-frame SED fits: the data (red filled circle, \spitzer; red open pentagons, WISE; blue open
 squares, \sdss), 
 plotted as luminosity ($\lambda
 \pedix{L}{$\lambda$}$, normalized to the X-ray luminosity as in Table~\ref{tab:xbs}) vs. the rest-frame wavelength, 
 are superposed on the corresponding best-fitting template SEDs (blue dotted line, host galaxy; red dashed line, AGN; black
 solid line, total).
  In addition to the name (first row) and the redshift (second row) of the source, in the last row of the legend we summarize the main parameters of the modelling:
 the morphological type of the host, the QSO template, the adopted extinction curve (see Fig.~\ref{fig:templ}), and the dust extinction.
 See Sect.~\ref{sect:sed} for details.}%
\end{figure*}

\renewcommand{\thefigure}{\arabic{numfigsed}}

\begin{figure*}
 \centering
 \includegraphics[angle=0,height=6.5cm]{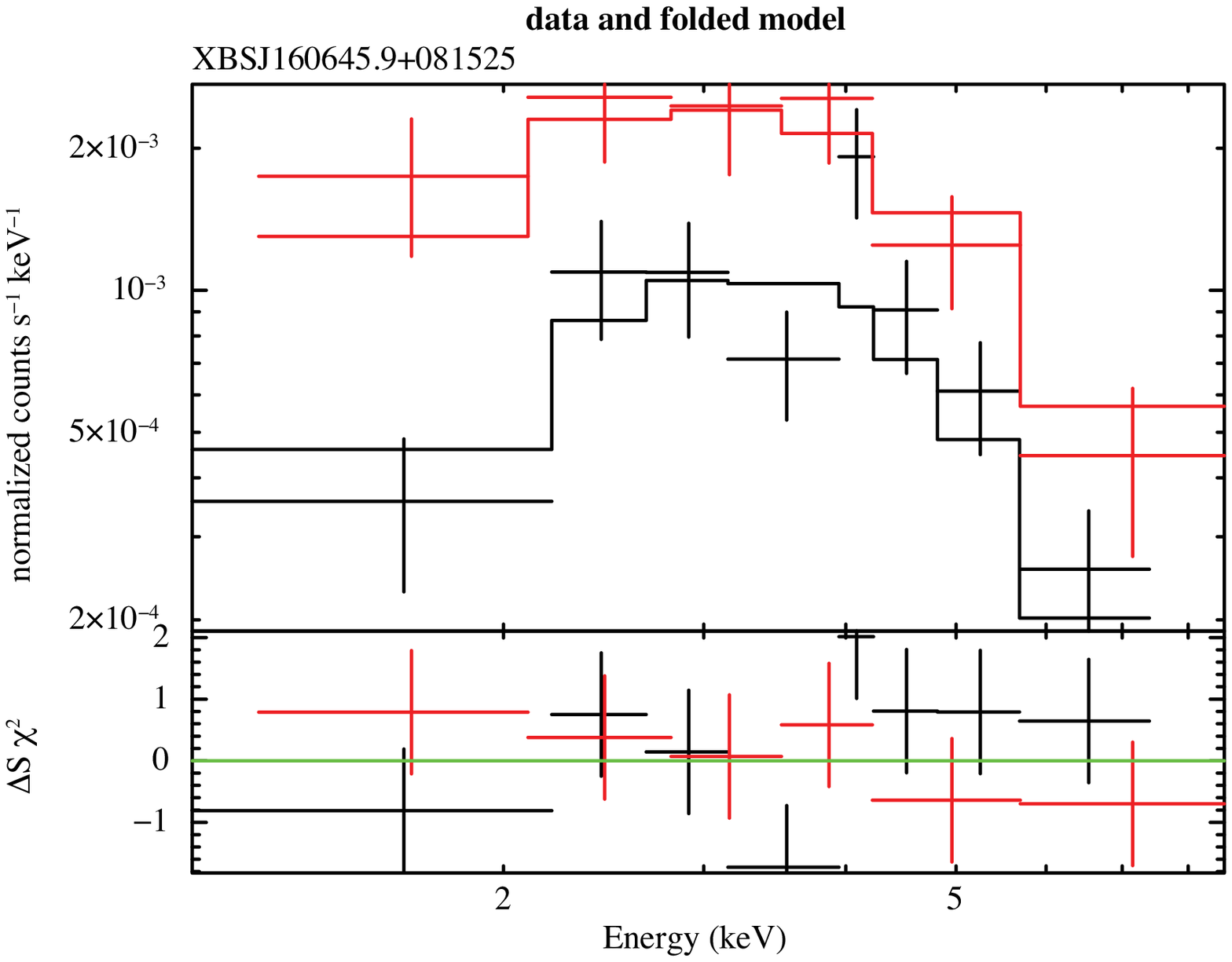}\\
 \vskip -0.5truecm
 \includegraphics[angle=-90,width=8.6cm]{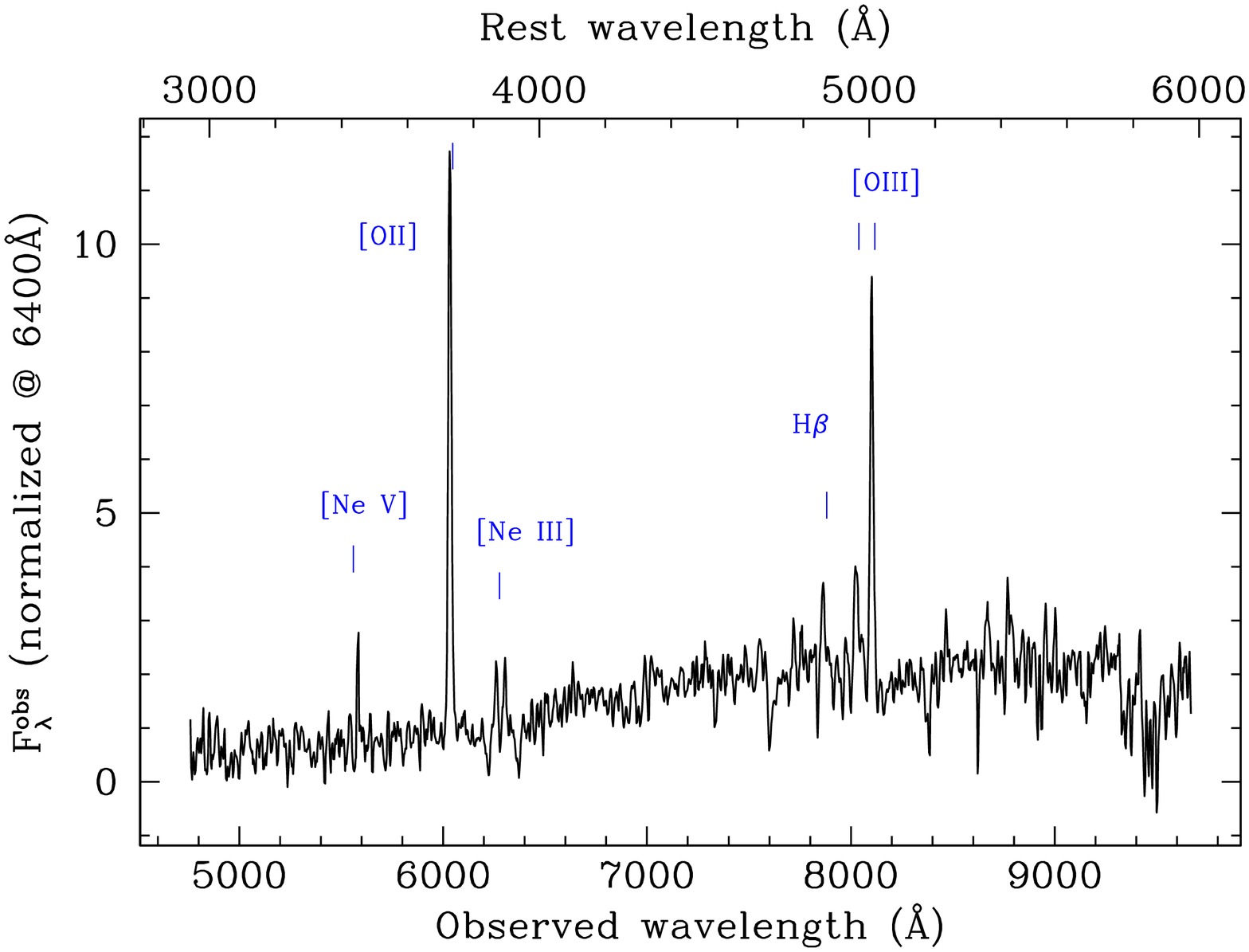}
 \vskip -0.5truecm
 \includegraphics[height=9cm]{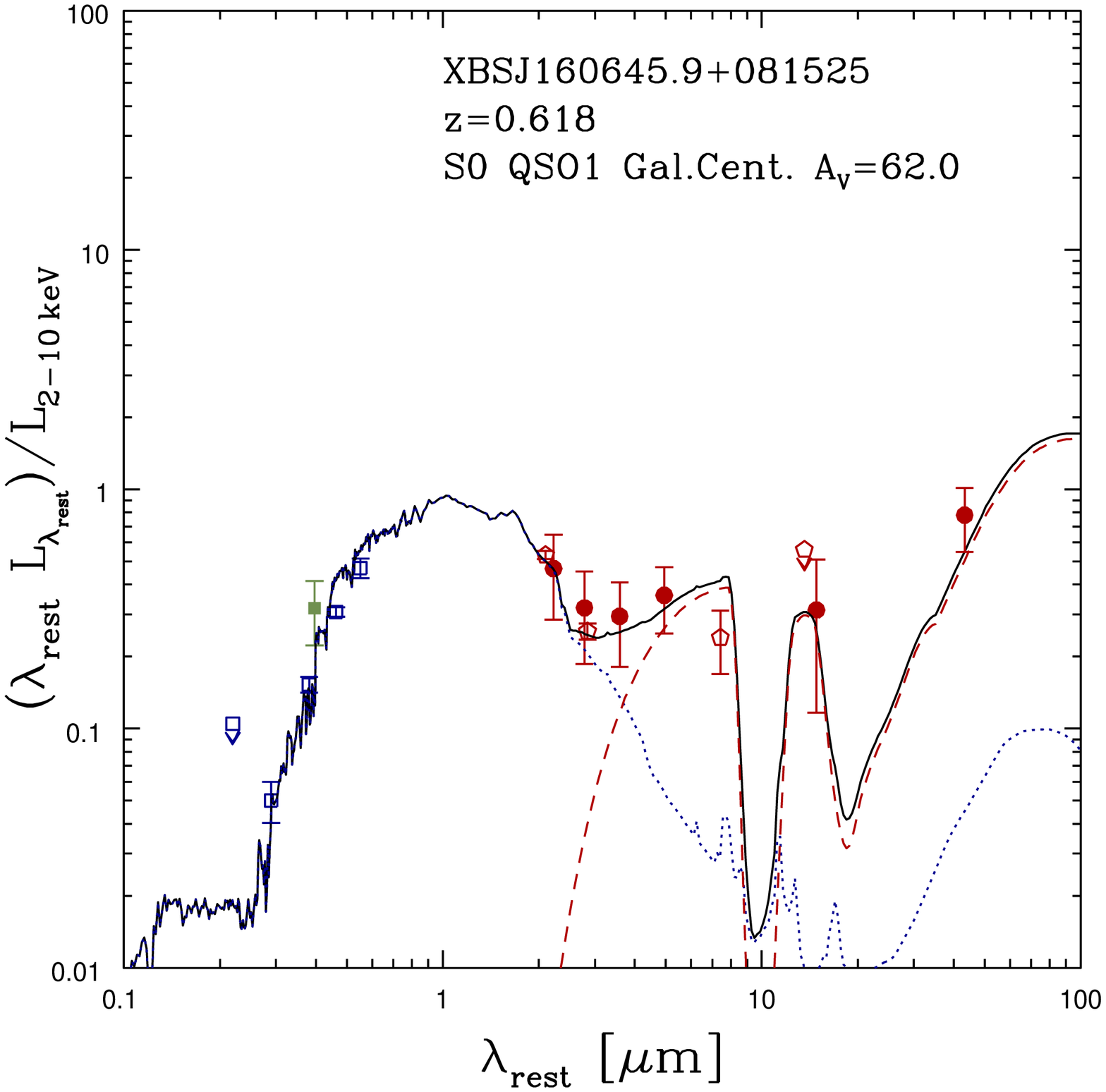}
\caption{ (cont.) {\it Top panel:} \xmm\ data and residuals (pn $+$ combined MOS; red and black crosses and lines, respectively), from \citet{corral11}. 
 {\it Central panel:} optical spectrum; the strongest emission lines are marked.
 {\it Bottom panel:} rest-frame SED fits: the data (red filled circles, \spitzer; red open pentagons, WISE;  green filled
 square, data in the $R$-band; blue open
 squares, \sdss), 
 plotted as luminosity ($\lambda
 \pedix{L}{$\lambda$}$, normalized to the X-ray luminosity as in Table~\ref{tab:xbs}) vs. the rest-frame wavelength, 
 are superposed on the corresponding best-fitting template SEDs (blue dotted line, host galaxy; red dashed line, AGN; black
 solid line, total).
  In addition to the name (first row) and the redshift (second row) of the source, in the last row of the legend we summarize the main parameters of the modelling:
 the morphological type of the host, the QSO template, the adopted extinction curve (see Fig.~\ref{fig:templ}), and the dust extinction.
 See Sect.~\ref{sect:sed} for details.}%
\end{figure*}

\renewcommand{\thefigure}{\arabic{numfigsed}}

\begin{figure*}
 \centering
 \includegraphics[angle=0,height=6.5cm]{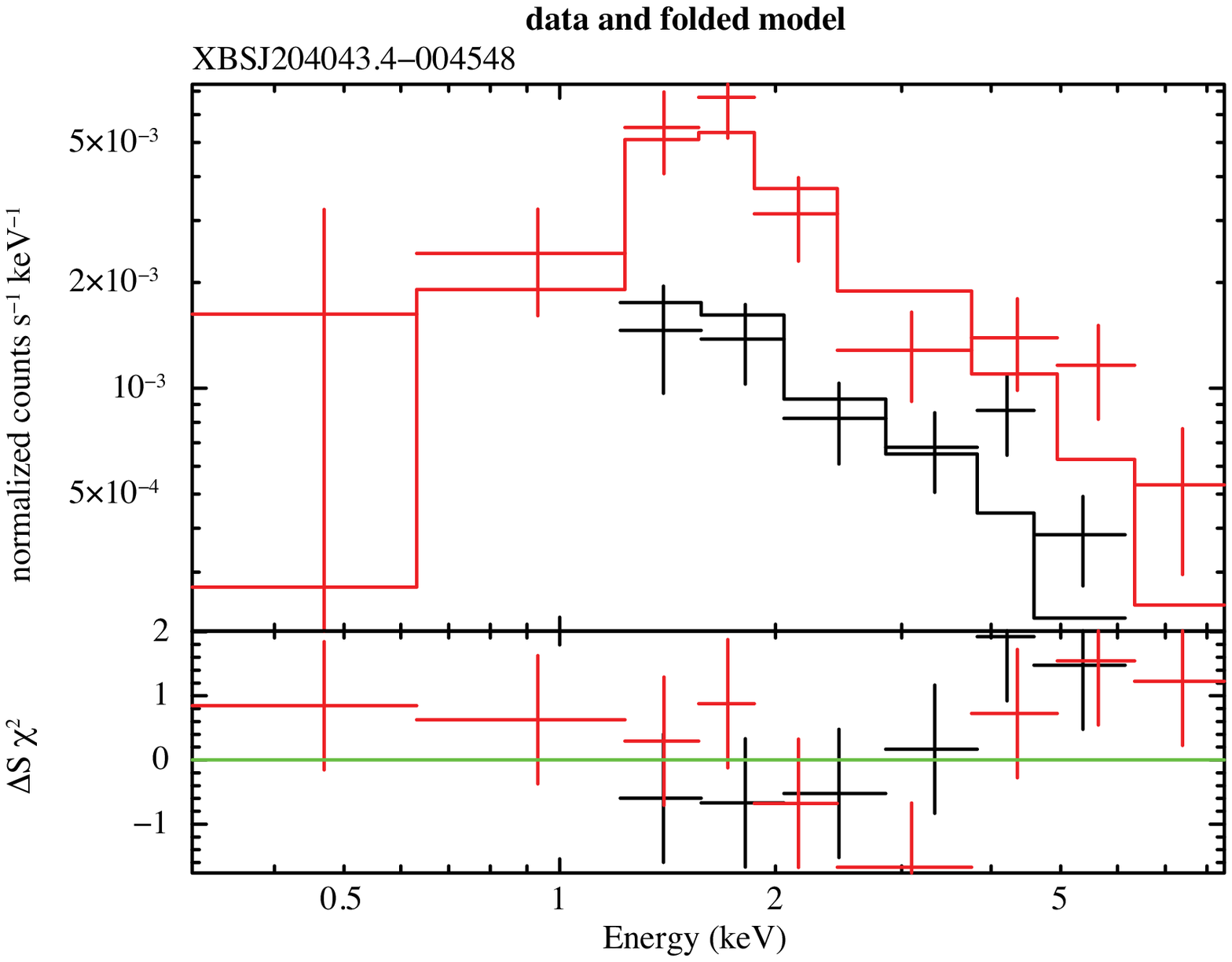}\\
 \vskip -0.5truecm
 \includegraphics[angle=-90,width=8.6cm]{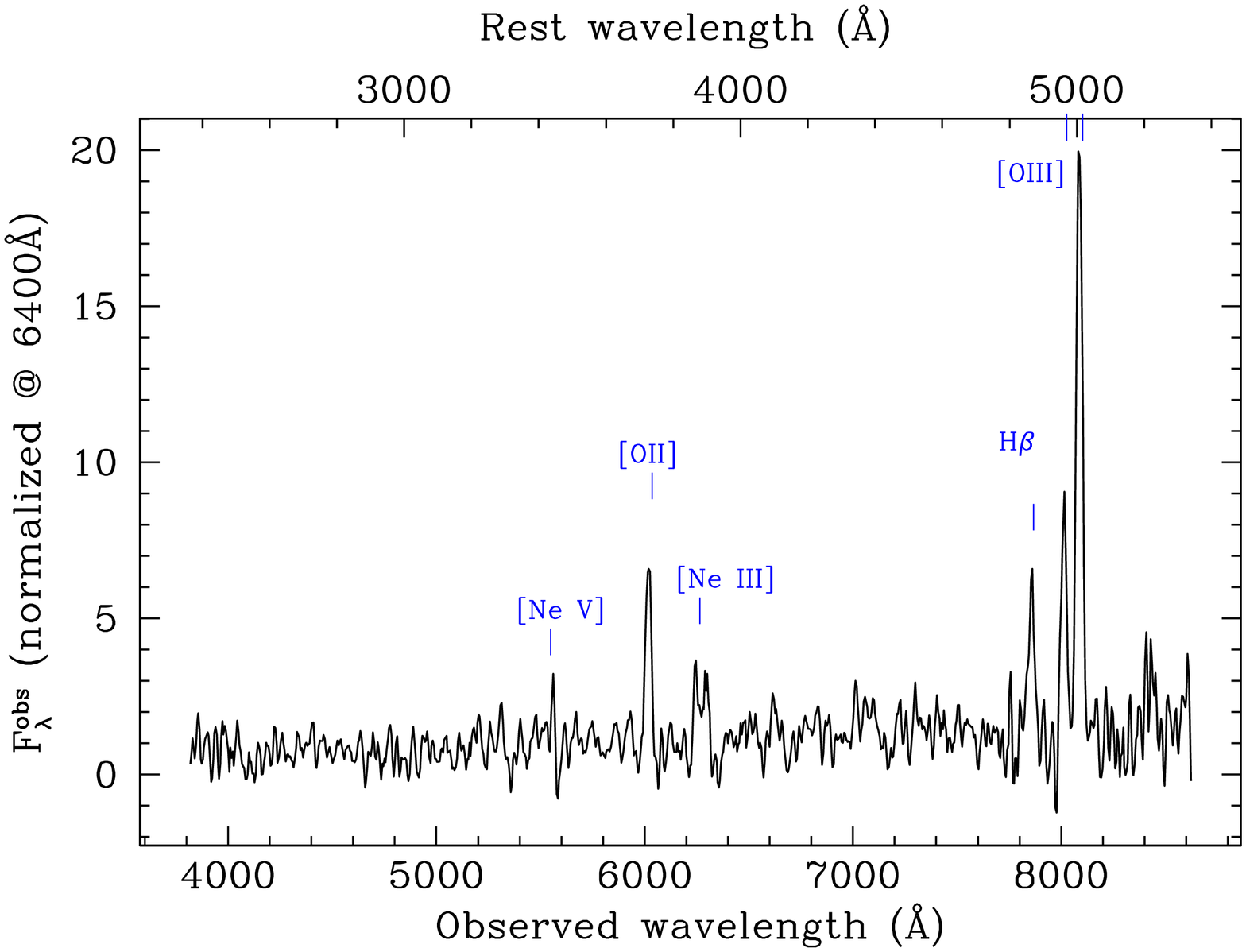}
 \vskip -0.5truecm
 \includegraphics[height=9cm]{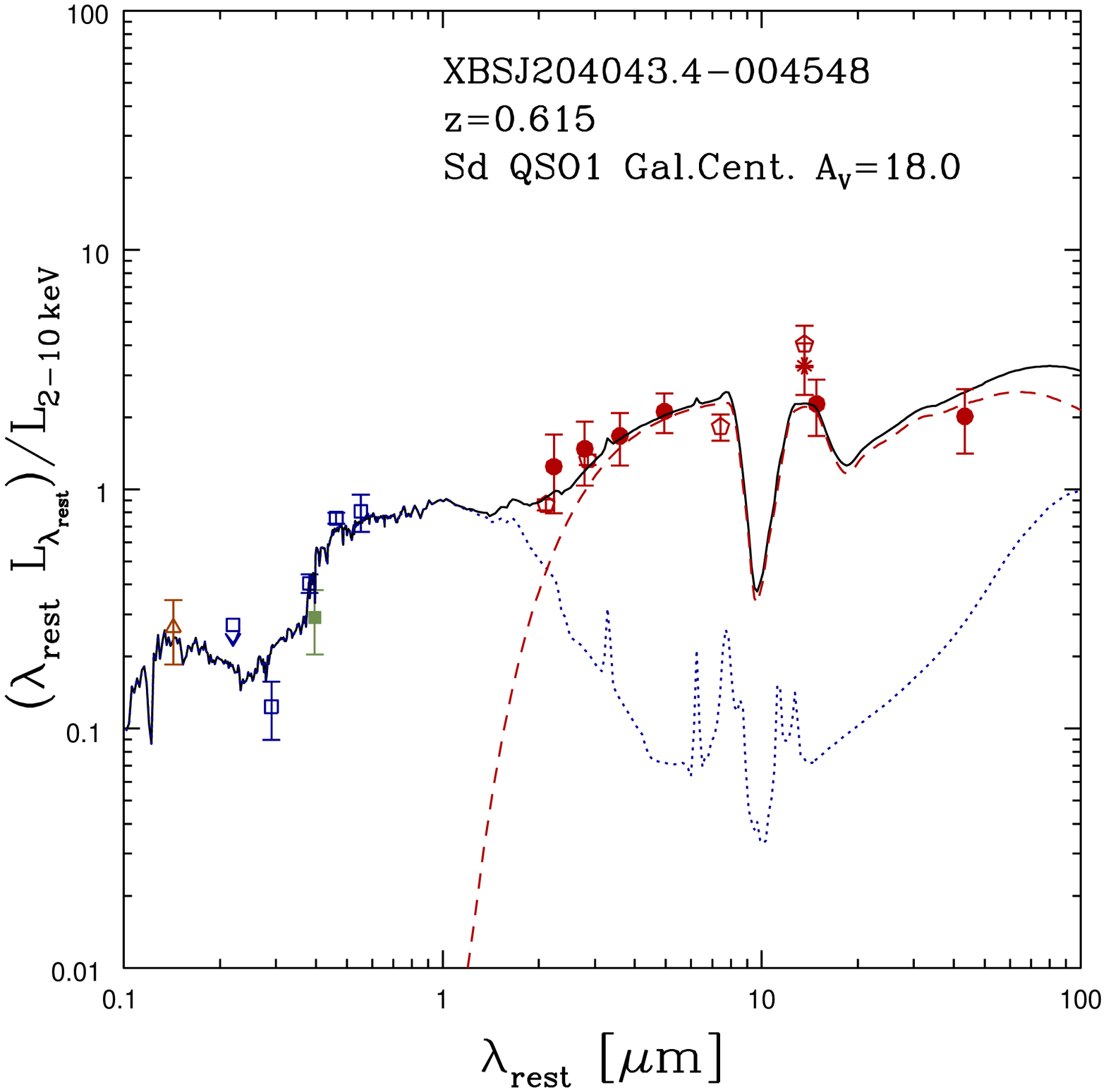}
\caption{ (cont.) {\it Top panel:} \xmm\ data and residuals (pn $+$ combined MOS; red and black crosses and lines, respectively), from \citet{corral11}. 
 {\it Central panel:} optical spectrum (from Caccianiga et al. 2004); the strongest emission lines are marked.
 {\it Bottom panel:} rest-frame SED fits: the data (red filled circles, \spitzer; red open pentagons, WISE; green filled
 square, data in the $R$-band; blue open
 squares, \sdss; brown open triangle, \galex\ NUV flux), 
 plotted as luminosity ($\lambda
 \pedix{L}{$\lambda$}$, normalized to the X-ray luminosity as in Table~\ref{tab:xbs}) vs. the rest-frame wavelength, 
 are superposed on the corresponding best-fitting template SEDs (blue dotted line, host galaxy; red dashed line, AGN; black
 solid line, total).
  In addition to the name (first row) and the redshift (second row) of the source, in the last row of the legend we summarize the main parameters of the modelling:
 the morphological type of the host, the QSO template, the adopted extinction curve (see Fig.~\ref{fig:templ}), and the dust extinction.
 See Sect.~\ref{sect:sed} for details.}%
\end{figure*}
\addtocounter{numfigsed}{1}


\begin{thebibliography}{99}

\bibitem[\protect\citeauthoryear{Abazajian et al.}{2009}]{abazajian09} 
Abazajian K.~N., et al., 2009, ApJS, 182, 543 

\bibitem[\protect\citeauthoryear{Antonucci}{1993}]{antonucci93} 
Antonucci R., 1993, ARA\&A, 31, 473 

\bibitem[\protect\citeauthoryear{Arnaud}{1996}]{xspec} 
Arnaud K.~A., 1996, ASPC,101, 17

\bibitem[\protect\citeauthoryear{Bianchi, Maiolino, \& Risaliti}{2012}]{bianchi12} 
Bianchi S., Maiolino R., Risaliti G., 2012, AdAst, 2012,  

\bibitem[\protect\citeauthoryear{Bohlin, Savage, \& Drake}{1978}]{bohlin78}
Bohlin R.~C., Savage B.~D., Drake J.~F., 1978, ApJ, 224, 132

\bibitem[\protect\citeauthoryear{Chiar \& Tielens}{2006}]{chiar06} 
Chiar J.~E., Tielens A.~G.~G.~M., 2006, ApJ, 637, 774

\bibitem[\protect\citeauthoryear{Caccianiga et al.}{2007}]{caccianiga07} 
Caccianiga A., Severgnini P., Della Ceca R., Maccacaro T., Carrera F.~J., Page M.~J., 2007, A\&A, 470, 557 

\bibitem[\protect\citeauthoryear{Caccianiga et al.}{2008}]{caccianiga08} 
Caccianiga A., et al., 2008, A\&A, 477, 735 

\bibitem[\protect\citeauthoryear{Caccianiga et al.}{2013}]{caccianiga13} 
Caccianiga A., Fanali R., Severgnini P., Della Ceca R., Marchese E., Mateos S., 2013, A\&A, 549, A119 

\bibitem[\protect\citeauthoryear{Cattaneo et al.}{2009}]{cattaneo09} 
Cattaneo A., et al., 2009, Natur, 460, 213 

\bibitem[\protect\citeauthoryear{Corral et al.}{2011}]{corral11} 
Corral A., Della Ceca R., Caccianiga A., Severgnini P., Brunner H., Carrera F.~J., Page M.~J., Schwope A.~D., 2011, A\&A, 530, A42 

\bibitem[\protect\citeauthoryear{Dadina}{2008}]{dadina08} 
Dadina M., 2008, A\&A, 485, 417 

\bibitem[\protect\citeauthoryear{Davies et al.}{2007}]{davies07} 
Davies R.~I., M{\"u}ller S{\'a}nchez F., Genzel R., Tacconi L.~J., Hicks E.~K.~S., Friedrich S., Sternberg A., 2007, ApJ, 671, 1388 

\bibitem[\protect\citeauthoryear{Della Ceca et al.}{2008}]{rdc08} 
Della Ceca R., et al., 2008, A\&A, 487, 119 

\bibitem[\protect\citeauthoryear{Della Ceca et al.}{2004}]{rdc04} 
Della Ceca R., et al., 2004, A\&A, 428, 383 

\bibitem[\protect\citeauthoryear{de Rosa et al.}{2008}]{derosa08} 
de Rosa A., Bassani L., Ubertini P., Panessa F., Malizia A., Dean A.~J., Walter R., 2008, A\&A, 483, 749 

\bibitem[\protect\citeauthoryear{de Rosa et al.}{2012}]{derosa12} 
de Rosa A., et al., 2012, MNRAS, 420, 2087 

\bibitem[\protect\citeauthoryear{Dickey \& Lockman}{1990}]{nh} 
Dickey J.~M., Lockman F.~J., 1990, ARA\&A, 28, 215 

\bibitem[\protect\citeauthoryear{Elvis et al.}{1994}]{elvis94} 
Elvis M., et al., 1994, ApJS, 95, 1 

\bibitem[\protect\citeauthoryear{Fabian, Celotti, \& Erlund}{2006}]{fabian06} 
Fabian A.~C., Celotti A., Erlund M.~C., 2006, MNRAS, 373, L16 

\bibitem[\protect\citeauthoryear{Fabian, Vasudevan, \& Gandhi}{2008}]{fabian08} 
Fabian A.~C., Vasudevan R.~V., Gandhi P., 2008, MNRAS, 385, L43 

\bibitem[\protect\citeauthoryear{Fabian et al.}{2009}]{fabian09} 
Fabian A.~C., Vasudevan R.~V., Mushotzky R.~F., Winter L.~M., Reynolds 
C.~S., 2009, MNRAS, 394, L89 

\bibitem[\protect\citeauthoryear{Fanali et al.}{2013}]{fanali13} 
Fanali R., Caccianiga A., Severgnini P., Della Ceca R., Marchese E., 
Carrera F.~J., Corral A., Mateos S., 2013, MNRAS, 433, 648 

\bibitem[\protect\citeauthoryear{Ferrarese \& Merritt}{2000}]{ferrarese00} 
Ferrarese L., Merritt D., 2000, ApJ, 539, L9 

\bibitem[\protect\citeauthoryear{Ferrarese et al.}{2006}]{ferrarese06} 
Ferrarese L., et al., 2006, ApJS, 164, 334 

\bibitem[\protect\citeauthoryear{Fiore et al.}{2003}]{fiore03} 
Fiore F., et al., 2003, A\&A, 409, 79 

\bibitem[\protect\citeauthoryear{Francis et al.}{1991}]{francis91} 
Francis P.~J., Hewett P.~C., Foltz C.~B., 
Chaffee F.~H., Weymann R.~J., Morris S.~L., 1991, ApJ, 373, 465 

\bibitem[\protect\citeauthoryear{Graham}{2007}]{graham07} 
Graham A.~W., 2007, MNRAS, 379, 711 

\bibitem[\protect\citeauthoryear{Graham \& Worley}{2008}]{graham08} 
Graham A.~W., Worley C.~C., 2008, MNRAS, 388, 1708 

\bibitem[\protect\citeauthoryear{Granato et al.}{2004}]{granato04} 
Granato G.~L., De Zotti G., Silva L., 
Bressan A., Danese L., 2004, ApJ, 600, 580 

\bibitem[\protect\citeauthoryear{Heckman \& Best}{2014}]{heckman14} 
Heckman T., Best P., 2014, arXiv, arXiv:1403.4620 

\bibitem[\protect\citeauthoryear{Heymann \& Siebenmorgen}{2012}]{heymann12} 
Heymann F., Siebenmorgen R., 2012, ApJ, 751, 27 

\bibitem[\protect\citeauthoryear{Hoenig et al.}{2006}]{hoenig06} 
Hoenig S.~F., Beckert T., Ohnaka K., Weigelt G., 2006, A\&A, 452, 459 

\bibitem[\protect\citeauthoryear{Hoenig}{2013}]{hoenig13} 
Hoenig S.~F., 2013, arXiv, arXiv:1301.1349 

\bibitem[\protect\citeauthoryear{Hopkins et al.}{2006}]{hopkins06} 
Hopkins P.~F., Hernquist L., Cox T.~J., Di Matteo T., Robertson B., Springel V., 2006, ApJS, 163, 1 

\bibitem[\protect\citeauthoryear{Kawaguchi \& Mori}{2010}]{kawaguchi10} 
Kawaguchi T., Mori M., 2010, ApJ, 724, L183 

\bibitem[\protect\citeauthoryear{Kormendy \& Richstone}{1995}]{kormendy95} 
Kormendy J., Richstone D., 1995, ARA\&A, 33, 581 

\bibitem[\protect\citeauthoryear{Kormendy \& Gebhardt}{2001}]{kormendy01} 
Kormendy J., Gebhardt K., 2001, AIPC, 586, 363 

\bibitem[\protect\citeauthoryear{Kormendy \& Ho}{2013}]{kormendy13} 
Kormendy J., Ho L.~C., 2013, ARA\&A, 51, 511 

\bibitem[\protect\citeauthoryear{Krolik \& Begelman}{1988}]{krolik88} 
Krolik J.~H., Begelman M.~C., 1988, ApJ, 329, 702 

\bibitem[\protect\citeauthoryear{Krolik}{1998}]{krolik98} 
Krolik J.~H., 1998,  ``Active Galactic Nuclei: From the Central Black Hole to the Galactic Environment'', 
Princeton: Princeton University Press

\bibitem[\protect\citeauthoryear{Laor \& Netzer}{1989}]{laor89} 
Laor A., Netzer H., 1989, MNRAS, 238, 897 

\bibitem[\protect\citeauthoryear{Lapi et al.}{2014}]{lapi14} 
Lapi A., Raimundo S., Aversa R., Cai Z.-Y., Negrello M., Celotti A., De 
Zotti G., Danese L., 2014, ApJ, 782, 69 

\bibitem[\protect\citeauthoryear{Lonsdale et al.}{2003}]{lonsdale03} 
Lonsdale C.~J., et al., 2003, PASP, 115, 897 

\bibitem[\protect\citeauthoryear{Lusso et al.}{2011}]{lusso11} 
Lusso E., et al., 2011, A\&A, 534, A110 

\bibitem[\protect\citeauthoryear{Lynden-Bell}{1969}]{lyndenbell69} 
Lynden-Bell D., 1969, Natur, 223, 690 

\bibitem[\protect\citeauthoryear{McLure \& Dunlop}{2002}]{mclure02} 
McLure R.~J., Dunlop J.~S., 2002, MNRAS, 331, 795 

\bibitem[\protect\citeauthoryear{Maiolino et al.}{2001}]{maiolino01} 
Maiolino R., Marconi A., Salvati M., Risaliti G., Severgnini P., Oliva E., La Franca F., Vanzi L., 2001, A\&A, 365, 28 

\bibitem[\protect\citeauthoryear{Makovoz \& Marleau}{2005}]{makovoz05} 
Makovoz D., Marleau F.~R., 2005, PASP, 117, 1113 

\bibitem[\protect\citeauthoryear{Malizia et al.}{2014}]{malizia14} 
Malizia A., Molina M., Bassani L., Stephen J.~B., Bazzano A., Ubertini P., Bird A.~J., 2014, ApJ, 782, L25 

\bibitem[\protect\citeauthoryear{Marchese et al.}{2012}]{marchese12} 
Marchese E., Della Ceca R., Caccianiga A., Severgnini P., Corral A., Fanali R., 2012, A\&A, 539, A48 

\bibitem[\protect\citeauthoryear{Marconi \& Hunt}{2003}]{marconi03} 
Marconi A., Hunt L.~K., 2003, ApJ, 589, L21 

\bibitem[\protect\citeauthoryear{Mason et al.}{2001}]{om} 
Mason K.~O., et al., 2001, A\&A, 365, L36

\bibitem[\protect\citeauthoryear{Markowitz, Krumpe, \& Nikutta}{2014}]{markowitz14} 
Markowitz A.~G., Krumpe M., Nikutta R., 2014, MNRAS, 439, 1403 

\bibitem[\protect\citeauthoryear{Mateos et al.}{2010}]{mateos10} 
Mateos S., et al., 2010, A\&A, 510, A35 

\bibitem[\protect\citeauthoryear{Mateos et al.}{2012}]{mateos12} 
Mateos S., et al., 2012, MNRAS, 426, 3271 

\bibitem[\protect\citeauthoryear{Mateos et al.}{2013}]{mateos13} 
Mateos S., Alonso-Herrero A., Carrera F.~J., Blain A., Severgnini P., 
Caccianiga A., Ruiz A., 2013, MNRAS, 434, 941 

\bibitem[\protect\citeauthoryear{Merloni et al.}{2014}]
{merloni14} Merloni A., et al., 2014, MNRAS, 437, 3550 

\bibitem[\protect\citeauthoryear{Mignoli et al.}{2013}]{mignoli13} 
Mignoli M., et al., 2013, A\&A, 556, A29 

\bibitem[\protect\citeauthoryear{Nandra et al.}{2007}]{pexmon} 
Nandra K., O'Neill P.~M., George I.~M., Reeves J.~N., 2007, MNRAS, 382, 194 

\bibitem[\protect\citeauthoryear{Nenkova, Ivezi{\'c}, \& Elitzur}{2002}]{nenkova02} 
Nenkova M., Ivezi{\'c} {\v Z}., Elitzur M., 2002, ApJ, 570, L9 

\bibitem[\protect\citeauthoryear{Nenkova et al.}{2008a}]{nenkova08a} 
Nenkova M., Sirocky M.~M., Ivezi{\'c} {\v Z}., Elitzur M., 2008, ApJ, 685, 147 

\bibitem[\protect\citeauthoryear{Nenkova et al.}{2008b}]{nenkova08b} 
Nenkova M., Sirocky M.~M., Nikutta R., Ivezi{\'c} {\v Z}., Elitzur M., 2008, ApJ, 685, 160 

\bibitem[\protect\citeauthoryear{Ness et al.}{2010}]{xmmhb} 
Ness, J.-U., (Revision editor) \& ESA: \xmm\ SOC\ 2010, ``\xmm\ Users Handbook'', Issue 2.8.1 \\
http://xmm.esac.esa.int/external/xmm\_user\_support/\\
documentation/uhb/XMM\_UHB.html

\bibitem[\protect\citeauthoryear{Park et al.}{2012}]{park12} 
Park D., et al., 2012, ApJ, 747, 30 

\bibitem[\protect\citeauthoryear{Pier \& Krolik}{1992}]{pier92} 
Pier E.~A., Krolik J.~H., 1992, ApJ, 401, 99 

\bibitem[\protect\citeauthoryear{Piconcelli et al.}{2005}]{piconcelli05} 
Piconcelli E., Jimenez-Bail{\'o}n E., Guainazzi M., Schartel N., Rodr{\'{\i}}guez-Pascual P.~M., 
Santos-Lle{\'o} M., 2005, A\&A, 432, 15 

\bibitem[\protect\citeauthoryear{Polletta et al.}{2007}]{polletta07} 
Polletta M., et al., 2007, ApJ, 663, 81

\bibitem[\protect\citeauthoryear{Pozzi et al.}{2007}]{pozzi07} 
Pozzi F., et al., 2007, A\&A, 468, 603 

\bibitem[\protect\citeauthoryear{Pozzi et al.}{2010}]{pozzi10} 
Pozzi F., et al., 2010, A\&A, 517, A11 

\bibitem[\protect\citeauthoryear{R Core Team}{2013}]{rstat} 
R Core Team, 2013, ``R: A Language and Environment for Statistical Computing'', R Foundation for Statistical Computing, Vienna, Austria;
http://www.R-project.org/

\bibitem[\protect\citeauthoryear{Salpeter}{1964}]{salpeter64} 
Salpeter E.~E., 1964, ApJ, 140, 796 

\bibitem[\protect\citeauthoryear{Schartmann et al.}{2008}]{schartmann08} 
Schartmann M., Meisenheimer K., Camenzind M., Wolf S., Tristram K.~R.~W., Henning T., 2008, A\&A, 482, 67 

\bibitem[\protect\citeauthoryear{Severgnini et al.}{2006}]{severgnini06} 
Severgnini P., et al., 2006, A\&A, 451, 859 

\bibitem[\protect\citeauthoryear{Shen et al.}{2011}]{shen11} 
Shen Y., et al., 2011, ApJS, 194, 45 

\bibitem[\protect\citeauthoryear{Silva et al.}{1998}]{silva98} 
Silva L., Granato G.~L., Bressan A., Danese L., 1998, ApJ, 509, 103 

\bibitem[\protect\citeauthoryear{Spinoglio \& Malkan}{1989}]{spinoglio89} 
Spinoglio L., Malkan M.~A., 1989, ApJ, 342, 83 

\bibitem[\protect\citeauthoryear{Stalin et al.}{2010}]{stalin10} 
Stalin C.~S., Petitjean P., Srianand R., Fox A.~J., Coppolani F., Schwope A., 2010, MNRAS, 401, 294 

\bibitem[\protect\citeauthoryear{Str{\"u}der et al.}{2001}]{pn} 
Str{\"u}der L., et al., 2001, A\&A, 365, L18

\bibitem[\protect\citeauthoryear{Sun \& Malkan}{1989}]{sun89} 
Sun W.-H., Malkan M.~A., 1989, ApJ, 346, 68 

\bibitem[\protect\citeauthoryear{Tacconi et al.}{1994}]{tacconi94} 
Tacconi L.~J., Genzel R., Blietz M., Cameron M., Harris A.~I., Madden S., 1994, ApJ, 426, L77 

\bibitem[\protect\citeauthoryear{Tacconi}{1996}]{tacconi96} 
Tacconi L.~J., 1996, IAUS, 178, 489 

\bibitem[\protect\citeauthoryear{Trouille et al.}{2009}]{trouille09} 
Trouille L., Barger A.~J., Cowie L.~L., Yang Y., Mushotzky R.~F., 2009, ApJ, 703, 2160 

\bibitem[\protect\citeauthoryear{Turner et al.}{2001}]{mos} 
Turner M.~J.~L., et al., 2001, A\&A, 365, L27 

\bibitem[\protect\citeauthoryear{Vasudevan et al.}{2010}]{vasudevan10} 
Vasudevan R.~V., Fabian A.~C., Gandhi P., 
Winter L.~M., Mushotzky R.~F., 2010, MNRAS, 402, 1081 

\bibitem[\protect\citeauthoryear{Watson et al.}{2001}]{watson01} 
Watson M.~G., et al., 2001, A\&A, 365, L51 

\bibitem[\protect\citeauthoryear{Wilms, Allen, \& McCray}{2000}]{wilms} 
Wilms J., Allen A., McCray R., 2000, ApJ, 542, 914 

\bibitem[\protect\citeauthoryear{Wright et al.}{2010}]{wise} 
Wright E.~L., et al., 2010, AJ, 140, 1868 

\bibitem[\protect\citeauthoryear{Zakamska et al.}{2006}]{zakamska06} 
Zakamska N.~L., et al., 2006, AJ, 132, 1496 

\end{thebibliography}
\end{document}